\documentclass{aa}
\usepackage{graphicx}
\usepackage{bm,natbib,url,wasysym,color}
\usepackage[varg]{txfonts}
\bibpunct{(}{)}{;}{a}{}{,} 
\graphicspath{{./fig/}{./png/}}

\graphicspath{{./figs/}{./png/}}

\newcommand{\EQ}{\begin{equation}}
\newcommand{\EN}{\end{equation}}
\newcommand{\EQA}{\begin{eqnarray}}
\newcommand{\ENA}{\end{eqnarray}}

\newcommand{\Eq}[1]{Equation~(\ref{#1})}

\newcommand{\Eqss}[2]{Equations~(\ref{#1})--(\ref{#2})}

\newcommand{\Sec}[1]{Section~\ref{#1}}

\newcommand{\Fig}[1]{Fig.~\ref{#1}}

\newcommand{\Figs}[2]{Figs.~\ref{#1} and \ref{#2}}

\newcommand{\Tab}[1]{Table~\ref{#1}}

\newcommand{\bra}[1]{\langle #1\rangle}

\newcommand{\fluc}[1]{#1^\prime}
\newcommand{\meanf}[1]{\overline #1{}^{\rm fil}}
\newcommand{\mean}[1]{\overline #1}
\newcommand{\meanrho}{\overline{\rho}}

{}
{}
{}

{}
{}
{}
{}
{}
{}
{}
{}
{}
\newcommand{\meanBB}{\overline{\mbox{\boldmath $B$}}{}}{}
{}
{}
{}
{}
{}
{}
{}
{}

{}
{}
{}

{}

{}
{}

\newcommand{\gggg}{\mbox{\boldmath $g$} {}}

\newcommand{\PP}{\mbox{\boldmath $\Pi$} {}}

\newcommand{\uu}{\mbox{\boldmath $u$} {}}
\newcommand{\UU}{\mbox{\boldmath $U$} {}}

\newcommand{\bb}{\mbox{\boldmath $b$} {}}
\newcommand{\BB}{\mbox{\boldmath $B$} {}}

\newcommand{\JJ}{\mbox{\boldmath $J$} {}}

\newcommand{\AAA}{\mbox{\boldmath $A$} {}}

\newcommand{\ff}{\mbox{\boldmath $f$} {}}

\newcommand{\nab}{\mbox{\boldmath $\nabla$} {}}

\newcommand{\SSSS}{\mbox{\boldmath ${\sf S}$} {}}

\newcommand{\erf}{{\rm erf}}

\newcommand{\dd}{{\rm d} {}}

\def\Pm{\mbox{\rm Pr}_{\rm M}}
\def\Rm{\mbox{\rm Re}_{\rm M}}
\def\Peff{{\cal P}_{\rm eff}}

\def\Rey{\mbox{\rm Re}}

\def\cs{c_{\rm s}}
\def\csq{c_{\rm s}^2}

\def\vA{v_{\rm A}}

\def\kf{k_{\rm f}}
\def\kc{k_{\rm c}}

\def\urms{u_{\rm rms}}

\def\etatz{\eta_{\rm t0}}

\def\Beq{B_{\rm eq}}
\def\Beqz{B_{\rm eq0}}

\def\half{{\textstyle{1\over2}}}

\def\onethird{{\textstyle{1\over3}}}

\newcommand{\G}{\,{\rm G}}

\newcommand{\kG}{\,{\rm kG}}

\newcommand{\kms}{\,{\rm km/s}}

\def\tautd{\tau_{\rm td}}
\def\tautdm{\tilde{t}_{\rm max}}
\def\Bfm{\overline{B}{}_z^{\rm fil\,max}}
\def\Hr{H_{\rho}}
\begin{document}

\titlerunning{Bipolar regions in a two-layer model}

\authorrunning{Warnecke et al.}
\title{Bipolar region formation in stratified two-layer turbulence}

\author{J. Warnecke\inst{1,2} \and I. R. Losada \inst{2,3}
  \and A. Brandenburg \inst{2,3,4,5} \and N. Kleeorin \inst{6,2}
  \and I. Rogachevskii \inst{6,2}}
\institute{Max-Planck-Institut für Sonnensystemforschung,
  Justus-von-Liebig-Weg 3, D-37077 G\"ottingen, Germany\\
\email{warnecke@mps.mpg.de}\label{inst1}
\and Nordita, KTH Royal Institute of Technology and Stockholm University,
Roslagstullsbacken 23, SE-10691 Stockholm, Sweden \label{inst2}
\and Department of Astronomy, AlbaNova University Center,
Stockholm University, SE-10691 Stockholm, Sweden \label{inst3}
\and JILA and Department of Astrophysical and Planetary Sciences,
University of Colorado, Boulder, CO 80303, USA \label{inst4}
\and Laboratory for Atmospheric and Space Physics,
University of Colorado, Boulder, CO 80303, USA \label{inst5}
\and Department of Mechanical
Engineering, Ben-Gurion University of the Negev, POB 653,
Beer-Sheva 84105, Israel \label{inst6}}

\date{Received 12 February  2015/ Accepted 14 February 2016}
\abstract{}{%
This work presents an extensive study of the previously discovered
formation of bipolar flux concentrations in a two-layer model.
We interpret the formation process in terms of negative effective magnetic pressure
instability (NEMPI), which is a possible mechanism to explain
the origin of sunspots.
}{%
In our simulations, we use a Cartesian domain of isothermal stratified
gas that is divided into two layers.
In the lower layer, turbulence is forced with
transverse nonhelical random waves, whereas in the upper layer no flow is
induced.
A weak uniform magnetic field is imposed in the entire domain at all times.
In most cases, it is horizontal, but  a vertical and an inclined field
are also considered.
In this study we vary the stratification by changing the
gravitational acceleration, magnetic Reynolds number,
strength of the imposed magnetic field, and size of the domain to
investigate their influence on the formation process.
}{%
Bipolar magnetic structure formation takes place over a large range of parameters.
The magnetic structures become more intense for higher
stratification until the density contrast becomes around $100$ across
the turbulent layer.
For the fluid Reynolds numbers considered, magnetic flux concentrations
are generated at magnetic Prandtl number between 0.1 and 1.
The magnetic field in bipolar regions increases with higher imposed
field strength until the field becomes comparable to the
equipartition field strength of the turbulence.
A larger horizontal extent enables the flux concentrations to become
stronger and more coherent.
The size of the bipolar structures turns out to be independent of the
domain size.
A small imposed horizontal field component is necessary to generate
bipolar structures.
In the case of bipolar region formation, we find an exponential
growth of the large-scale magnetic field, which is indicative of a
hydromagnetic instability.
Additionally, the flux concentrations are correlated with strong large-scale
downward and converging flows.
These findings imply that NEMPI is responsible for magnetic
flux concentrations.}
{}

\maketitle

\section{Introduction}

One of the main manifestations of solar activity is the occurrence
of sunspots on the surface of the Sun, showing cyclic behavior with
a period of 11 years.
Sunspots are concentrations
of strong magnetic field suppressing the
convective heat transport from the interior of the Sun to its surface.
This causes sunspots to be cooler and to appear darker on the solar
disk.
Sunspots were observed and counted by Galileo Galilei more than
400 years ago, and their magnetic origin was discovered by
\cite{Hale:1908} over 100 years ago.
However, the formation mechanism of sunspots is still
the subject of active discussions and investigations.

For a long time it was believed that the solar dynamo produces strong
magnetic fields at the bottom of the convection zone \citep{Par75,SW80,GW81}.
At this location, called the tachocline \citep{SZ92}, there is
a strong shear layer \citep{Schouea98} that might be able to produce a strong toroidal magnetic field.
This field is believed to become unstable and rise
upward in the form of flux tubes, which reach the surface to form
bipolar structures, including sunspot pairs \citep[e.g.,][]{CMS95}.
However, this picture has been questioned.
Global simulations of self-consistent convectively driven dynamos are
able to produce strong magnetic fields without the presence of a
tachocline \citep[e.g.,][]{Racine11,KMB12,ABMT14,KKOBWKP15}.
These simulations are also able to reproduce the equatorward migration
of the toroidal field as observed in the Sun.
The magnetic field
is strongest in the middle of the convection zone and
propagates from there both toward the surface and the bottom of the convection
zone \citep{KMCWB13}.
Furthermore, \cite{WKKB14} found that the equatorward migration
occurring in their global simulations of self-consistent convectively
driven dynamos can be explained entirely by the Parker-Yoshimura rule
\citep{P55,Yos75} of a propagating $\alpha \, \Omega$ dynamo wave,
where $\alpha$ is related to the kinetic helicity and $\Omega$ is the
local rotation rate of the Sun.
With a positive $\alpha$, the radial gradient of $\Omega$ has to be
negative for equatorward migration to occur.
The Parker-Yoshimura rule was also recently verified for these
simulations using $\alpha_{\phi\phi}$ determined with the test-field
method \citep{WRKKB16}.
In the Sun, $\dd \Omega/\dd r$ is negative in the near-surface shear
layer \citep{Thometal96,BSG14}.
This suggests that  in the Sun the toroidal field can also be generated in the upper
layers of the convection zone owing to the near-surface shear
\citep{B05}.
Additionally, the magnetic field, if generated at the bottom of the convection
zone, might become unstable at field strengths of around $1\kG$
\citep{ASF07,ASR07}. 
This instability would occur  much before the magnetic field is
amplified to $10^5\G$, which is needed for a coherent flux tube to
reach the surface without strong distortion
\citep{CG87,DC93}.
The generation of strong coherent magnetic flux tubes
has not yet been seen in self-consistent dynamo simulations
\citep{GK11}.
What has been seen, however, are flux tubes that appear in hydromagnetic
turbulence \citep{NBJRRST92,BJNRST96}, analogously to vortex tubes in
hydrodynamic turbulence \citep{SJO90}.
They appear as short strands when visualized through field vectors at
places where the field exceeds a certain threshold, but can display
a serpentine tube-like structure when visualized as field lines regardless
of the local field strength \citep{NBBMT11,FF14}.
Furthermore, the flux bundles found in these two papers rise because of
a combination of advection and magnetic buoyancy.
Given their size and further expansion when ascending to the surface,
their role in sunspot formation remains inconclusive.
An alternative to producing spots in a global dynamo simulation of rapidly
rotating stars was found by \cite{YGCR15}.
 These authors were able to generate a single polar spot without
the help of rising tubes.
However, the simulations began with a large-scale dipolar field, which might
have contributed to the formation process.

Results from helioseismology concerning the importance of the
tachocline in the global dynamo do not support
a deeply rooted flux tube scenario in that
the shear at the bottom of the convection zone
has not shown the periodic variations found in the bulk of
convection zone \citep{Howe00,AB11}, where the period is the same as that of
the activity cycle of the Sun \citep[see, e.g.,][]{Howe09}.
One would expect that a strong magnetic field generated
in the tachocline would also backreact on the differential rotation.
Furthermore, no signs of rising flux tubes have yet been found
in helioseismology.
\cite{BBF10} computed the expected signatures and observational
limits of detecting the retrograde motion from the rising flux tube
model of \cite{Fan08}.
\cite{BBLK13} were unable to detect any signatures larger than 20 $\kms$.
However, they could exclude models of \cite{Cheung_etal10} and \cite{RC14}, but
other rising flux tube models might still be possible.
From statistical studies of emerging active regions, \cite{KS08} and
\cite{SK12} conclude that the tilt angle of bipolar regions with respect to the
east--west direction (Joy's law) evolves after the emergence occurs and is,
therefore, unlikely to be caused by the Coriolis force acting on a rising flux
tube.

If the toroidal magnetic field of the Sun is generated throughout the
convection zone, it is reasonable to assume that
there is a local mechanism that forms magnetic flux concentrations,
which then leads to sunspots seen at the solar surface.
\cite{SN12} identify the convective downward flows
associated with the supergranulation as one such location
where magnetic flux can be concentrated self-consistently; this causes
the formation of bipolar magnetic
structures of the size of pores.

Another possible mechanism is the negative effective
magnetic pressure instability (NEMPI).
In this instability,
the total (hydrodynamic plus magnetic) turbulent pressure is
reduced by a large-scale magnetic field so that the effective large-scale magnetic pressure
(the sum of turbulent and nonturbulent contributions) becomes negative.
This causes the surrounding plasma to flow into regions of low gas
pressure, which leads to downflows and vertical fields
that are concentrated further. 
This enhances the suppression of turbulent pressure, which results in the excitation of a large-scale magnetohydrodynamic
instability (NEMPI) and the formation of large-scale magnetic flux concentrations.
The original idea goes back to early work by
\cite{Kleeorin89,KRR90}, and has been established in theoretical
\citep{KMR93,KMR96,KR94,RK07} and numerical studies
\citep{BKR10,BKKMR11,BKKR12,LBKMR12,LBKR13,LBKR14,JBKMR13,JBLKR14,JBKMR15}.

The first  magnetic flux concentrations
of superequipartition strength produced by NEMPI were unipolar spots in the presence
of an imposed vertical field \citep{BKR13,BGJKR14}.
\cite{WLBKR13} were for the first time able to produce bipolar magnetic
regions with NEMPI using a two-layer setup
with a weak imposed horizontal magnetic field.
Turbulence is driven by a forcing function within the lower layer,
while in the upper unforced layer, called the coronal envelope, all motions
are a consequence of overshooting and magnetic field tension.
This approach was developed by \cite{WB10} and was used to produce
dynamo-driven coronal ejections \citep{WBM11,WBM12,WKMB12}.
These studies suggest that the dynamo operating in a two-layer model
becomes stronger and more easily excited than that in a one-layer model
\citep{WB14}.
Furthermore, in global simulations of a convectively driven dynamo, the
presence of a coronal layer on top of the convection zone leads to
spoke-like differential rotation together with a near-surface shear layer
\citep{WKMB13,WKKB15}, instead of otherwise mainly cylindrical contours
of angular velocity.

\cite{MBKR14} use a different two-layer setup in which turbulence is
present in both layers, but in the lower layer it is
driven helically, leading to large-scale dynamo action, while
in the upper layer, it is driven nonhelically.
This spatially separates  the dynamo from the formation of
magnetic flux concentrations.
With this setup, they were able to produce intense bipolar structures.
Recently, bipolar structures have also been studied in a similar
setup of spherical shells \citep{JBKMR15}.

In the present work, we extend the studies of \cite{WLBKR13}
concerning the detailed dependence on
density stratification (\Sec{sec:strat}), magnetic
Reynolds number (\Sec{sec:rm}), imposed magnetic field strength
(\Sec{sec:impB}), size of the computational domain
(\Sec{sec:size}),
and magnetic field inclination (\Sec{sec:incl})
to investigate and classify the formation mechanisms
of bipolar magnetic regions (\Sec{sec:form}).

\section{Model}
\label{model}

\begin{table*}[t!]\caption{
Summary of runs.
}\vspace{12pt}\centerline{\begin{tabular}{lccccccccccccc}
Run & Resolution  & Size & $g H_\rho/\csq$ & $\!\!{\rho_{\rm bot}/\rho_{\rm
  surf}}\!\!$ &$\Rey$& $\Pm$ & $\Beqz/B_0$ & $\Peff^{\rm min}$  & $B_z^{\rm
  max}/B_0$& $\Bfm/B_0$
&$\tautdm$&BR\\
\hline
\hline
A1 & $512^2\times1024$ & $(2\pi)^2\times3\pi$ & 0.1 & 1.4 & 38.0 & 0.5 & 40
&-0.021& 39 & 1.6 &-&NO\\
A2 & $512^2\times1024$  & $(2\pi)^2\times3\pi$ & 0.5 & 4.8 & 38.0 & 0.5 & 41
&-0.023&52&3.8&2.1&WEAK\\
A3 & $512^2\times1024$  & $(2\pi)^2\times3\pi$ & 0.7 & 8.9 & 38.1 & 0.5 & 42
&-0.026&56&5.4&1.9&YES\\
A4  &$512^2\times1024$  & $(2\pi)^2\times3\pi$ & 0.85 & 14 & 38.1 & 0.5 & 42
& -0.020&56&5.5&1.6&YES\\
A5  &$512^2\times1024$  & $(2\pi)^2\times3\pi$ & 1.00 & 23 & 38.2 & 0.5 & 43
&-0.022&67&9.2&1.1&YES\\
A6  & $512^2\times1024$ & $(2\pi)^2\times3\pi$& 1.20 & 42 & 38.4 & 0.5 & 44
&-0.023&74&8.1&1.2&YES\\
A7  & $512^2\times1024$ &$(2\pi)^2\times3\pi$ & 1.40 & 79 & 38.6 & 0.5 & 46
& -0.017&72&8.8&1.1&YES\\
A8  & $512^2\times1024$ &$(2\pi)^2\times3\pi$ & 1.50 & 108 & 38.7 & 0.5 & 48
& -0.017 & 58 & 6.6&0.7&YES\\
\hline
R1 & $512^2\times1024$ & $(2\pi)^2\times3\pi$ & 1.00 & 23 & 38.3 & 0.0625 & 43
&-0.018&8.7&2.9&-&NO\\
R2 & $512^2\times1024$ & $(2\pi)^2\times3\pi$ & 1.00 & 23 & 38.3 & 0.125 & 43
&-0.015&19&4.4&1.8&WEAK\\
R3 & $512^2\times1024$  & $(2\pi)^2\times3\pi$ & 1.00 & 23 & 38.3 & 0.25 & 43
&-0.020&31&6.0&1.5&YES\\
R4 &$512^2\times1024$ & $(2\pi)^2\times3\pi$ & 1.00 & 23 & 38.2 & 0.5 & 43
&-0.022&67&9.2&1.1&YES\\
R5 & $512^2\times1024$  & $(2\pi)^2\times3\pi$ & 1.00 & 23 & 35.7 & 1 & 40
&-0.028&91&4.3&2.0&WEAK\\
\hline
B1 &$512^2\times1024$ & $(2\pi)^2\times3\pi$ & 1.00 & 23 & 38.3 & 0.5 & 431
&-0.019&202&13&2.6&WEAK\\
B2 &$512^2\times1024$ & $(2\pi)^2\times3\pi$ & 1.00 & 23 & 38.3 & 0.5 & 173
&-0.023&138&14&3.6&YES\\
B3 &$512^2\times1024$  & $(2\pi)^2\times3\pi$ & 1.00 & 23 & 38.3 & 0.5 & 86
&-0.022&92&6.1&2.3&YES\\
B4 &$512^2\times1024$ & $(2\pi)^2\times3\pi$ & 1.00 & 23 & 38.2 & 0.5 & 43
&-0.022&67&9.2&1.1&YES\\
B5 & $512^2\times1024$  & $(2\pi)^2\times3\pi$ & 1.00 & 23 & 38.1 & 0.5 & 17
&-0.030&30&4.8&0.9&YES\\
B6 & $512^2\times1024$  & $(2\pi)^2\times3\pi$ & 1.00 & 23 & 37.7 & 0.5 & 8.5
&-0.059&16&3.2&0.8&YES\\
B7 & $512^2\times1024$ & $(2\pi)^2\times3\pi$ & 1.00 & 23 & 36.1 & 0.5 & 1.6
&-0.125&3.3 &0.2&-&NO\\
\hline
S1  &$512^3$ & $(2\pi)^2\times2\pi$ & 1.00 & 23 & 38.2 & 0.5 & 42
& -0.030&52&7.8&1.0&YES\\
S2  &$512^2\times1024$& $(2\pi)^2\times3\pi$& 1.00& 512& 38.9& 0.5 &  49
&-0.024&57&4.9&1.1&YES\\
S3 & $1024^3$ & $(4\pi)^2\times3\pi$ & 1.00 & 23 & 38.2 & 0.5 & 43
&-0.013&80&17&1.8&YES\\
\hline
THW  &$512^2\times1024$  & $(2\pi)^2\times3\pi$ & 1.00 & 23 & 38.2 & 0.5 & 45
&-0.022&56&5.3&0.6&YES\\
V  &$512^2\times1024$  & $(2\pi)^2\times3\pi$ & 1.00 & 23 & 38.1 & 0.5 & 43
&-0.021&107&30&3.9&SP\\
INC  &$512^2\times1024$  & $(2\pi)^2\times3\pi$ & 1.00 & 23 & 38.2 & 0.5 & 43
&-0.022&86&21&3.9&YES\\
\hline
F  &$512^2\times1024$  & $(2\pi)^2\times3\pi$ & 1.00 & 23 & 38.7 & 0.5 & 45
&-0.049&50&10&1.7&YES\\
\hline
\label{runs}\end{tabular}}\tablefoot{
Here, $g H_\rho/\csq$ is the normalized gravitational acceleration and
$\rho_{\rm bot}$ and $\rho_{\rm surf}$ are the horizontally averaged
densities at the bottom
and surface ($z=0$) of the domain, respectively.
$\Rey$ is the fluid Reynolds number, $\Pm$ is the magnetic Prandtl
number, $B_0$ is the imposed field, $\Beqz=\Beq$($z=0$) is the equipartition value
at the surface ($z=0$), and
$\Peff^{\rm min}$ is the minimum of the averaged effective magnetic
pressure $\Peff$ defined by \Eq{peff}; see also bottom row of
\Fig{strat}.
$B_z^{\rm max}$ is the maximum value of vertical magnetic field,
$\Bfm$ is the maximum value of the Fourier-filtered
vertical magnetic field; both are taken at the surface ($z=0$).
$\tautdm$ is the time when $\Bfm$ is taken in terms of
turbulent-diffusive time.
BR indicates whether or not there are bipolar regions or a single spot
(SP).
The runs~R4 and B4 are the same as A5.
}
\end{table*}

The model is essentially the same as that of \cite{WLBKR13}, but in this
work we vary the stratification, the imposed magnetic field, and the magnetic Reynolds number.
We use a Cartesian domain ($x,y,z$),
which has the size $L_x \times L_y \times L_z$,
where $L_x=L_y=2\pi$ and $L_z=3\pi$, except for Runs~S1
(where $L_z=2\pi$) and S3 (where $L_x=L_y=4\pi$).
We solve the magnetohydrodynamic equations in the presence
of vertical gravity $\gggg=(0,0,-g)$.
We apply the two-layer model of \cite{WB10},
which consists of a turbulent lower layer ($z\le 0$)
and a laminar upper layer ($z\ge0$), which is referred to as coronal envelope.
The extent of the turbulent layer is $-\pi\le z\le0$, except for Run~S2,
where it is $-2\pi\le z\le0$.
The main difference between these two layers is
the presence of the forcing function $\ff(x,y,z,t)$ in the lower layer, which is
called the turbulent layer.
For a smooth transition between the two layers, we apply a modulation
of the forcing function similar to \cite{WB10},
\begin{equation}
\theta_w(z)=\half\left(1-\erf{z\over w}\right),
\end{equation}
where $w$ is the width of the transition, which is chosen to be $0.05$
for all runs except Run~THW, where $w=0.02$.
We solve the compressible magnetohydrodynamic (MHD) equations
\begin{equation}
{D\uu\over D t}=\gggg +\theta_w(z)\ff
+{1\over\rho}\left[-\csq\nab\rho+\JJ\times\BB+\nab\cdot(2\nu\rho\SSSS)\right],
\end{equation}
\begin{equation}
{\partial\AAA\over\partial t}=\uu\times\BB+\eta\nabla^2\AAA,
\label{eq:ind}
\end{equation}
\begin{equation}
{D \ln\rho\over D t}=-\nab\cdot\uu,
\end{equation}
where $\rho$ is the density and $\cs$ is the sound speed, which is
constant in the entire domain. The convective
derivative is
$D/D t = \partial/\partial t + \uu\cdot \nab$.
The magnetic field is given by
$\BB=\BB_{\rm imp} +\nab\times\AAA$, where
$\BB_{\rm imp}=(0,B_0,0)$ is a weak uniform field in the
$y$ direction and $\BB$ is divergence free by construction.
For Run~V, we choose $\BB_{\rm imp}=(0,0,B_0)$ and for Run~INC
$\BB_{\rm imp}=(0,B_0,B_0)/\sqrt{2}$.
$B_0$ is kept constant during the simulation.
Here, $\JJ=\nab\times\BB/\mu_0$ is the current density, $\mu_0$ is the
vacuum permeability, $\nu$ is the kinematic viscosity, $\eta$ is the
magnetic diffusivity, 
\begin{equation}
{\sf S}_{ij}=\half(u_{i,j}+u_{j,i})-\onethird\delta_{ij}\nab\cdot\uu
\end{equation}
is the trace-free strain tensor,
and commas denote partial spatial differentiation.
For an isothermal equation of state, the pressure $p$ is related to the
density $\rho$ via $p=\csq\rho$.
The forcing function $\ff$ consists of random plane transverse
white-in-time, nonpolarized waves \citep[see][for details]{Hau04}.
The wavenumbers lie in a band around an average forcing number $\kf=30 \
k_1$, where $k_1=2\pi/L_x$ ($\kf=60 \ k_1$ for Run~S3) is the lowest
wavenumber possible in the domain.
The amplitude of the forcing is the same in all runs and is chosen to
yield a constant $\urms\approx 0.1\cs$ in the bulk of the turbulent layer,
where the rms velocity is defined as
\begin{equation}
 \urms=\bra{\uu^2}_{xy;z\le 0}^{1/2}\ ,
\end{equation}
and $\bra{.}_{xy}$ denotes a horizontal average and $\bra{.}_{z\le 0}$
denotes a vertical average over the turbulent layer ($z\le 0$).
We also use horizontal averaging to describe the mean of a
quantity, i.e.,\ $\bra{F}_{xy}=\mean{F}$.
However, to describe the large-scale field, we use a horizontal 2D
Fourier-filtered field with a cut-off wavenumber $\kc\le \kf/6$ and
use the notation $\meanf{F}$.
The density scale height $\Hr$ is chosen such that $k_1\Hr=1$
($k_1\Hr=2$ for Run~S3).

For classification and analysis, we use nondimensional and
dimensional numbers characterizing the physical properties of the MHD turbulence.
We define the fluid and magnetic Reynolds numbers of the system as
$\Rey\equiv\urms/\nu\kf$ and $\Rm\equiv\urms/\eta\kf$, respectively.
Therefore,
the magnetic Prandtl number is given by $\Pm\equiv\Rm/\Rey=\nu/\eta$.
To characterize the local strength of the magnetic field, we define an
equipartition field strength as
$\Beq(z)=(\mu_0\meanrho\,\overline{\uu^2})^{1/2}$, which is a function
of $z$,
or at the surface $\Beqz=\Beq(z=0)$.
Time is measured in turbulent-diffusive times,
$\tautd=\Hr^2/\etatz$, where $\etatz=\urms/3\kf$ is the estimated
turbulent diffusivity.
In the following we use units such that $\mu_0=1$.

We use horizontal periodic boundary conditions for all dependent
variables.
The top and bottom boundaries are stress-free
and the magnetic field is vertical.
The kinematic viscosity $\nu$ and magnetic diffusion $\eta$
are constant throughout the whole domain.
However, we employ higher values near the top boundary in high
stratification runs to stabilize the code, which becomes important
in regions of low density.
Except for Runs~S1 and S3, we apply a resolution of $512\times512\times1024$ grid points in
$x$, $y$, and $z$ directions; see second column of \Tab{runs}.
The difference from
the runs of \cite{WLBKR13} is that we double the
resolution and arithmetic precision to increase numerical accuracy.
The simulations are performed with the {\sc Pencil Code}%
\footnote{{\tt http://github.com/pencil-code/}},
which uses sixth-order explicit finite differences in space and a
third-order accurate time stepping method.

\begin{figure}[t!]
\begin{center}
\includegraphics[width=\columnwidth]{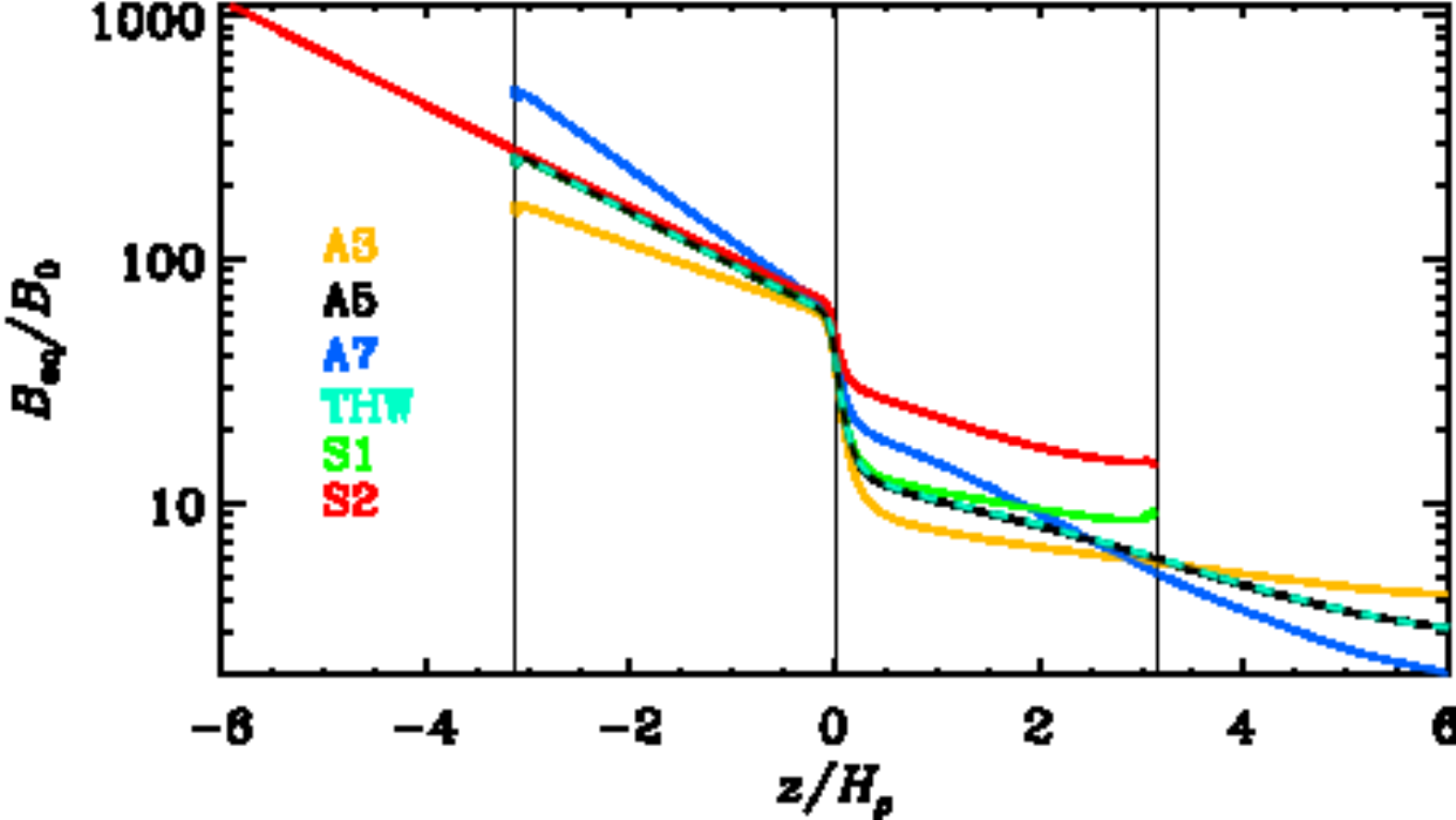}
\end{center}\caption[]{
Vertical profiles of equipartition magnetic field strengths $\Beq$ for
Runs~A3, A5, A7, THW,
S1, and S2 as a function of height $z/H_{\rho}$.
$\Beq$ is normalized by the imposed magnetic field $B_0$.
The vertical lines indicate $z=-\pi,\ 0, \ \pi$.
}\label{pbeq}
\end{figure}

\section{Results}

\begin{figure*}[t!]
\begin{center}
\includegraphics[width=0.33\textwidth]{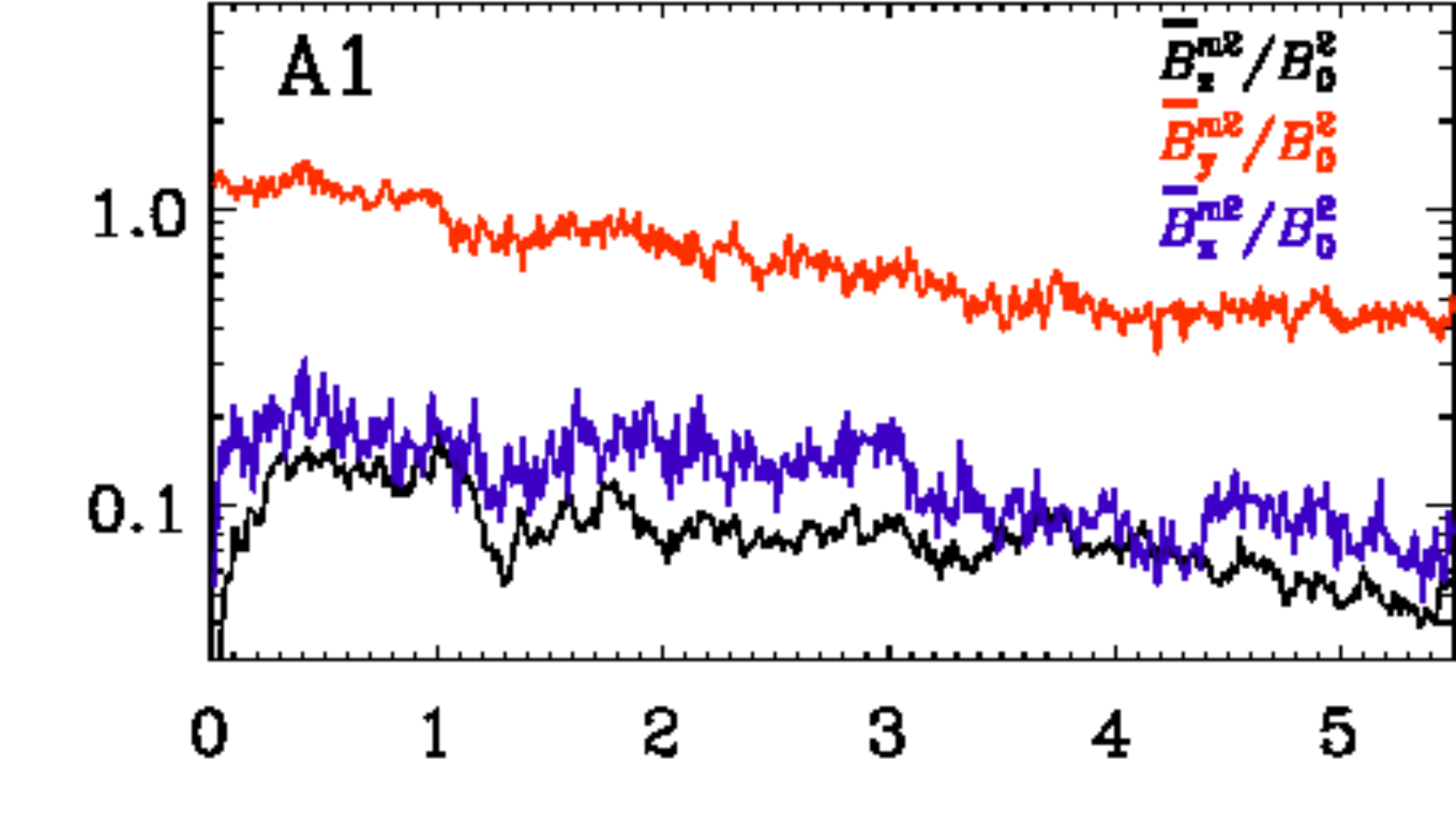}
\includegraphics[width=0.33\textwidth]{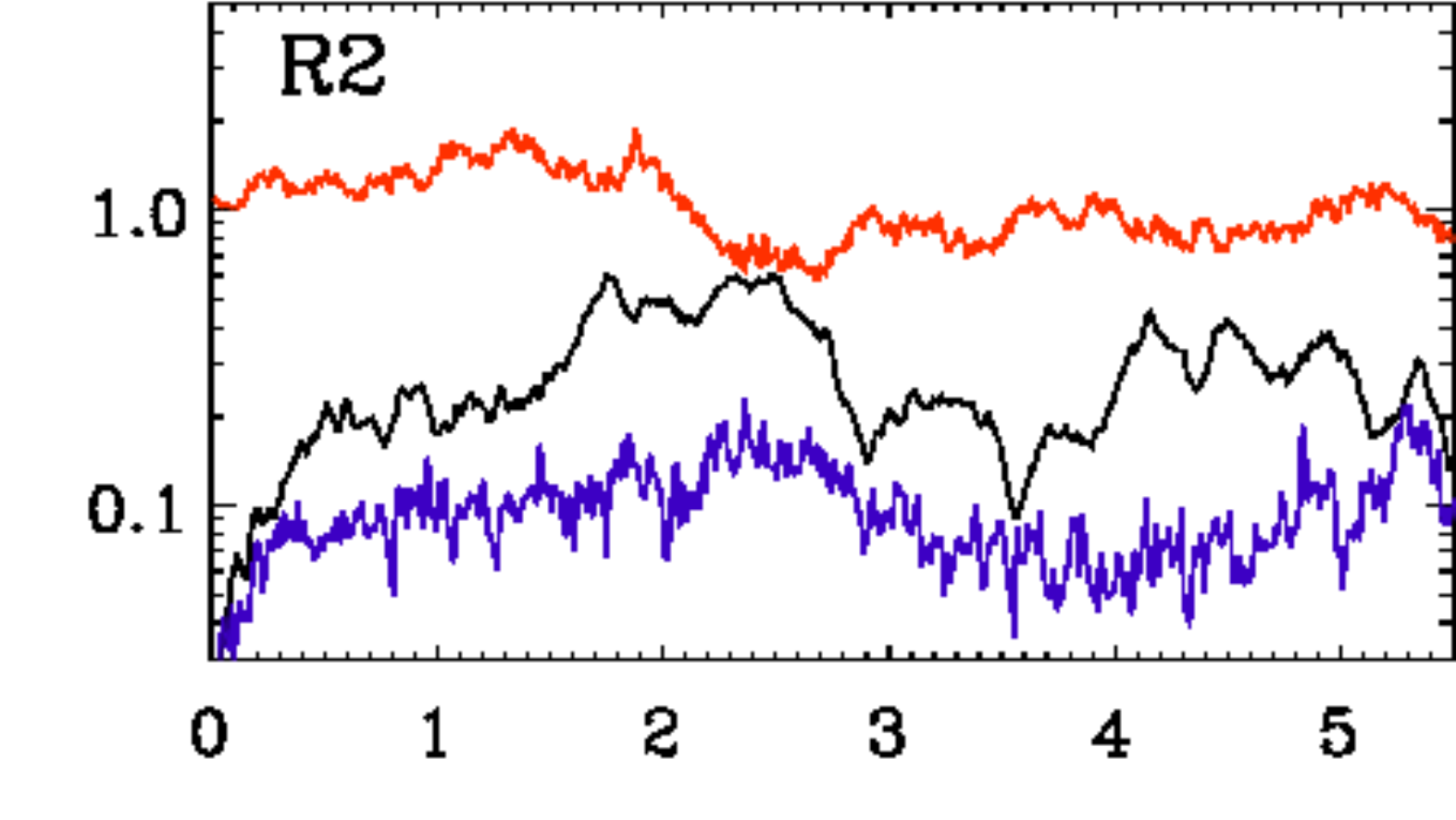}
\includegraphics[width=0.33\textwidth]{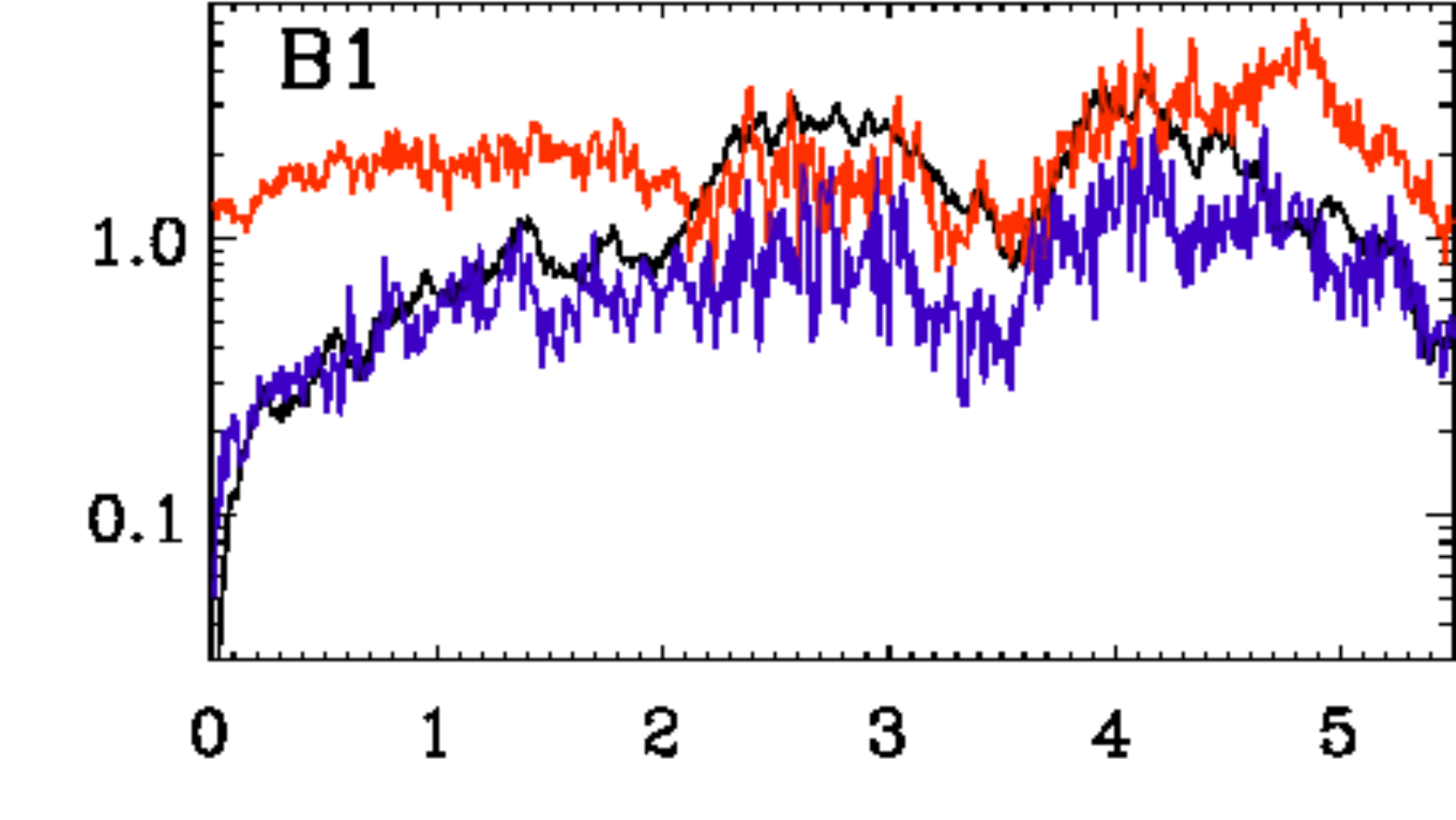}
\includegraphics[width=0.33\textwidth]{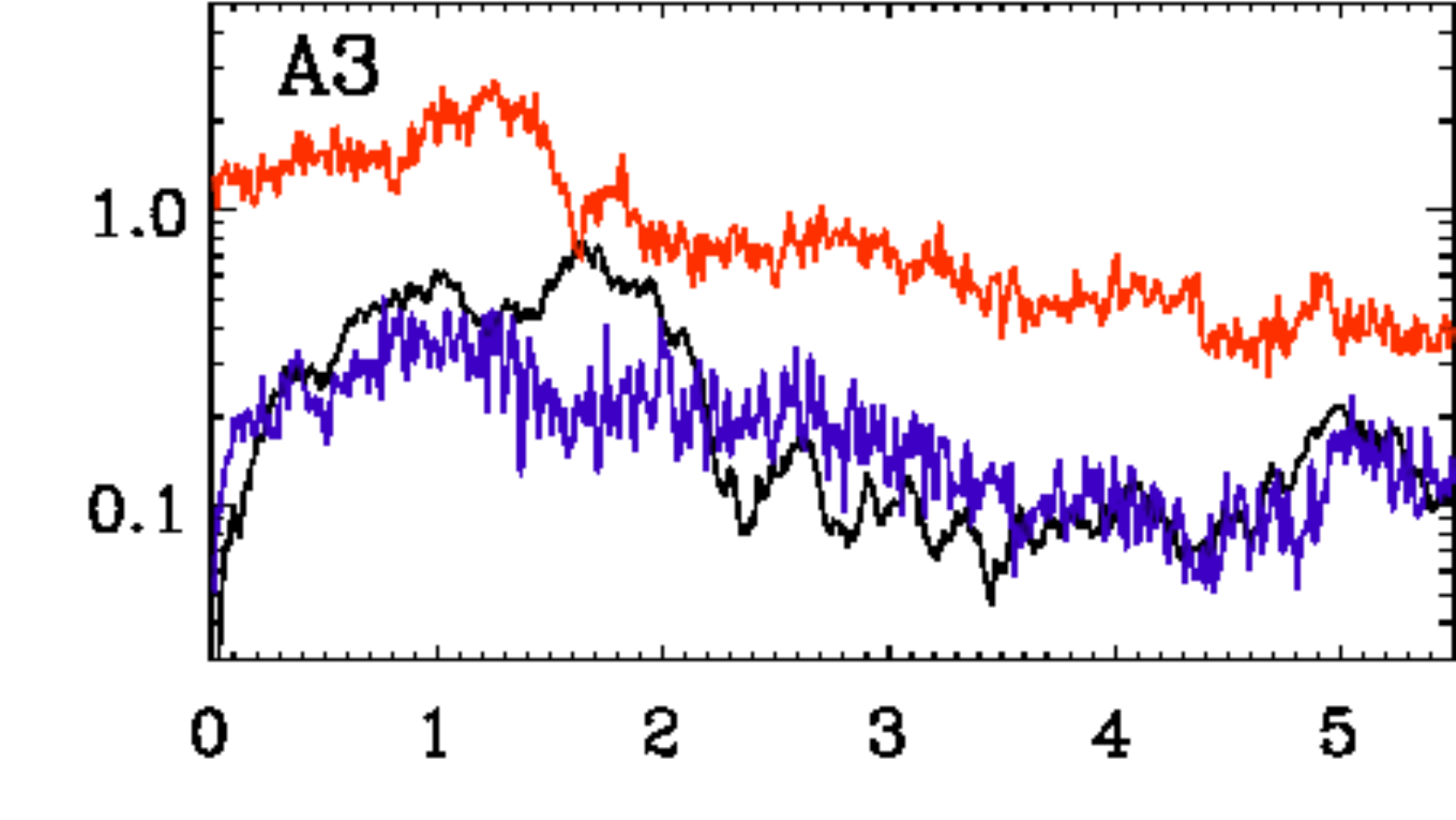}
\includegraphics[width=0.33\textwidth]{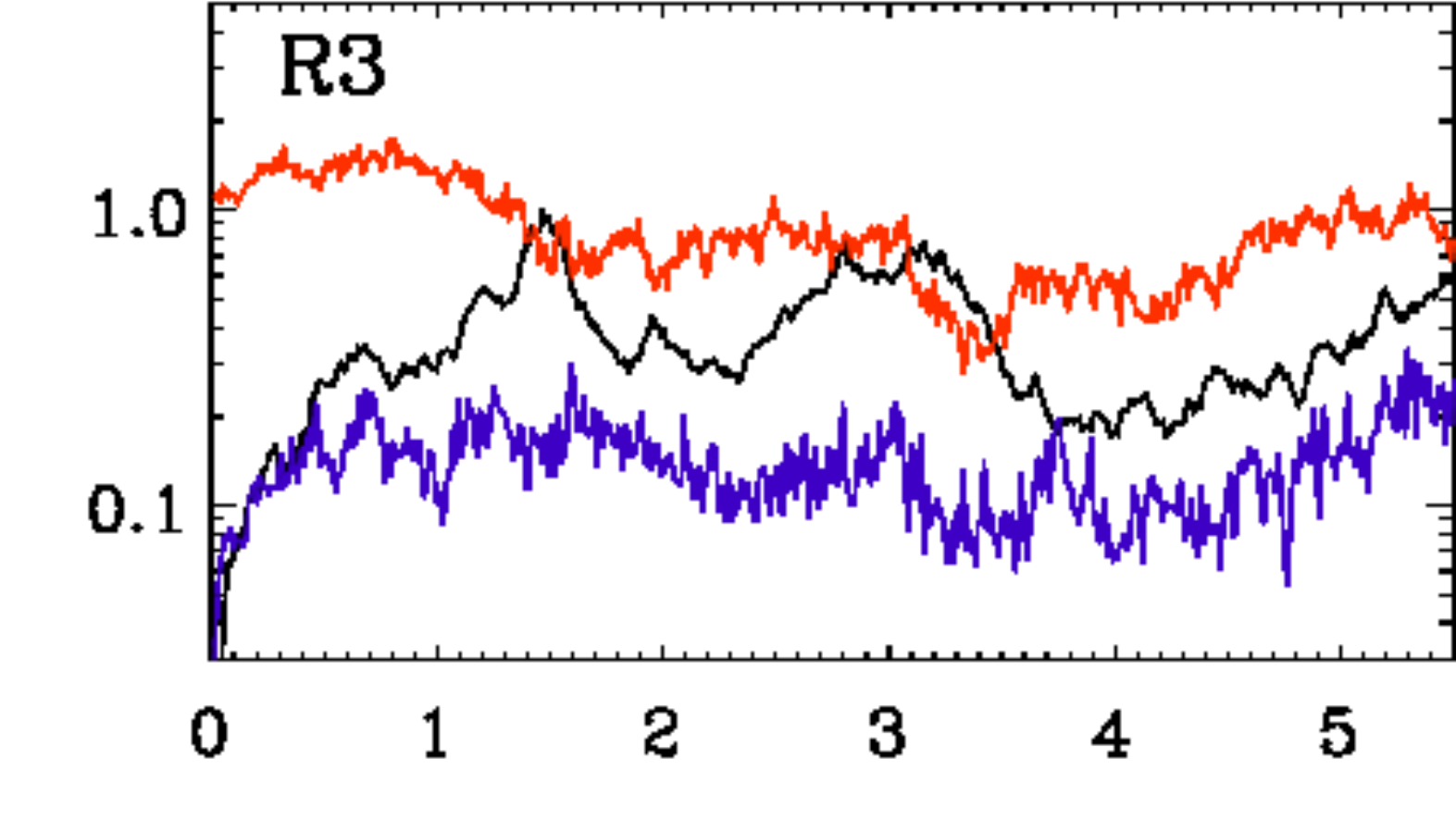}
\includegraphics[width=0.33\textwidth]{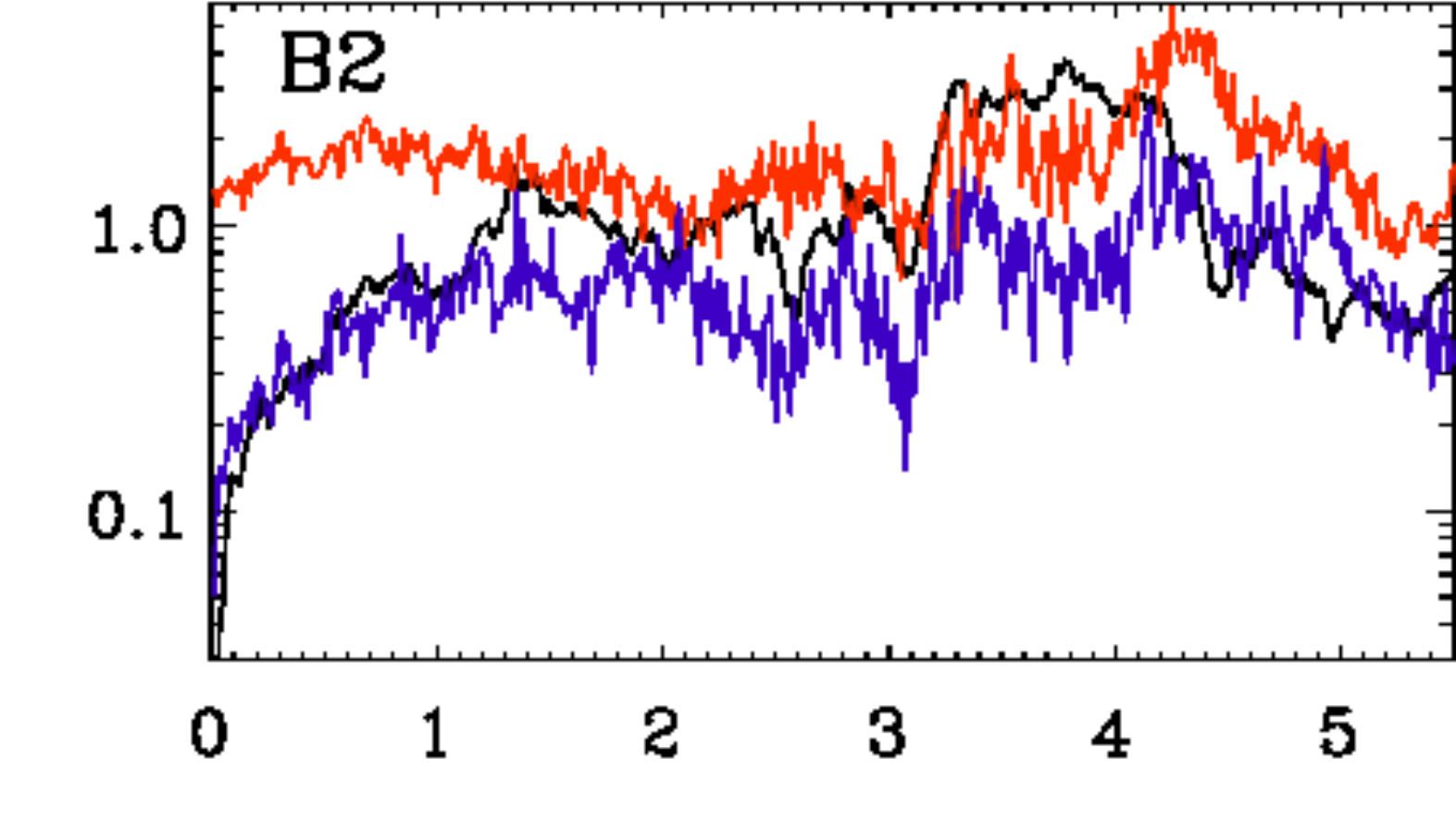}
\includegraphics[width=0.33\textwidth]{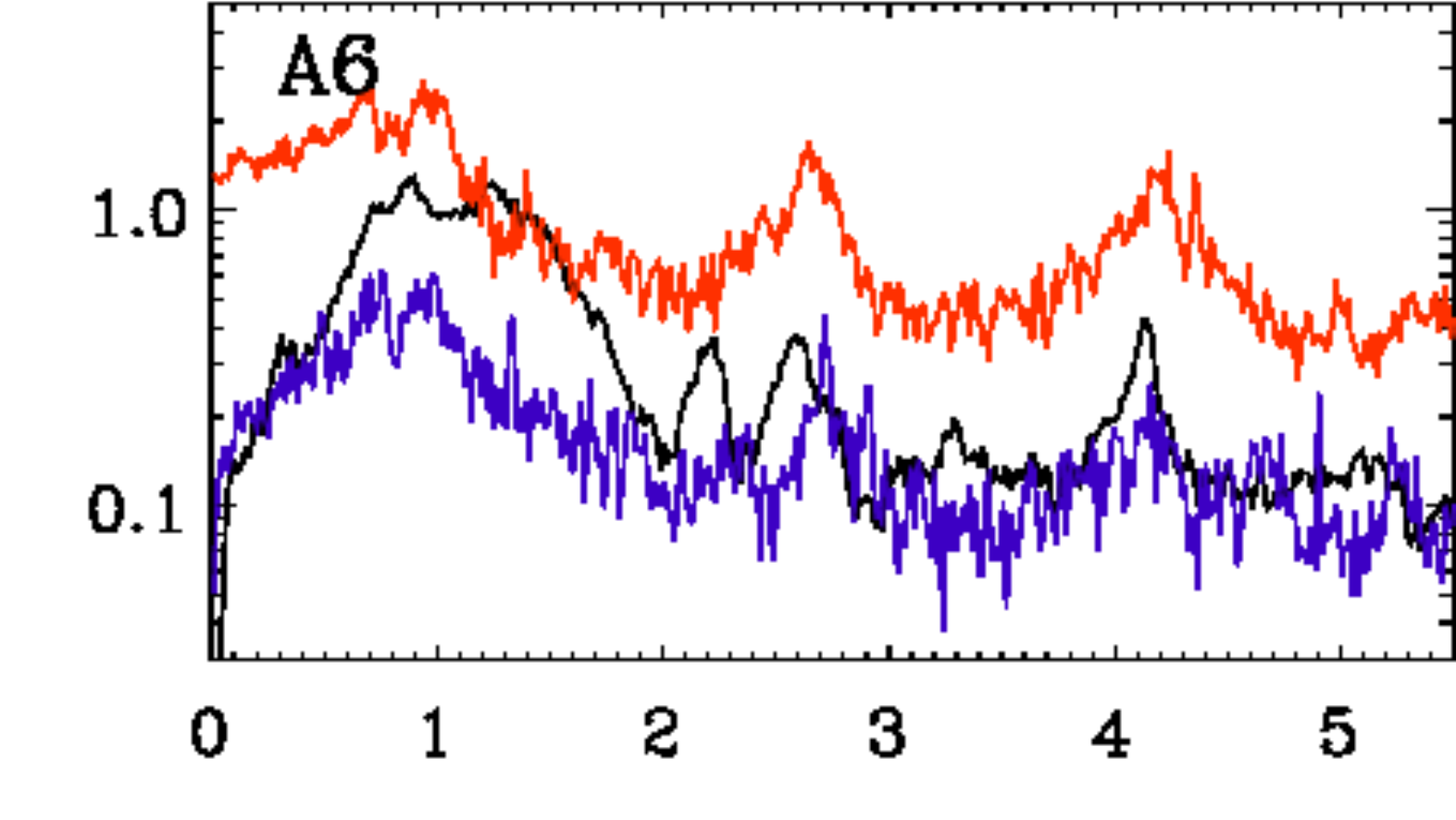}
\includegraphics[width=0.33\textwidth]{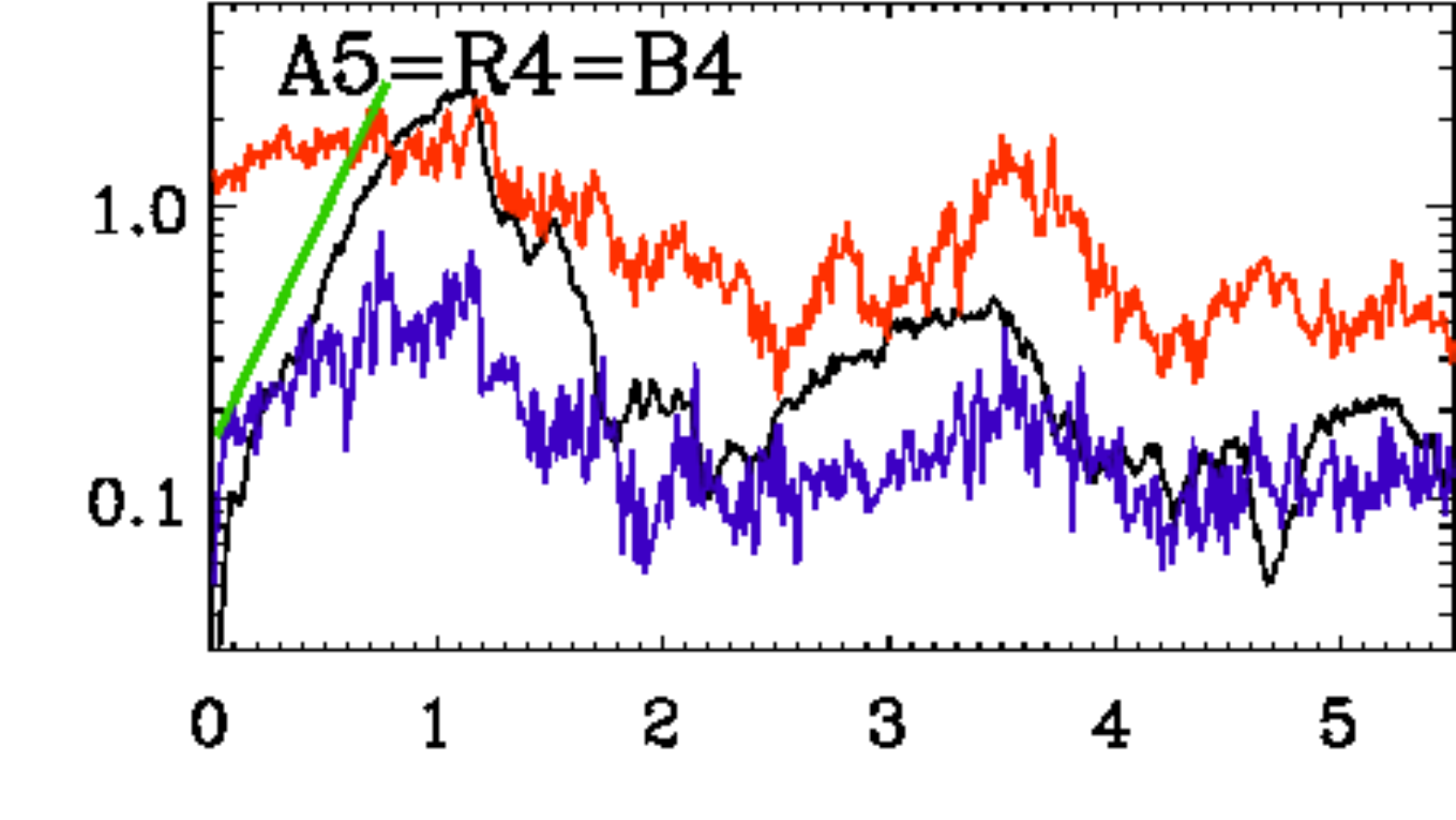}
\includegraphics[width=0.33\textwidth]{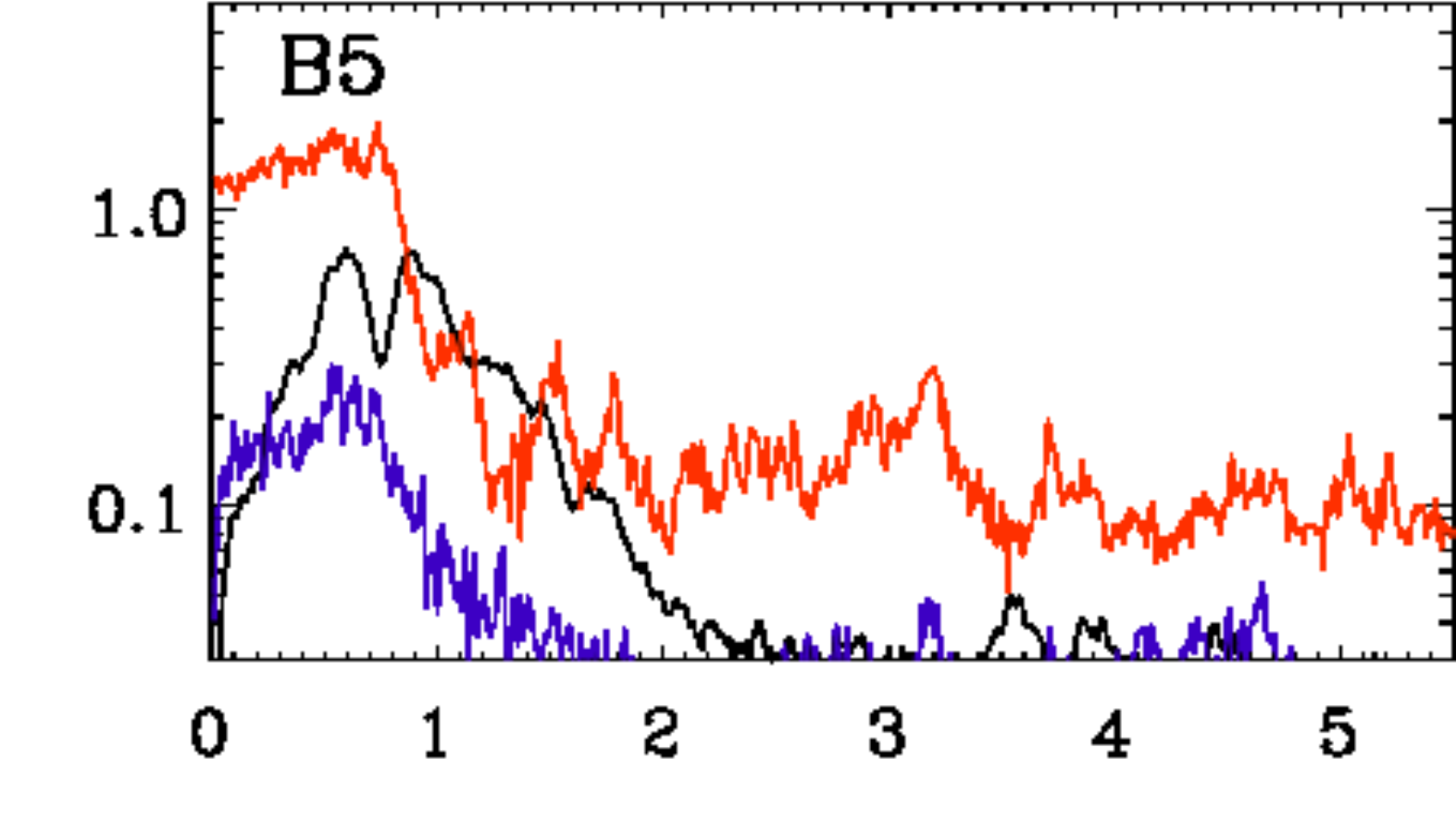}
\includegraphics[width=0.33\textwidth]{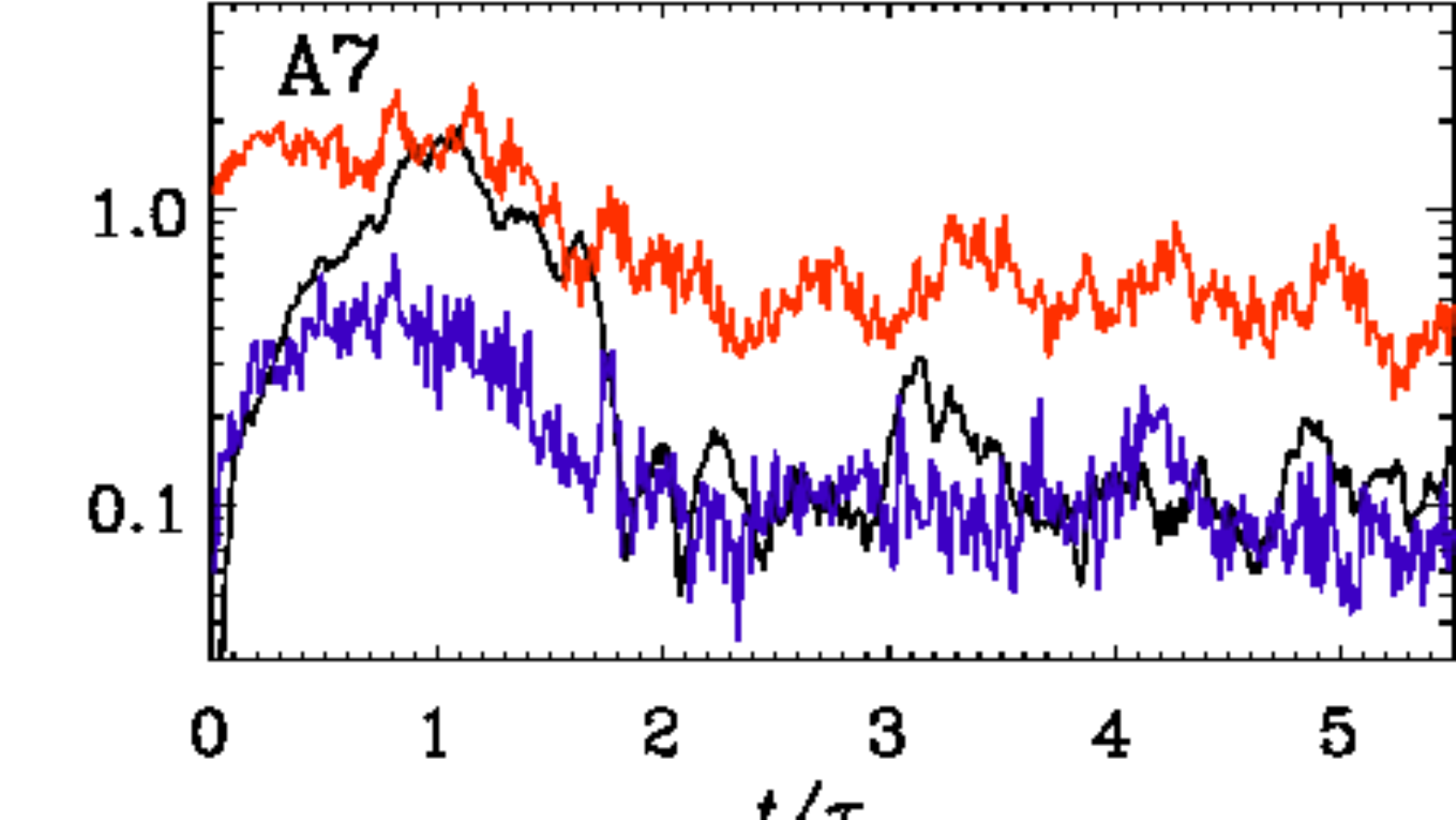}
\includegraphics[width=0.33\textwidth]{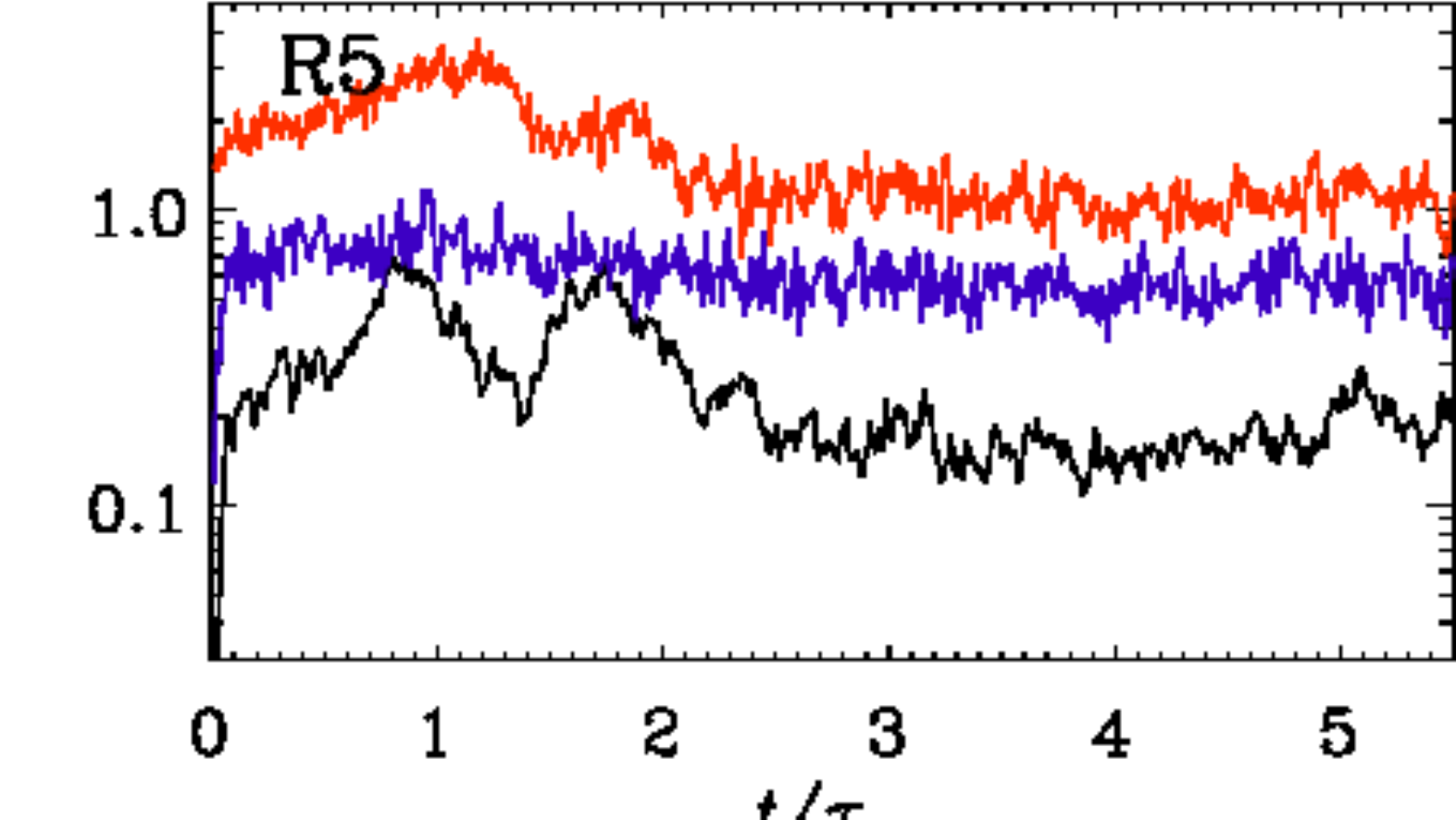}
\includegraphics[width=0.33\textwidth]{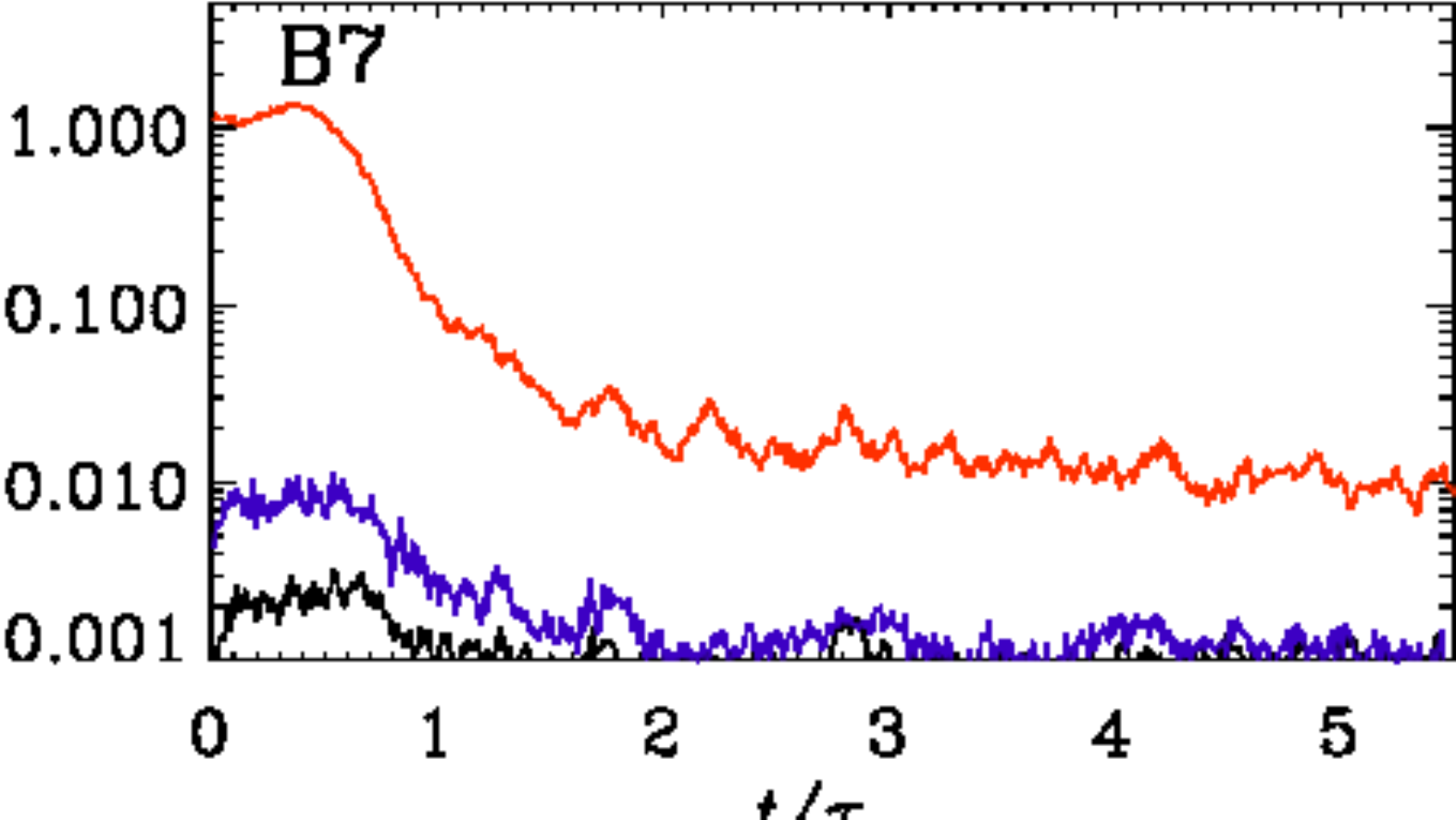}
\end{center}\caption[]{
Temporal evolution of the horizontally averaged, magnetic energy density
of the large-scale field at the surface ($z=0$) $\bra{{\meanBB^{\rm
      fil\,2}}}_{xy}$ for Sets~A (first column), R (second column), and
B (third column).
The three components are shown in blue ($x$), red ($y$), and black ($z$).
All values are normalized by the imposed field strength $B_0^2$.
The straight green line in the panel for Run~A5 shows the estimated
growth rate of $1.4/\tautd$ for vertical large-scale magnetic field.
}\label{pbtsm}
\end{figure*}

We comprehensively study the formation mechanism of the bipolar regions
found in \cite{WLBKR13} by changing the density stratification,
the magnetic Reynolds number, and the imposed magnetic field.
For each parameter we perform five to eight runs in various sets: Set~A for
the density study, Set~R for the magnetic Reynolds number study, and Set~B for
the imposed magnetic field study; see \Tab{runs}.
Furthermore, we use three different additional domain sizes to
investigate their influence on the formation process; see Set~S in
\Tab{runs} and
two additional runs with vertical (Run~V) and 45 degrees inclined
(Run~INC) imposed magnetic field.

The various stratifications and box sizes give rise to different
vertical profiles of equipartition field strength $\Beq$, which are
plotted in \Fig{pbeq}.
As a result of the transition from intense turbulence to
small velocities in the coronal envelope, $\Beq$ experiences a steep
decrease at the surface ($z=0$).

We start by investigating the evolution of the magnetic field at the
surface.
We therefore calculate the averaged magnetic energy density of the
large-scale field $\bra{{\meanBB^{\rm fil\,2}}(z=0)}_{xy}$; see
\Fig{pbtsm} for all three components.
Strong flux concentrations with high values for the large-scale
magnetic field are obtained (see \Tab{runs}) when the $z$ components
(black lines) are similar or larger than the $y$ component (red), as in
Runs~A5, A6, A7, R3, B2, and B5.
Furthermore, \Fig{pbtsm} shows a clear exponential
growth of the large-scale vertical magnetic field in those cases where
bipolar regions occur (compare with last column of \Tab{runs}).
This confirms that a hydromagnetic instability is responsible for the
formation of the bipolar regions found in these simulations.
In the second to last column of \Tab{runs}, $\tautdm=t_{\rm max}/\tautd$
is the time when $\Bfm$ is taken in terms of turbulent-diffusive time.
In Set~A, the formation of bipolar regions is connected to a growth
of magnetic energies in all components, but the $z$ component grows
exponentially during the first turbulent diffusion time for all runs,
except Run~A1.
Our estimated growth rate for Run~A5 is $1.4/\tautd$, which is plotted
as a straight line in \Fig{pbtsm}.
This growth rate is well in agreement of earlier studies with imposed
vertical and horizontal magnetic fields, i.e., those without a coronal envelope
\citep{BKR13,KBKMR12}.

The $x$ component of $\bra{{\meanBB^{\rm fil\,2}}(z=0)}_{xy}$ also
shows an exponential growth, but with a lower growth rate.
In Set~R, runs with both a lower and a higher magnetic Prandtl number than
Run~R4=A5 have a smaller growth rate, although Run~R3 also shows
bipolar regions.
In Runs~B1 and B2, there are also exponential increases of the energy of the
vertical magnetic field, which are related to the formation of bipolar
magnetic regions.
These increases tend to occur later and have higher energies than Run~A5.
In Run~B7, the vertical magnetic field is too weak to produce a magnetic
flux concentration, as is also indicated by the lack of exponential growth.
In the following, we  study these behaviors in more detail.

\subsection{Dependence on stratification}
\label{sec:strat}

\begin{figure}[t!]
\begin{center}
\includegraphics[width=\columnwidth]{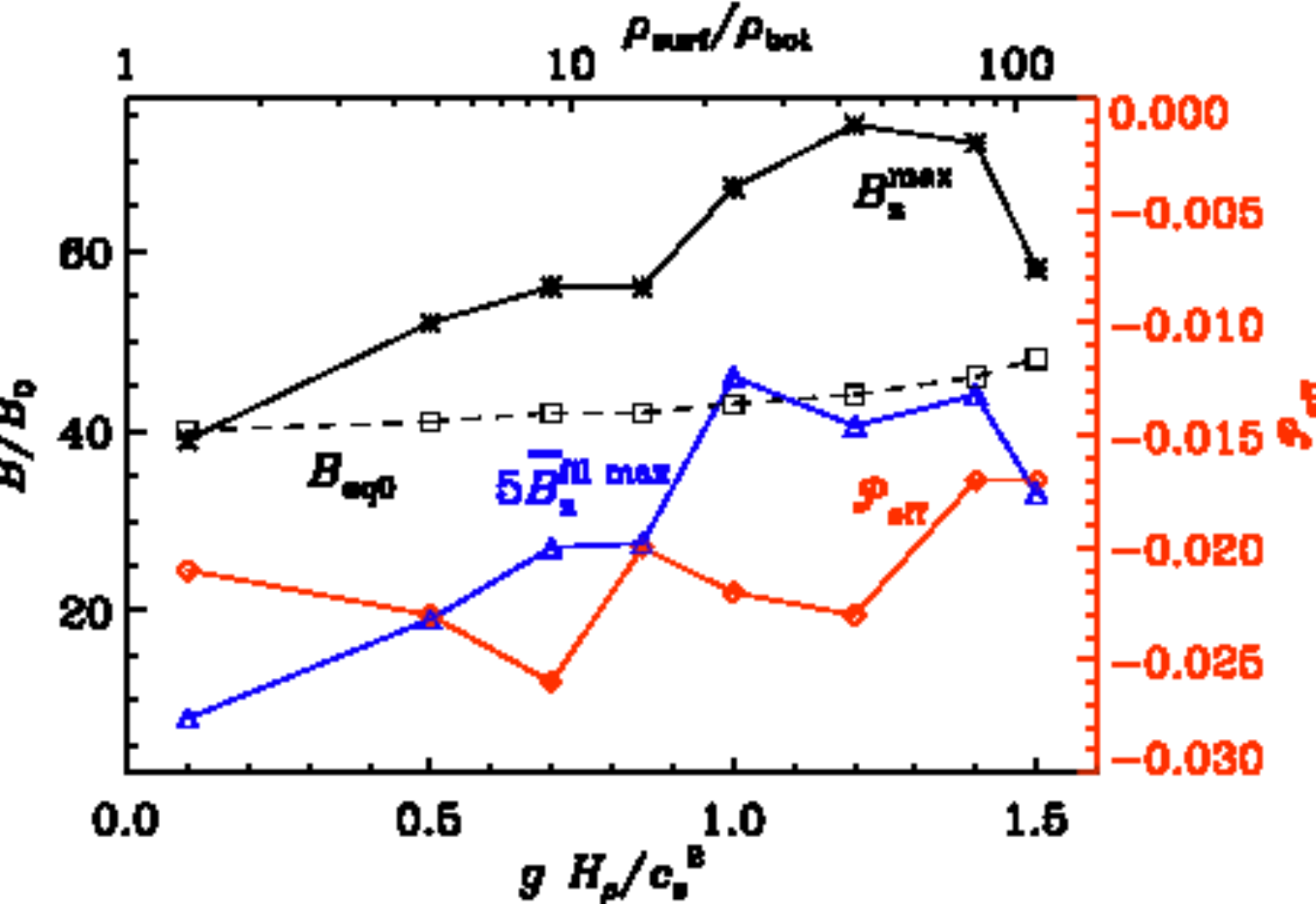}
\end{center}\caption[]{Dependence of magnetic field amplification and
  effective magnetic pressure on stratification.
Maximum vertical magnetic field $B_z^{\rm max}/B_0$ (solid black)
at the surface, maximum of the
large-scale vertical magnetic field $5\Bfm/B_0$ (blue)
at the surface, minimum of the effective magnetic pressure $\Peff$
(red), and equipartition field strength at the surface $\Beqz/B_0$
(dashed black) as a function of $g H_\rho/\csq$ and density
  contrast $\rho_{\rm surf}/\rho_{\rm bot}$ for Set~A.
}\label{pmB_strat}
\end{figure}

\begin{figure*}[t!]
\begin{center}
\includegraphics[width=0.33\textwidth]{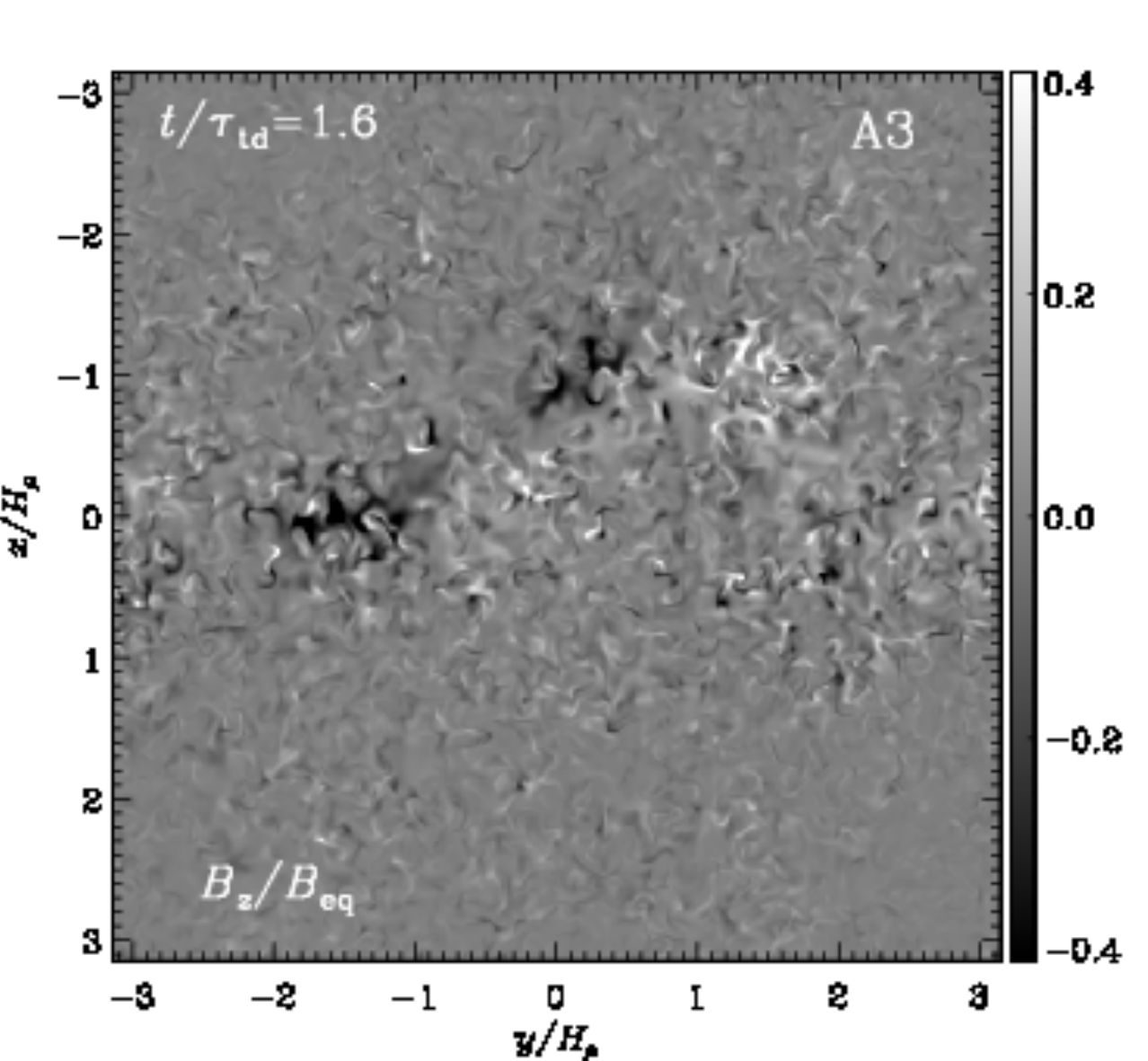}
\includegraphics[width=0.33\textwidth]{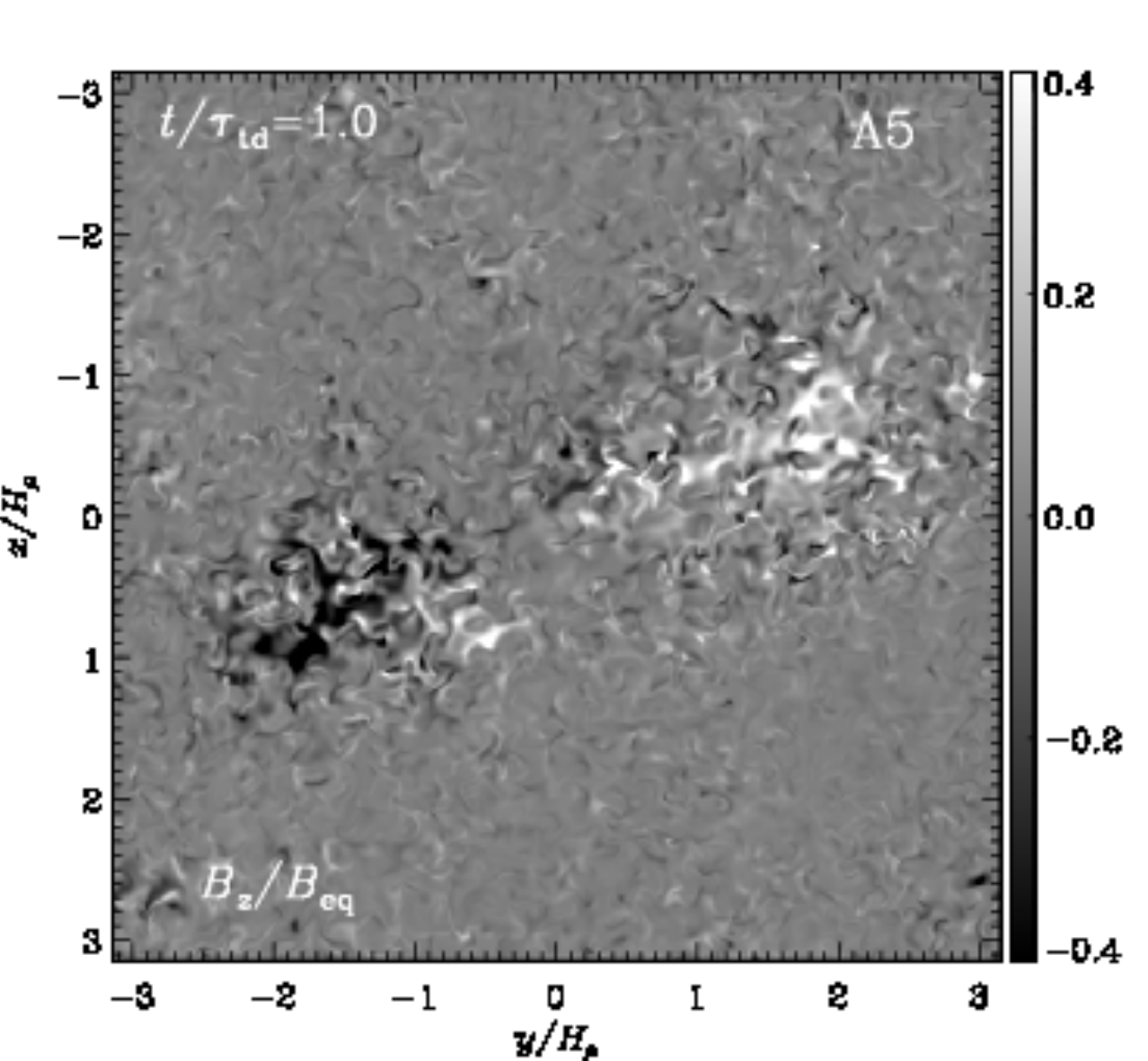}
\includegraphics[width=0.33\textwidth]{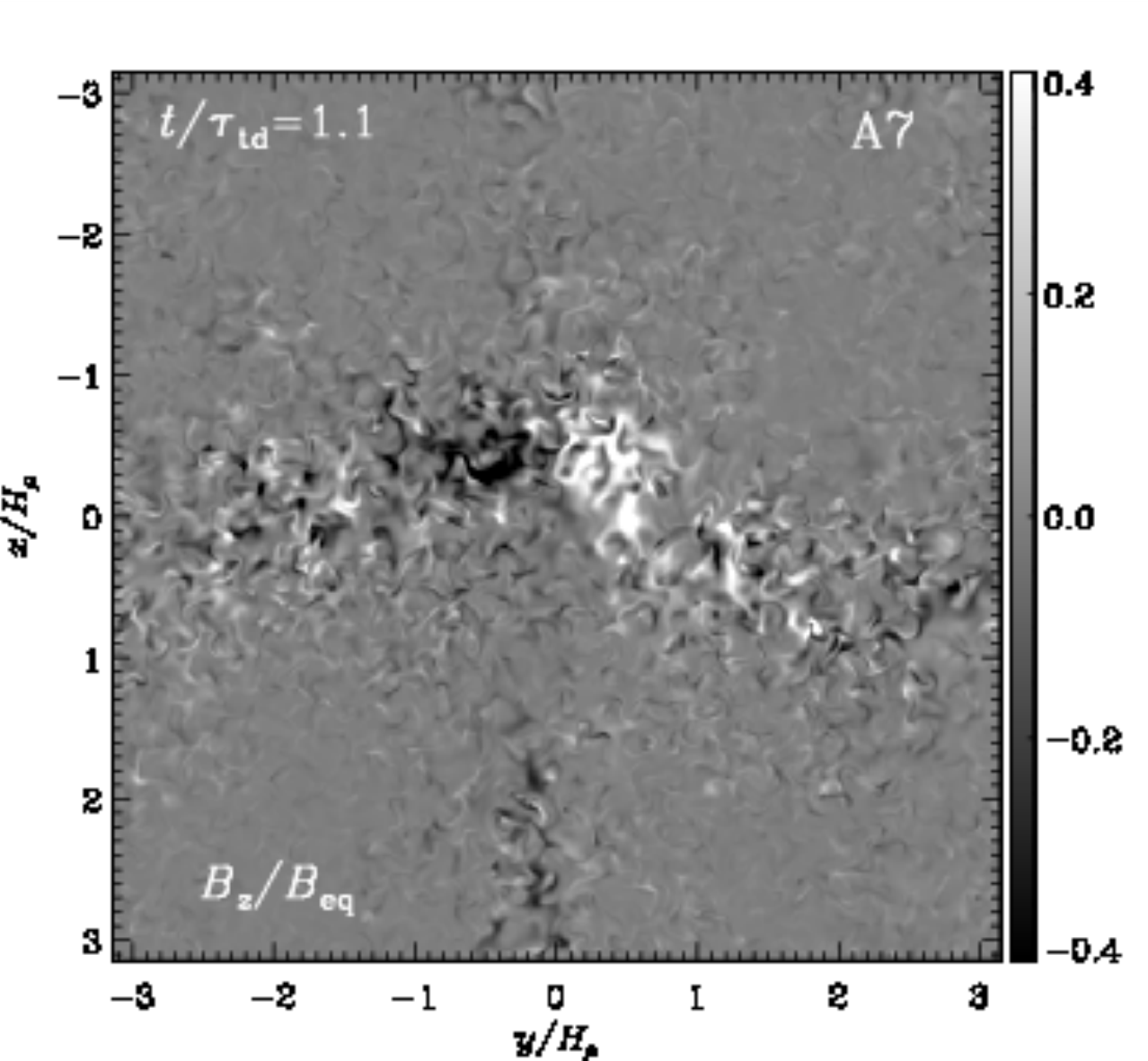}
\includegraphics[width=0.33\textwidth]{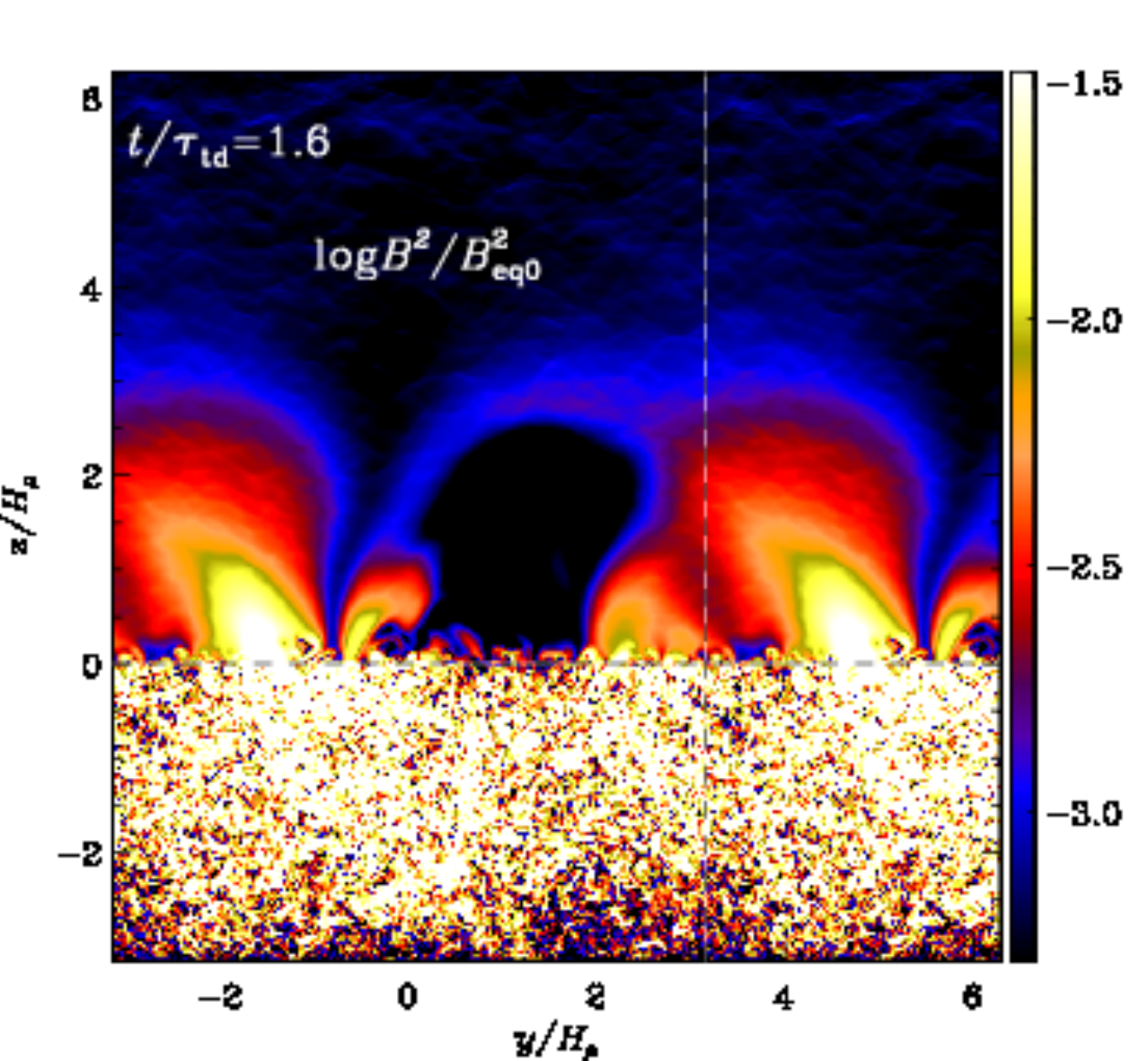}
\includegraphics[width=0.33\textwidth]{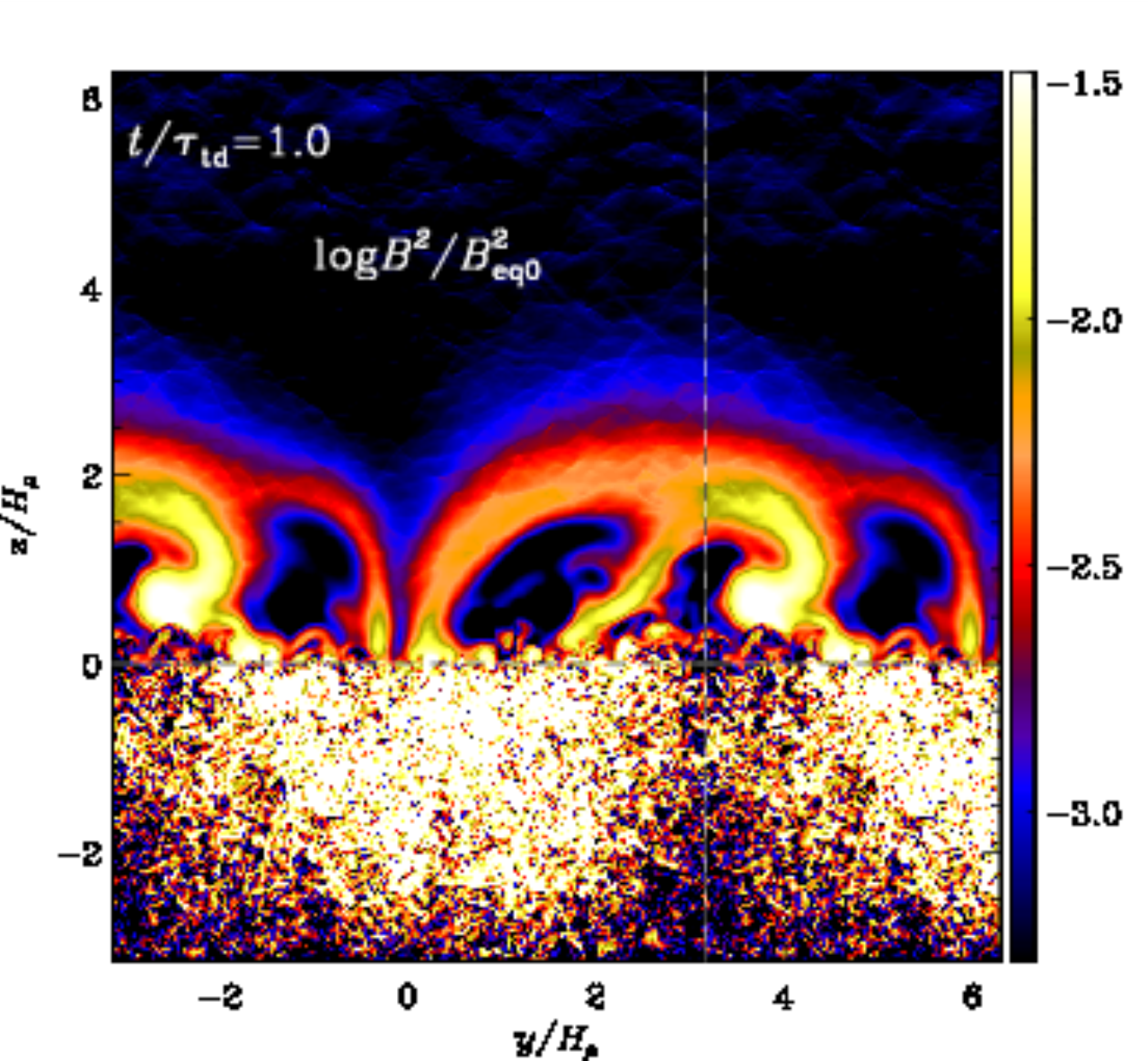}
\includegraphics[width=0.33\textwidth]{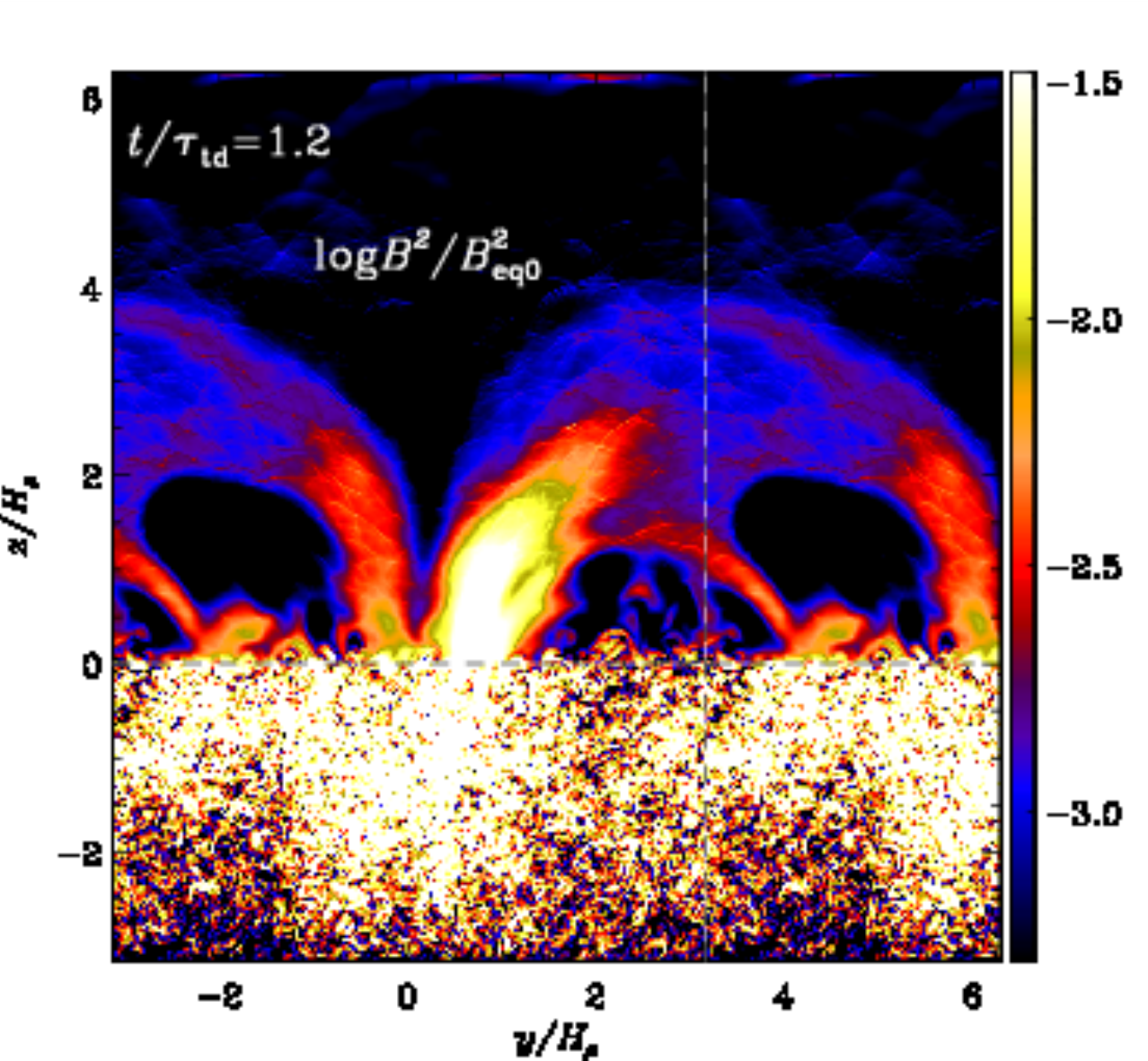}
\includegraphics[width=0.33\textwidth]{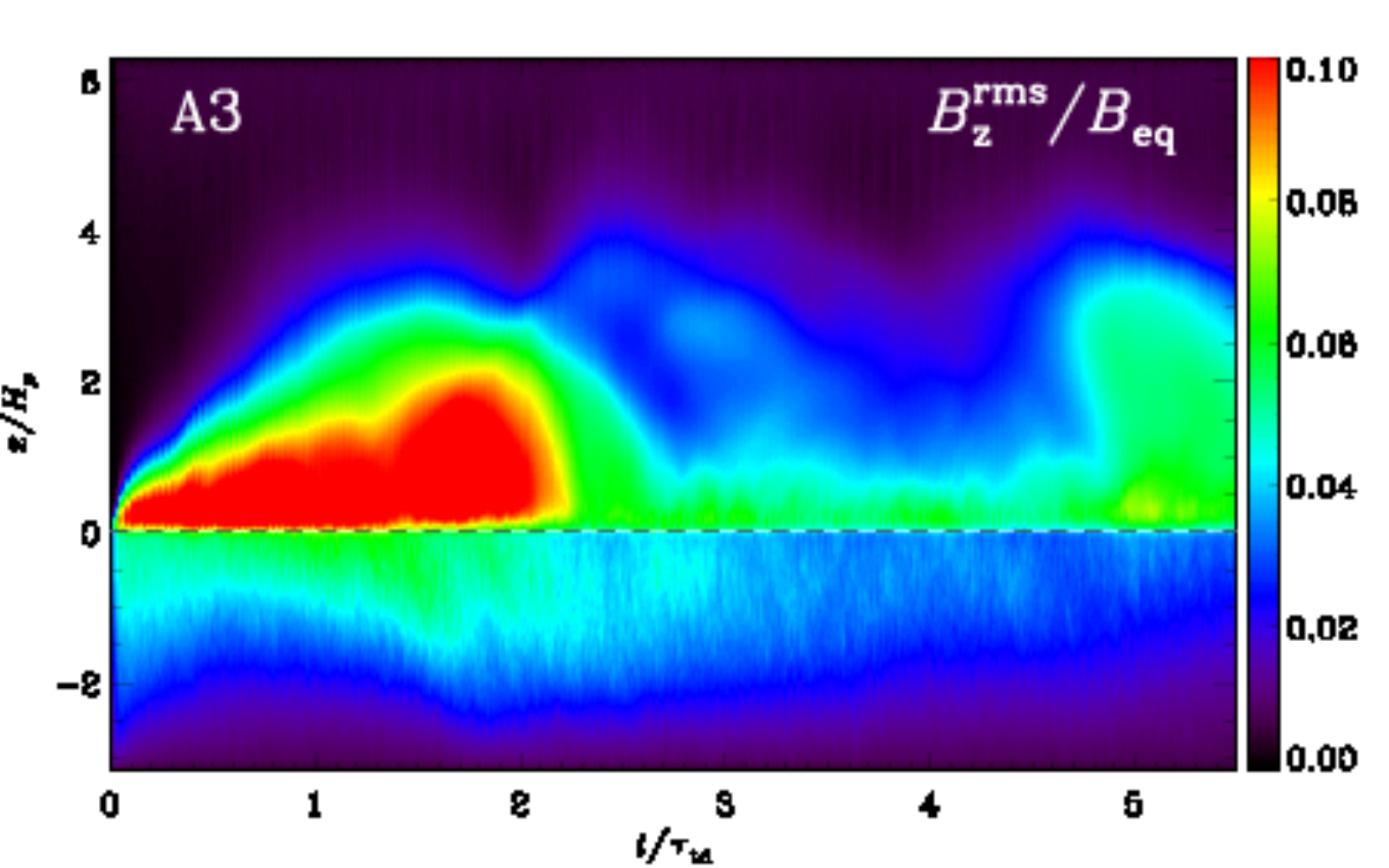}
\includegraphics[width=0.33\textwidth]{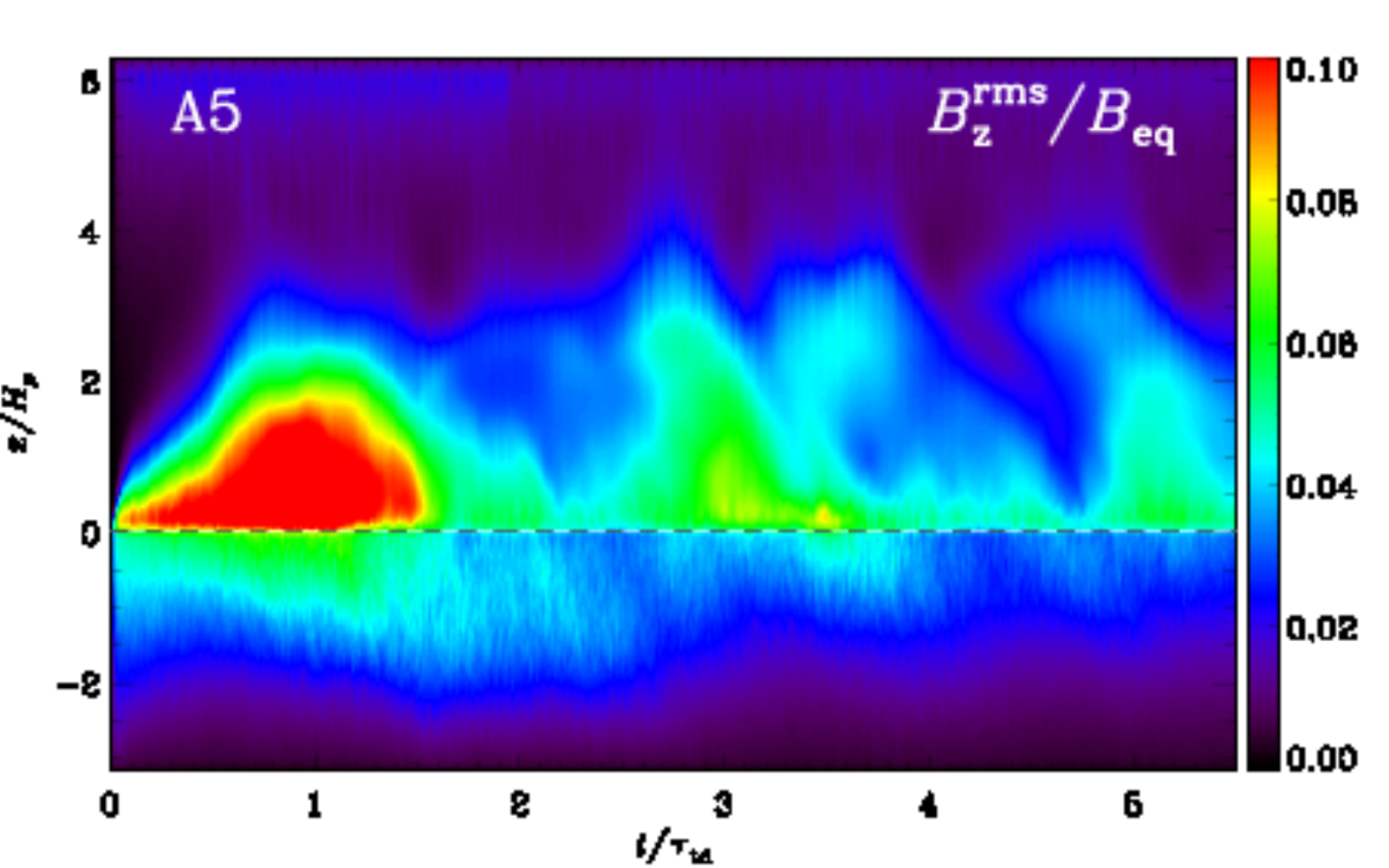}
\includegraphics[width=0.33\textwidth]{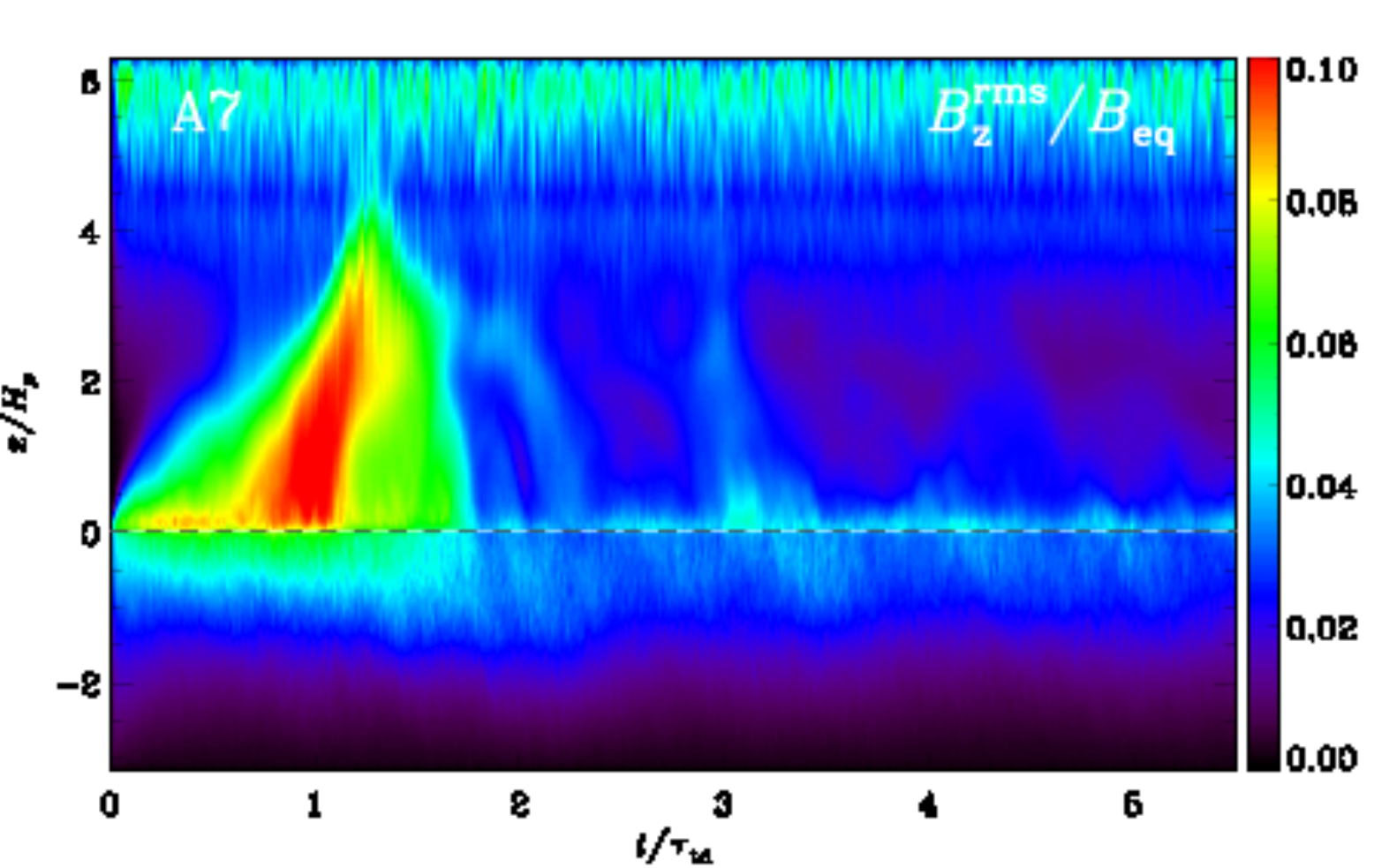}
\includegraphics[width=0.33\textwidth]{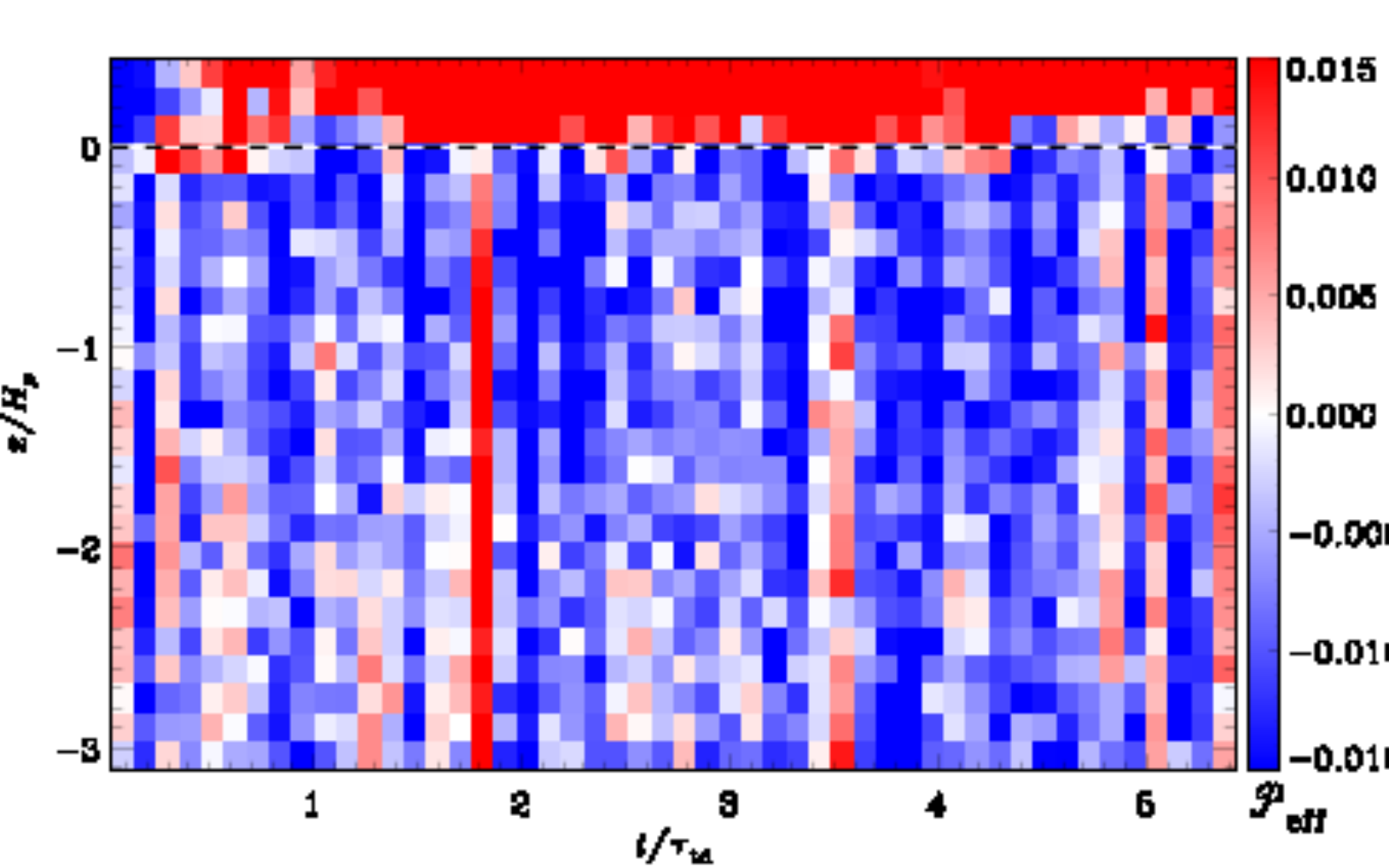}
\includegraphics[width=0.33\textwidth]{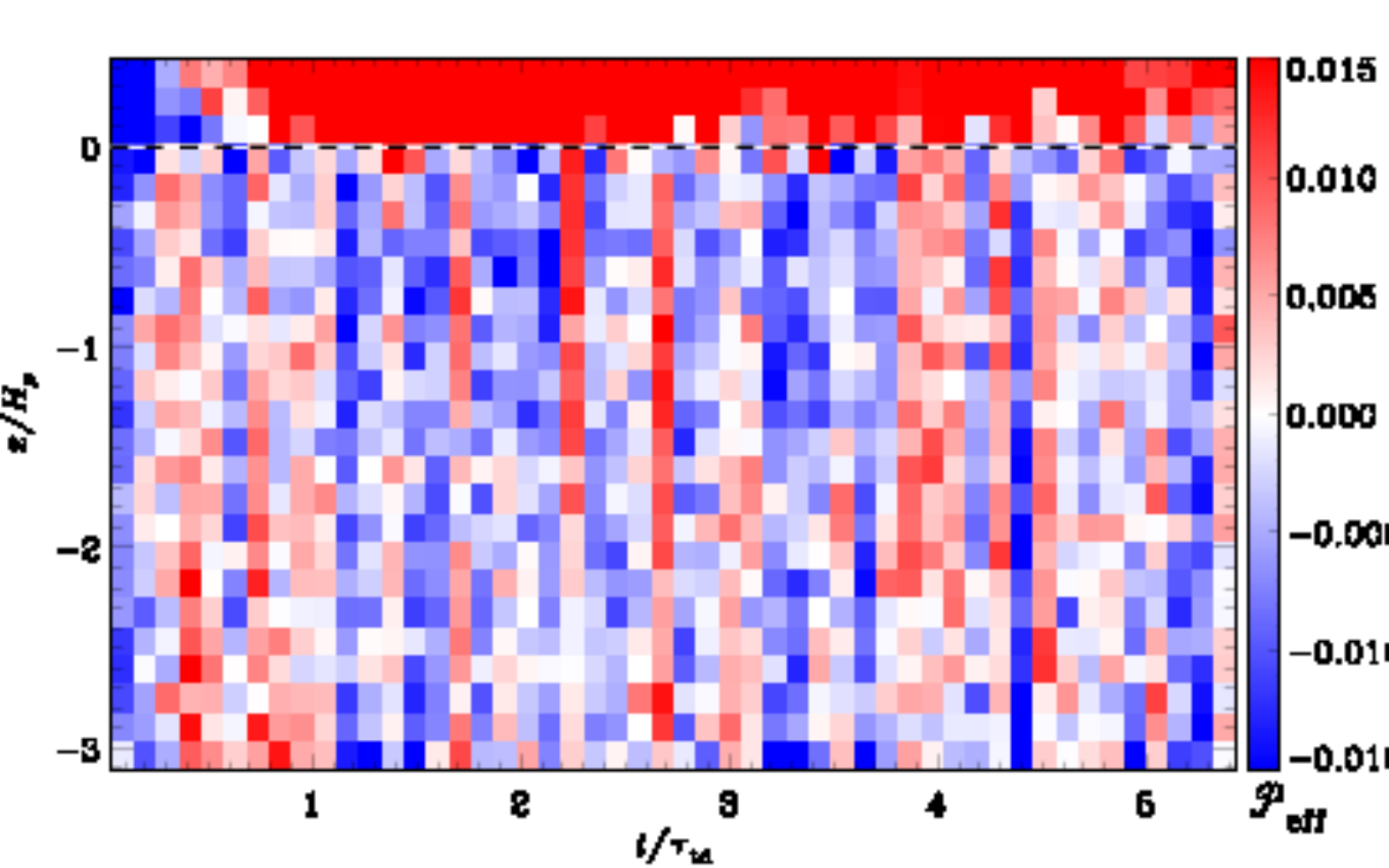}
\includegraphics[width=0.33\textwidth]{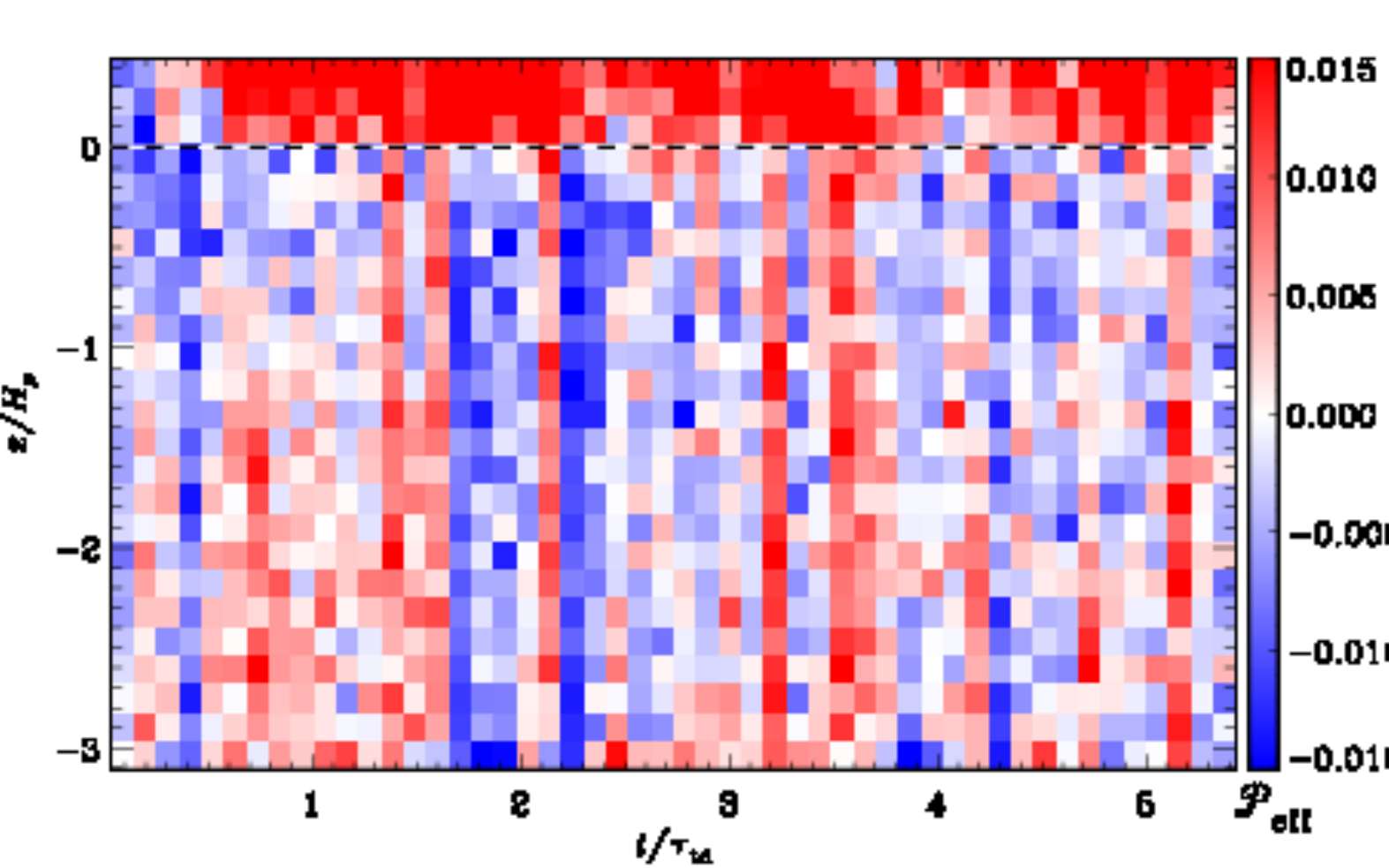}
\end{center}\caption[]{
Formation of bipolar regions for three different stratifications
({\it left column}: A3, {\it middle}: A5, {\it right}: A7).
{\it Top row:} normalized vertical magnetic field $B_z/\Beq$ plotted
at the $xy$ surface ($z=0$) at times when the bipolar regions are the
clearest.
{\it Second row:} normalized magnetic energy density plotted in the
$yz$ plane as a vertical cut through the bipolar region at $x=0$.
We replicated the domain by 50\% in the $y$ direction
(indicated by the vertical dashed lines) to give a
more complete impression about spot separation and arch length.
The black-white dashed lines indicate the replicated part and in the
last three rows the surface ($z=0$).
{\it Third row:} vertical rms magnetic field $B_z^{\rm
  rms}/\Beq=\bra{B_z^2}_{xy}^{1/2}/\Beq$ normalized by the local
equipartition value (see \Fig{pbeq} for vertical profiles) as a
function of $t/\tautd$ and $z/H_\rho$.
{\it Bottom row:} smoothed effective magnetic pressure $\Peff$
as a function of $t/\tautd$ and $z/H_\rho$.
Blue shades correspond to negative and red to positive values.
}\label{strat}
\end{figure*}

In Runs~A1--A8, we vary the density stratification in the
turbulent layer from $\rho_{\rm bot}/\rho_{\rm surf}=1.5$ to $108$
by changing the normalized gravity $g H_\rho/\csq$,
where $\rho_{\rm bot}$ and $\rho_{\rm surf}$ are
the horizontally averaged densities at the bottom
($z=-\pi$) and  at the surface ($z=0$) of the domain, respectively.
This is related to an overall stratification range from
$\rho_{\rm bot}/\rho_{\rm top}=2.6$ (Run~A1) to $1.2\times 10^6$
(Run~A8), where $\rho_{\rm top}$ is the horizontally averaged density at
the top of the domain ($z=2\pi$).
The formation of a bipolar region depends strongly on the
stratification.
For a small density contrast, as in Run~A1,
the amplification of vertical magnetic field is too weak to form magnetic
structures, its maximum is below the equipartition value at the surface; see
\Fig{pmB_strat}. The vertical magnetic field
in the flux concentrations
can reach superequipartition field strengths and an amplification of over 50 of
the imposed field strength already for a density contrast of $\rho_{\rm bot}/\rho_{\rm
surf}\approx5$, as in Run~A2.
However, the bipolar structures are still weak compared to those for higher
stratifications.
The field amplification
inside the flux concentrations
grows with increasing stratification.
The maximal vertical field strength reaches values of over 70$B_0$,
which is nearly twice the equipartition field strength at the surface.
The maximum field strength peaks at $\rho_{\rm bot}/\rho_{\rm surf}=42$
and decreases for even higher stratification (Run~A8).
This limits strong field concentrations to a range between
$\rho_{\rm bot}/\rho_{\rm surf}=23$ and $80$.
Field concentrations are still possible for higher and lower
stratifications,
but the strengths of the large-scale field inside the bipolar region
are smaller.

The density stratification also has an influence on the formation of
the bipolar region.
This is shown in the top row of \Fig{strat}, where we plot the
vertical magnetic field strength
at the surface at the time of strongest bipolar region formation.
Run~A3 with moderate stratification shows a magnetic field
concentration that has multiple poles, and the structure of the
bipole in Run A3 is not as clear as in Runs~A5 and A7.
In Run~A7, the bipolar region is more coherent
and magnetic spots are closer to each other than in Run~A5.
Furthermore, the maximum of the large-scale magnetic field
$\Bfm/B_0$, which is an indication
of the strength of bipolar regions, increases with higher stratification, as
shown by the blue line in \Fig{pmB_strat}.
A maximum of the large-scale magnetic field above about $5\,B_0$ seems
to indicate bipolar flux concentrations.
The inclination of the two polarities is most of the time
aligned with the imposed field direction.
However, in some cases, as in Run~A5, an alignment with the surface
diagonal is also possible.
Unfortunately, we cannot find any clear criteria that determine
the alignment.

In the second row of \Fig{strat}, we show how the magnetic field continues
above the surface.
Here we plotted $\log_{10} B^2/\Beqz^2$ at a time when the bipolar
regions are the clearest.
 The loop structures connecting the two
polarities are more pronounced for high stratification (Run~A7) than for moderate stratification
(Run~A3).
Furthermore, in Runs~A5 and A7, the magnetic energy in the turbulent
region is much more concentrated and structured than in Run~A3.
These plots indicate that with higher stratification, it is easier to
form loop-like structures in the coronal envelope.
However, the inclination of the bipolar region as in Run~A3 seems to
form more complex loops structures than what is shown in Fig.~5 of
\cite{WLBKR13}.

In the third row of \Fig{strat}, we plot the horizontally averaged rms
value of the vertical magnetic field $B_z^{\rm rms}=\bra{B_z^2}_{xy}^{1/2}$,
which is normalized by the local equipartition value, as a function of
time and height.
In the coronal envelope, where turbulent forcing is absent,
$\Beq$ is much lower than in the turbulent layer; see \Fig{pbeq}
for the vertical profiles of $\Beq$.
This leads to high values of $B_z^{\rm rms}/\Beq$ in the coronal envelope.
We chose this normalization using $\Beq$ instead of $\Beqz$
because of the better visibility of the concentration of vertical
flux.
For all three cases, which are shown in the third row of \Fig{strat}, the field
emerges from the turbulent layer, forming a bipolar region and then
generating loop-like structures in the coronal envelope.
After $t/\tautd\approx2$, the vertical field decays, and new strong
flux concentrations are not able to form.
This is related to a persistent change of the average stratification
after the magnetic field is applied.

An indicator of structure formation through the negative effective
magnetic pressure instability (NEMPI) is the effective
magnetic pressure $\Peff$.
For its derivation, we start with the definition of the turbulent
stress tensor $\PP$, i.e.,
\begin{equation}
\Pi_{ij}^{(\mean{B})}\equiv\overline{\rho \fluc{u}_i\fluc{u}_j}
+\half\delta_{ij}\mu_0^{-1}\overline{\bb^2}-\mu_0^{-1}\overline{b_ib_j},
\end{equation}
where the first term is the Reynolds stress tensor and the
last two terms are the turbulent magnetic pressure and turbulent
Maxwell stress tensors.
The superscript $(\mean{B})$ indicates the turbulent stress tensor
under the influence of the
mean magnetic field; $\Pi_{ij}^{(0)}$ is the turbulent stress tensor without
mean magnetic field, where both, the turbulent Maxwell stress and the
Reynolds stress are free from the influence of the mean magnetic field.
Here we define mean and fluctuations through horizontal averages,
$\mean{\BB}\equiv\bra{\BB}_{xy}$, such that $\BB=\mean{\BB}+\bb$ and
$\uu=\mean{\UU}+\fluc{\uu}$.
Using symmetry arguments, we can express the difference in the
turbulent stress tensor $\PP$ for the magnetic and nonmagnetic case
in terms of the mean magnetic field \citep[see, e.g.,][]{BKKR12},
\begin{equation}
\Delta\Pi_{ij}=\Pi_{ij}^{(B)}-\Pi_{ij}^{(0)}=-q_p \delta_{ij} {\mean{\BB}^2\over 2}
+q_s \mean{B_i}\mean{B_j} -q_g {g_i g_j\over g^2} \mean{\BB}^2,
\label{pi-mean}
\end{equation}
where $q_p$, $q_s$, and $q_g$ are parameters expressing the importance of
the mean-field magnetic pressure,  mean-field magnetic stress, and vertical anisotropy caused by gravity.
They are to be determined in direct numerical simulations:
${g_i}$ are components of $\gggg$, which in our setup has only a
component in the negative $z$ direction.
The normalized effective magnetic pressure is then defined as
\begin{equation}
\Peff=\frac{1}{2} (1-q_p)\,\beta^2,\quad {\rm with}\quad
\beta^2={\mean{\BB}^2\over\Beq^2},
\label{peff}
\end{equation}
where we can calculate from \Eq{pi-mean}
\begin{equation}
q_p=-{1\over\mean{\BB}^2} \, \left(\Delta\Pi_{xx}+\Delta\Pi_{yy}-\left(\Delta\Pi_{xx}-\Delta\Pi_{yy}\right)\,{\mean{B_x}^2
  +\mean{B_y}^2\over\mean{B_x}^2 -\mean{B_y}^2} \right) ,
\label{q_p}
\end{equation}
\begin{equation}
q_s={\Delta\Pi_{xx}-\Delta\Pi_{yy}\over\mean{B_x}^2 -\mean{B_y}^2} ,
\label{q_s}
\end{equation}
\begin{equation}
q_g={1\over\mean{\BB}^2} \,
\left(-\Delta\Pi_{zz}-q_p{\mean{\BB}^2\over 2}
  +q_s\mean{B_z}^2\right).
\label{q_g}
\end{equation}
In the bottom row of \Fig{strat}, we show $\Peff$ for Runs~A3, A5, and A7, where
$\Peff$ was evaluated in $50\times20$ bins in time and
height within the turbulent layer.
From these maps, we deduct the minimum values $\Peff^{\rm min}$ and
list them in the ninth column of \Tab{runs}; see also
Figures~\ref{pmB_strat}, \ref{pmB_rm}, and \ref{pmB_impB}.

We find that the area with negative effective magnetic pressure
$\Peff$ decreases for stronger stratifications (see the bottom row of \Fig{strat}).
For Run~A3, the smoothed $\Peff$ is negative in basically all of the
turbulent layer at all times, except for some short time intervals.
The values are often below $-0.005$, but occasionally even below $-0.01$.
For higher stratification, the intervals of positive values of $\Peff$ become
longer and negative values become in general weaker.
In Run~A7, the smoothed $\Peff$ fluctuates around zero with
equal amounts of positive and negative values.

As $\Peff$ is plotted in the same time interval as $B_z^{\rm rms}$
(third row of \Fig{strat}),
it enables us to compare the time evolutions of structure
formation and $\Peff$.
For Run~A7, there seems to be a relation between the two,
i.e., structure formation occurs when $\Peff$ is negative.
When $B_z^{\rm rms}$ has a strong peak at around $t/\tautd\approx1$,
$\Peff$ has a minimum between $t/\tautd\approx0.5$ and 1 close to the
surface.
In Runs~A3 and A5, $\Peff$ is also weak when $B_z^{\rm  rms}$ is strong,
but this does not only happen when $B_z^{\rm  rms}$ is strong.
In general, the minimum value of the smoothened $\Peff$ does not indicate
the existence of NEMPI
as a possible formation mechanism of flux concentration in the context of dependency
on density stratification.
There is a weak opposite trend:
$\Peff$ becomes less negative for large stratification, even though
$\Bfm$ increases for larger stratification; see \Fig{pmB_strat}.

\begin{figure}[t!]
\begin{center}
\includegraphics[width=\columnwidth]{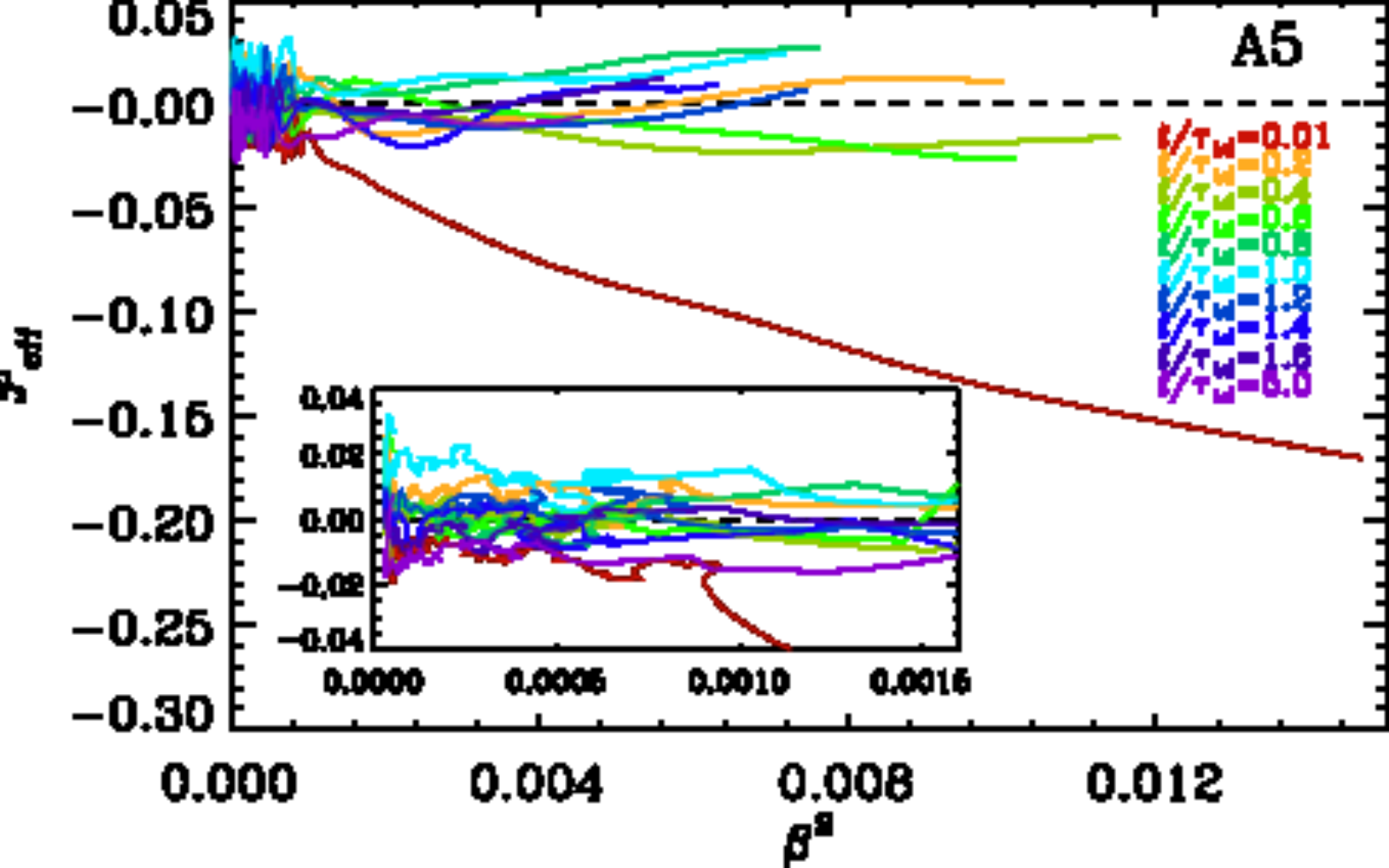}
\end{center}\caption[]{
Effective magnetic pressure $\Peff$ plotted
over $\beta^2=\meanBB^2/\Beq^2$ at ten different times for Run~A5.
The inlay shows a zoom-in to the lower values of $\beta^2$, where we
have averaged over 40 points to reduce the noise.
The shown values are limited to the turbulent layer ($z\le0$).
}\label{growth}
\end{figure}

Indeed, the value of $\Peff$ itself is not the decisive quantity, as
the growth rate $\lambda$ of NEMPI is given by
\citep{RK07,KBKMR13,BGJKR14}
\begin{equation}
\lambda= {\vA \over H_\rho} \,\left(-2 { \dd\Peff\over \dd\beta^2}
\right)^{1/2} \, {k_x \over k} -\etatz \,  k^2.
\end{equation}
See Appendix A of \cite{KBKMR13} for a detailed derivation
with an imposed horizontal field, and Sect. 2.1 of \cite{BGJKR14}
with a vertical field.
Here $\vA=B_0/\sqrt{\mu_0 \rho}$ is the Alfv\'{e}n speed.
To get an idea about the form of $\dd\Peff/\dd\beta^2$,
we plot in \Fig{growth} $\Peff$ versus
$\beta^2=\meanBB^2/\Beq^2$ at different times for Run~A5.
In the beginning of the simulation, the growth rate is positive for all
values of $\beta^2$ in the turbulent layer
because the derivative, $\dd\Peff /\dd\beta^2$ is negative.
As the simulation progresses, the growth rate become weaker and mainly
at larger values of $\beta^2$ it is positive.
After the formation of the largest and strongest concentrations at
around $t/\tautd=\tautdm=1.0$ (light blue), the growth rate is positive
only at low values of $\beta^2$, as shown in the inlay of \Fig{growth}.
However, even when the growth rate is positive, the actual values of
$\Peff$ are still positive.
This behavior of the growth rate fits well with the temporal
evolution of the large-scale magnetic field as shown in \Fig{pbtsm}.
There, $\bra{{\meanBB^{\rm fil\,2}}}_{xy}$ exhibits an exponential
growth until around $t/\tautd=1.0$, saturates and then decays after
$t/\tautd=1.2$.
At low values of $\beta^2$, $\Peff$ does not show a strong indication
of a negative slope;
it seems nearly constant and the growth rate is
close to zero; see inlay of \Fig{growth}.

We should note here that the mean field $\meanBB$ is in the
direction of the imposed magnetic field, i.e., the $y$ direction, while
the field in the spots points in the positive or negative $z$ direction.
Therefore, besides the usual formation of concentrations with the same
polarity as the imposed field, we have here an
additional mechanism to turn the field from horizontal to vertical.
One of these mechanisms can be magnetic buoyancy \citep[e.g.,][]{P55b},
which is actually visible in \Fig{growth}, where $\dd\Peff
/\dd\beta^2$ becomes positive.
Even though it is not easy to determine the growth rate for the
simulations, we can get a rough idea by looking at $\tautdm$ for
increasing stratification.
Interestingly, $\tautdm$ tends to decrease, implying a stronger growth
rate for larger stratification.

\begin{figure}[t!]
\begin{center}
\includegraphics[width=\columnwidth]{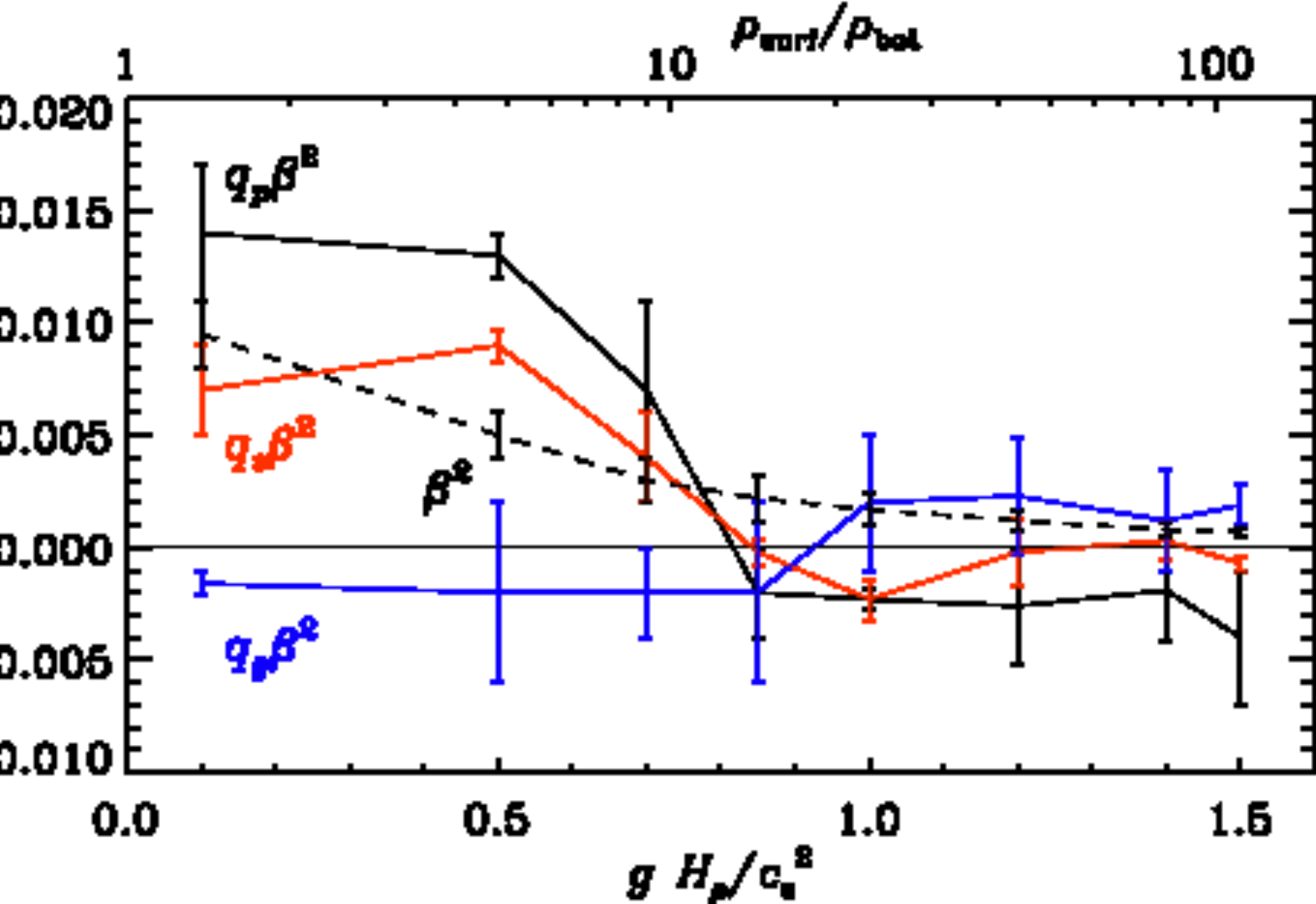}
\end{center}\caption[]{
Dependence of parameters $q_p$ (black), $q_s$
(red), and $q_g$ (blue) on stratification for Set~A.
We normalize the parameters by multiplying with $\beta^2$ (dashed
black).
The legend of the x-axis is the same as in \Fig{pmB_strat}.
The parameters are computed as a temporal and spatial mean over the
turbulent layer.
Error bars are estimated using the maximum difference of the total mean
with the means of each third of the time series.
}\label{qp}
\end{figure}

To understand the dependence on stratification, we analyze the three
parameters in the three terms of \Eq{pi-mean} defined in \Eqss{q_p}{q_g}.
They quantify the importance of the different contributions to the
turbulent stress tensor $\PP$.
In \Fig{qp}, we plot the parameters $q_p$, $q_s$, and $q_g$ as
functions of density stratification.
The errors are relatively large because
the parameters are strongly fluctuating in time and space.
Nevertheless, there are some systematic trends with increasing
density stratification.
The parameter $q_p\beta^2$ is related to $\Peff$ and shows a strong decrease
from low to moderate stratifications ($\rho_{\rm bot}/\rho_{\rm surf}<15$),
and it is even larger than the decrease in $\beta^2$ itself.
This means, the average $\Peff$ is only negative for
$\rho_{\rm bot}/\rho_{\rm surf}$ smaller than $\approx15$.
For larger stratifications, $\Peff$ is on average positive.
However, this also means that, as shown in the last row of \Fig{strat},
$\Peff$ can be negative at certain times and certain depths.
The parameter $q_g$, describing vertical anisotropy due to gravity, is
negative for low and moderate stratifications and becomes positive for
high stratification showing a steady increase.
Therefore, $q_g\beta^2>\beta^2$ can also decrease the turbulent pressure,
which is the trace of $\PP$.
This seems to be the case at least on average for high stratifications
($\rho_{\rm bot}/\rho_{\rm surf}>20$).
However, because of the direction of the gravity, only $\Pi_{zz}$ is
suppressed.
This might be related to the fact that we still find bipolar regions
for high stratification, but the field strength is weaker than for
moderate stratifications.
This behavior might explain the ``gravitational quenching'' found by
\cite{JBLKR14}.
The coefficient $q_s$, characterizing the importance of the off-diagonal
components of the turbulent
stress tensor, does not seem to have a strong influence on the result.
Furthermore, the sign is positive for low stratifications, close to
zero for higher stratifications, and therefore $q_s\beta^2<\beta^2$
for most of the runs.
Thus, the  $q_s$ terms could only suppress the turbulent pressure if
the components of the magnetic stress tensor themselves were negative.
The averaged coefficients $q_p$, $q_s$, and $q_g$ indicate that the
main mechanism for flux concentration for low and moderate
stratifications ($\rho_{\rm bot}/\rho_{\rm surf}\le15$) is related to
the negative effective magnetic pressure $\Peff$, whereas for high
stratifications ($\rho_{\rm bot}/\rho_{\rm surf}\ge15$), the
contribution of the vertical anisotropy due to gravity is more
important.
However, as discussed before,
the averaged quantities are strongly affected by fluctuations.
Comparing our values with previous works \citep{BKKR12,KapBKMR12},
we find broad agreement.
In \cite{BKKR12}, $q_g\beta^2$ is smaller and positive for
similar stratification, while $q_s\beta^2$ is close to zero.
In the present work $q_p$ is negative instead of positive for the same
stratification.
In \cite{KapBKMR12},
where turbulent convection is considered instead of
forced turbulence, $q_g$ turns out to be positive and $q_s$
negative, which is similar to our simulations with similar stratification.
A detailed comparison with \cite{WLBKR13} reveals that the structure
of the bipolar region and its $\tautdm$ of case A is not exactly the
same as in Run~A5, even though the only difference is the resolution
and precision.
This suggests that in the simulations of \cite{WLBKR13} the resolution
was not sufficient to model this highly turbulent medium.

In addition to the change in stratification, we also change the forcing width from $w=0.05$ to $w=0.02$ in one
case (Run~THW).
The resulting change in the vertical profile of the equipartition field
strength is small, as shown in \Fig{pbeq}.
The resulting bipolar regions, however, are slightly weaker, $B_z^{\rm
  max}/B_0=56$, and the large-scale field is significantly weaker than in
Run~A5.
This might also be related to the fact that the $\Beqz$ is slightly
higher.
Thus, in summary, the forcing width does not have a  strong influence
on the occurrence of bipolar regions.

\subsection{Dependence on magnetic Reynolds number}
\label{sec:rm}
\begin{figure}[t!]
\begin{center}
\includegraphics[width=\columnwidth]{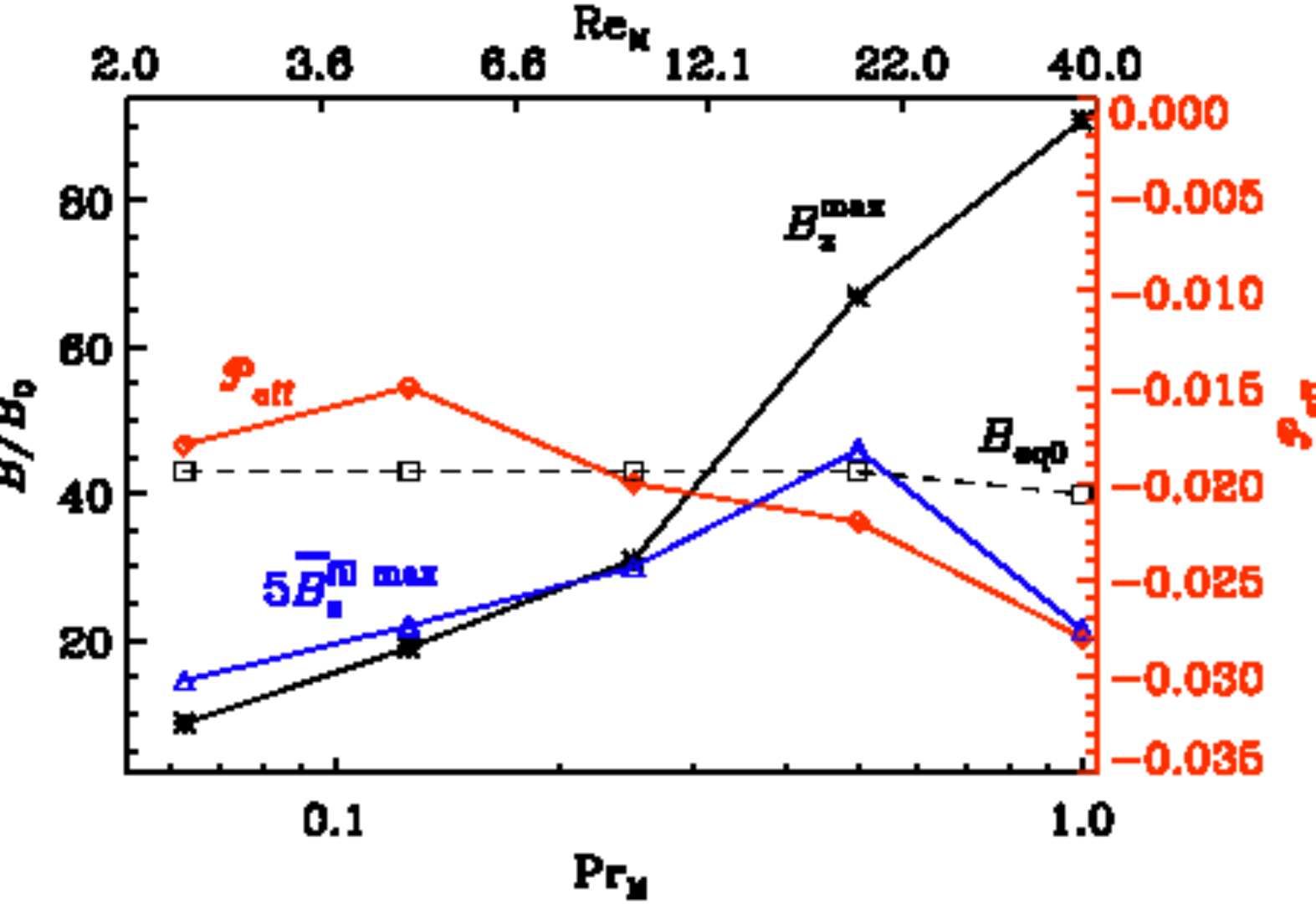}
\end{center}\caption[]{Dependence of magnetic field amplification and
  effective magnetic pressure on magnetic Prandtl number $\Pm$ and magnetic Reynolds
  number $\Rm$ for Set~R.
The legend is otherwise the same as in \Fig{pmB_strat}.
}\label{pmB_rm}
\end{figure}
\begin{figure}[t!]
\begin{center}
\includegraphics[width=\columnwidth]{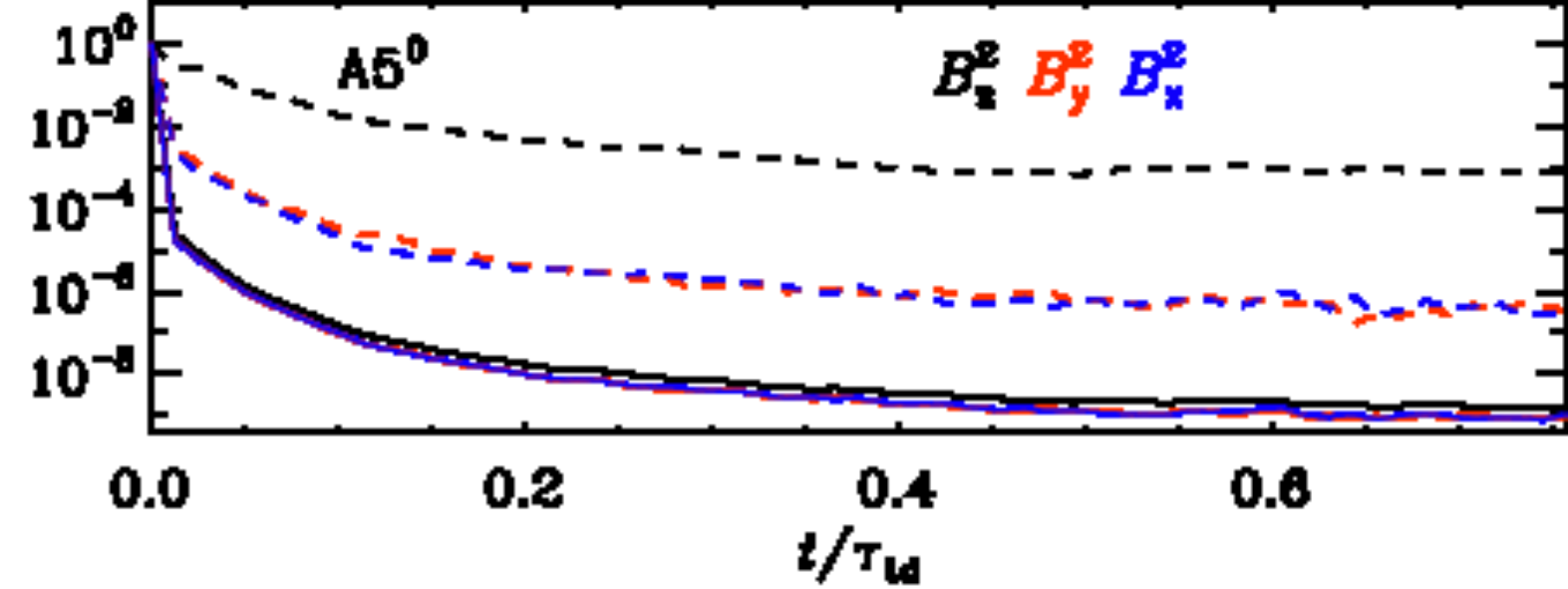}
\includegraphics[width=\columnwidth]{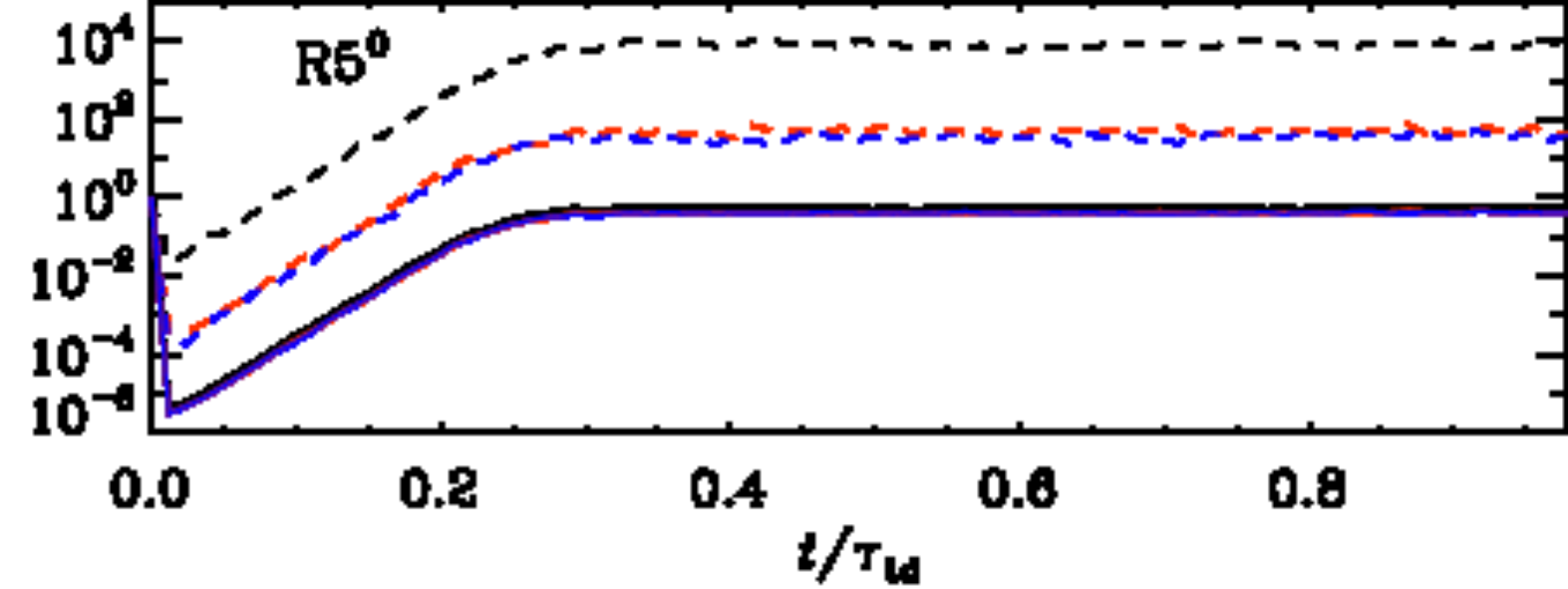}
\end{center}\caption[]{
Temporal evolution of the horizontally averaged magnetic energy density
at the surface ($z=0$) for Runs~A5$^0$ and R5$^0$, where $B_0=0$.
The three components are shown in blue ($x$), red ($y$), and black
($z$), where solid lines indicate the total magnetic energy and dashed lines the large-scale magnetic energy.
All values are normalized by their values at $t/\tautd=0$.
}\label{pbtsm_0}
\end{figure}

As a next step we investigate the dependency on magnetic Reynolds
number $\Rm$.
We keep $\Rey$ fixed (around 40) and change $\Pm$ by a factor of 16;
see the seventh column in \Tab{runs}.
Run~R1, has the lowest $\Pm$ and a magnetic Reynolds number of $\Rm=2.4$.
This implies that microscopic diffusion is of the same order as
turbulent diffusion.
The effect of negative magnetic pressure is weak for such low magnetic Reynolds numbers.
Indeed, the maximum amplification of the magnetic field due to
the flux concentration is around 5, which is
nearly ten times less than the equipartition value.
Also the amplification of the large-scale magnetic field is weak.
Even though the minimum value of $\Peff$ is similar to those of Set~A,
NEMPI cannot be excited, presumably because the growth rate of NEMPI is
smaller than the damping rate caused by turbulent and microscopic magnetic
diffusion.

Increasing $\Rm$ and $\Pm$ leads to larger field amplifications
and stronger large-scale fields inside the flux concentrations; see
\Figs{pbtsm}{pmB_rm}.
However, the vertical field can only reach
superequipartition values when $\Pm$ is above 0.5.
The dependence on $\Rm$ can also be seen from
the time $\tautdm$
(time instant when $\Bfm$ is reached).
Increasing $\Rm$ leads to a shorter $\tautdm$, but in
Run~R5, the instability is weakened and causes
$\tautdm$ to be longer.
This behavior can also be seen in the evolution of the
components of the magnetic energy; see \Fig{pbtsm}.
For $\Pm\le1$, the growth becomes steeper with increasing $\Pm$
until the maximum is reached for Run~A5=R4.
For $\Pm=0.5$, i.e.,\ for Run~R5, the growth rate is again smaller than
for Run~A5=R4.

In Run~R5, the magnetic Prandtl number is unity and a small-scale
dynamo is excited.
This is illustrated in \Fig{pbtsm_0}, where we plot the $x$, $y$, and $z$
components of the magnetic energy
as a function of $t/\tautd$ for Runs~A5$^0$ and R5$^0$.
These two simulations are identical to Runs~A5 and R5,
except that we set
the imposed field $B_0$ to zero and use a weak, white-noise seed magnetic field
instead.
For Run~A5$^0$ all components of the magnetic field decay as
expected because NEMPI needs a small imposed mean magnetic field to
operate.
In Run~R5$^0$ a small-scale dynamo operates and generates magnetic
field in all components, but their rms values stay constant after
exponential amplification.
Even though the magnetic field amplification is maximal in
Run~R5, small-scale dynamo action weakens the formation of
large-scale vertical magnetic structures.
Earlier work \citep{BKKR12} demonstrated that the relevant
mean-field parameter proportional to the growth rate is
reduced to 2/3 of it original value when $\Rm>60$.
Therefore, $\Bfm$ is smaller than in Run~R4 and the
bipolar magnetic region is weaker.
On the other hand, $\Peff$ is actually more negative than in Run~R4.
The magnetic field produced by the small-scale dynamo reduces $\urms$
and, therefore, $\Rey$ and $\Beqz$ also .

In the Sun, the fluid and magnetic Reynolds numbers are expected to be
very large, allowing therefore a small-scale dynamo to operate even
though the magnetic Prandtl number might be small \cite[see,
e.g.,][]{B11,Rempel14}.
This might weaken the formation of bipolar regions due to NEMPI in
the Sun, but large $\Rey$ and $\Rm$ could also enhance the NEMPI
growth rate due to lower diffusion.
However, a reliable extrapolation of the interaction of NEMPI and the
small-scale dynamo is not possible at the moment, as the simulations
of both NEMPI and small-scale dynamo are still operating
in a regime that is too diffusive compared with the Sun.

\subsection{Dependence on imposed magnetic field strength}
\label{sec:impB}

\begin{figure}[t!]
\begin{center}
\includegraphics[width=\columnwidth]{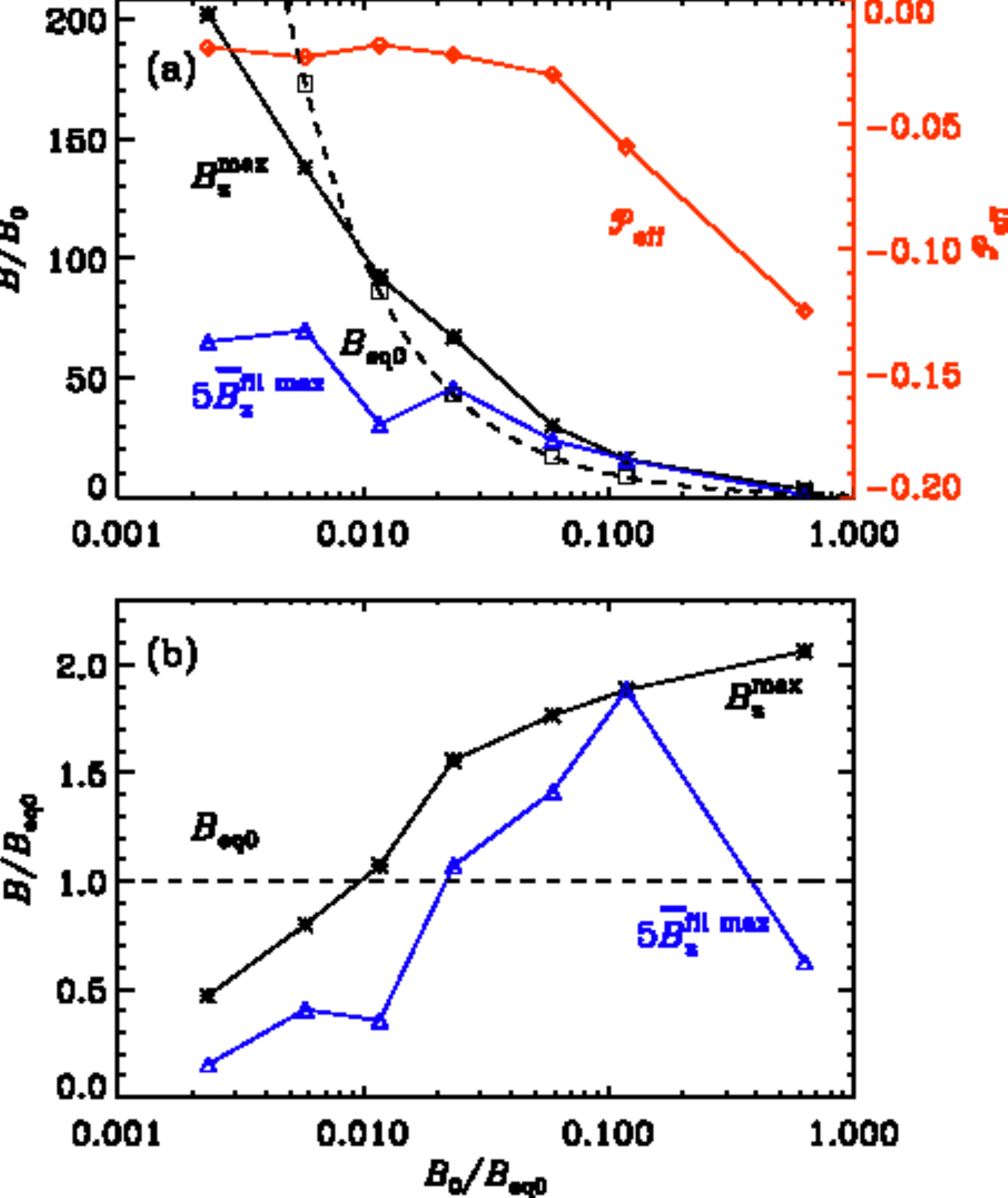}
\end{center}\caption[]{
Dependence of magnetic field amplification and effective magnetic pressure
on the imposed magnetic field $B_0/\Beqz$ for Set~B.
The magnetic field is normalized by the imposed magnetic field $B_0$
(a) or by the equipartition field strength at the surface $\Beqz$ (b).
Otherwise the legend is the same as in \Fig{pmB_strat}.
}\label{pmB_impB}
\end{figure}

In the runs of \cite{WLBKR13} and in all runs of Sets~A, R, and S, we
impose a weak horizontal magnetic field.
The strength of this field is less than $1/40$ of the equipartition
field strength at the surface, i.e., the ratio between it and
the equipartition field strength is more than $1/200$ at the bottom of
the domain in the case of Run~A5.
To investigate the dependence on the imposed magnetic field strength, we
vary the imposed field in the runs of Set~B from $B_0/\Beqz=1/430$ to
2/3; see the eighth column in \Tab{runs}.
In \Fig{pmB_impB}, we show the dependence of the magnetic field and
$\Peff$ on $B_0/\Beqz$.
In Run~B1, where $B_0$ is weak, the field strength is high
enough to serve as an initial magnetic field for NEMPI to work, but
only weak flux concentrations are formed.
Therefore the field amplification is around two times smaller than the
equipartition field strengths.
The large-scale field is even more than 30 times lower than the
equipartition field, therefore, preventing the formation of high flux
concentrations.
However,  here the large-scale magnetic energy also shows an exponential
growth; see \Fig{pbtsm}.
In Run~B2, the imposed field strength is sufficient to form bipolar
magnetic regions, even though the maximum vertical field strength is
just below the equipartition field strength.
An increase of the imposed field leads to a stronger magnetic field
inside the flux concentration compared with $\Beqz$, see
\Fig{pmB_impB}(b), but weaker fields compared to the imposed magnetic
field; see \Fig{pmB_impB}(a).
This is plausible:
if a weak field is imposed, just a small fraction of the turbulent
energy is used to concentrate and amplify the field to higher field
strength.
This leads to a high ratio of $B_z^{\rm max}/B_0$, but to a low
ratio of $B_z^{\rm max}/\Beqz$.
In Run~B6, where the imposed field is strong, a small concentration
and amplification of $B_z^{\rm max}/B_0=16$ can lead to strong
superequipartition field strengths of $B_z^{\rm max}/\Beqz=1.9$.
For a strong imposed magnetic field, when the derivative
$\dd\Peff/\dd\beta^2$ becomes positive, NEMPI cannot be excited
and magnetic spots are not expected to form \citep{KBKMR13}.
In particular, in Run~B7 the magnetic field becomes too strong, so
no bipolar magnetic region can be built up.
This leads us to conclude that there is an optimal imposed field strength,
which is between $B_0/\Beqz=0.012$ and $0.12,$
when superequipartition magnetic flux concentrations and bipolar
magnetic structures can be formed.

\begin{figure*}[t!]
\begin{center}
\includegraphics[width=0.33\textwidth]{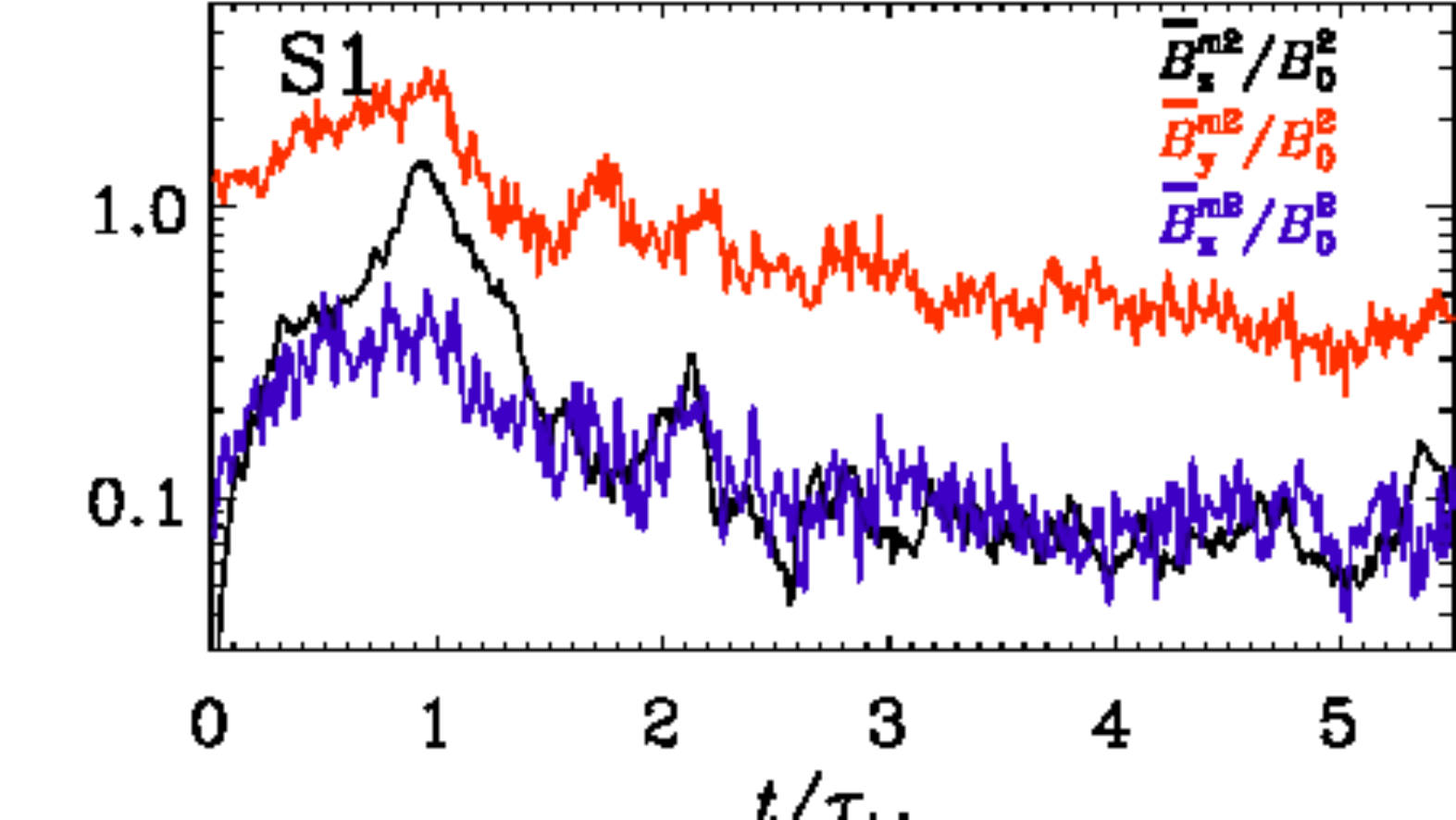}
\includegraphics[width=0.33\textwidth]{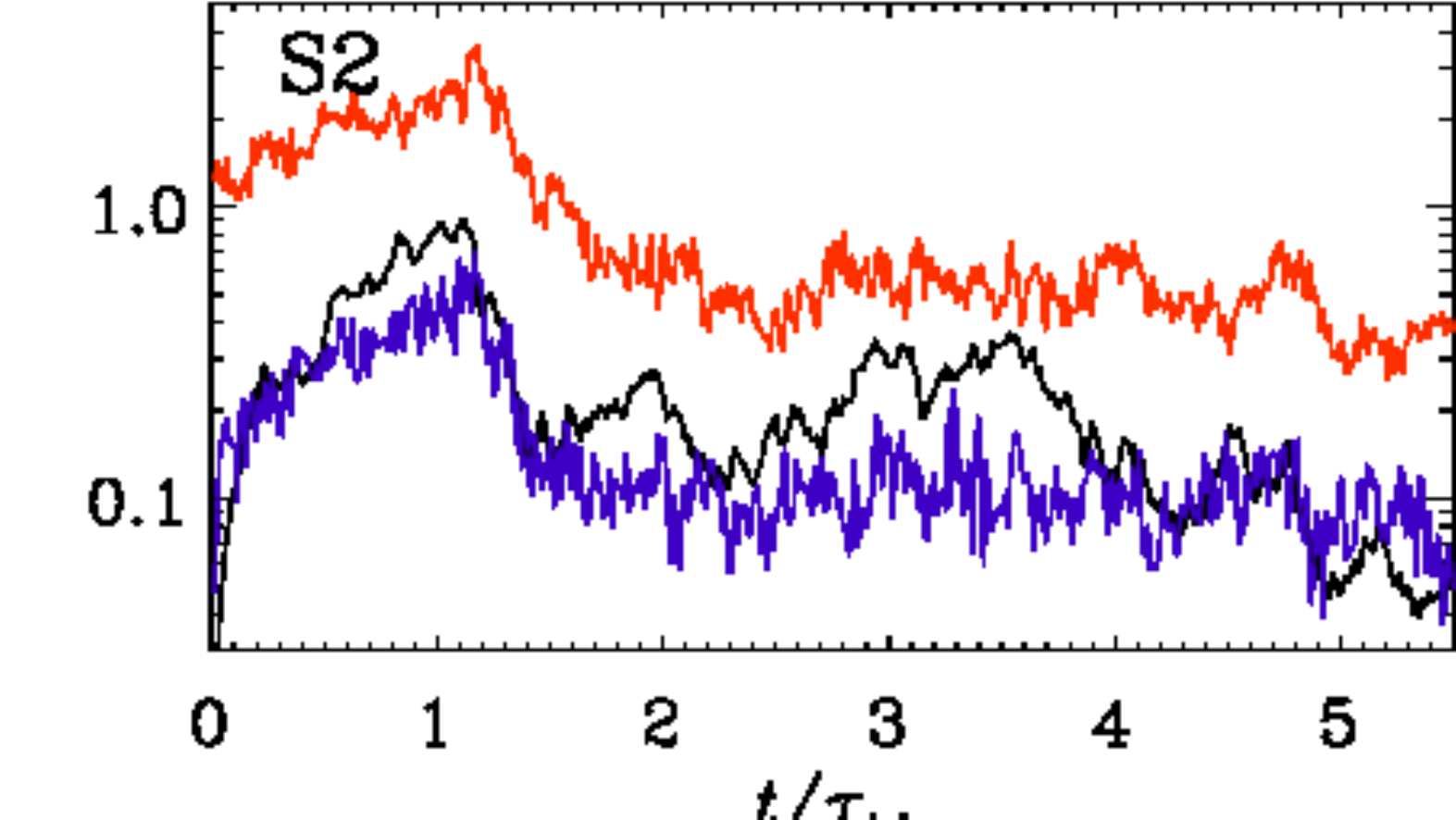}
\includegraphics[width=0.33\textwidth]{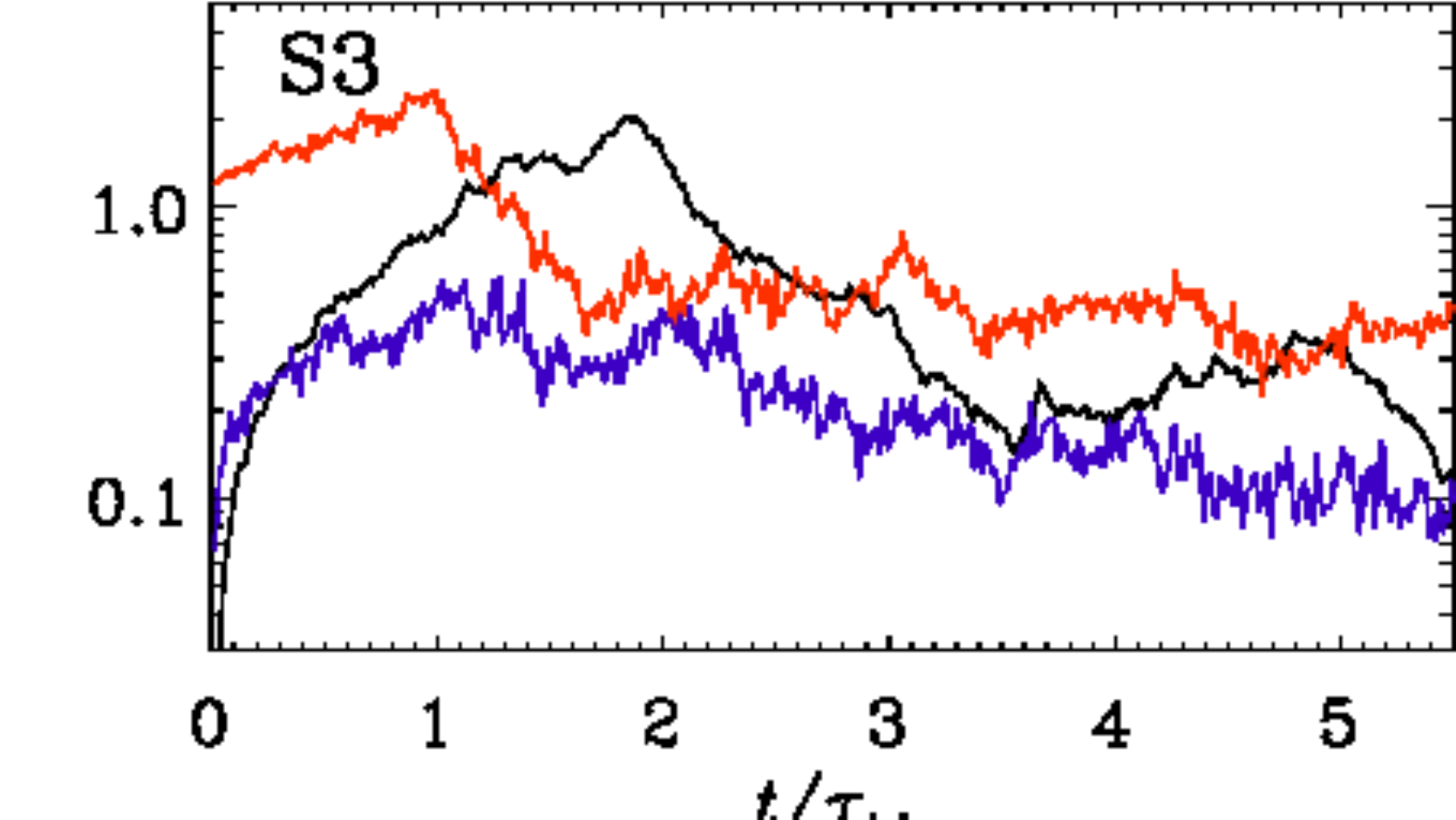}
\includegraphics[width=0.33\textwidth]{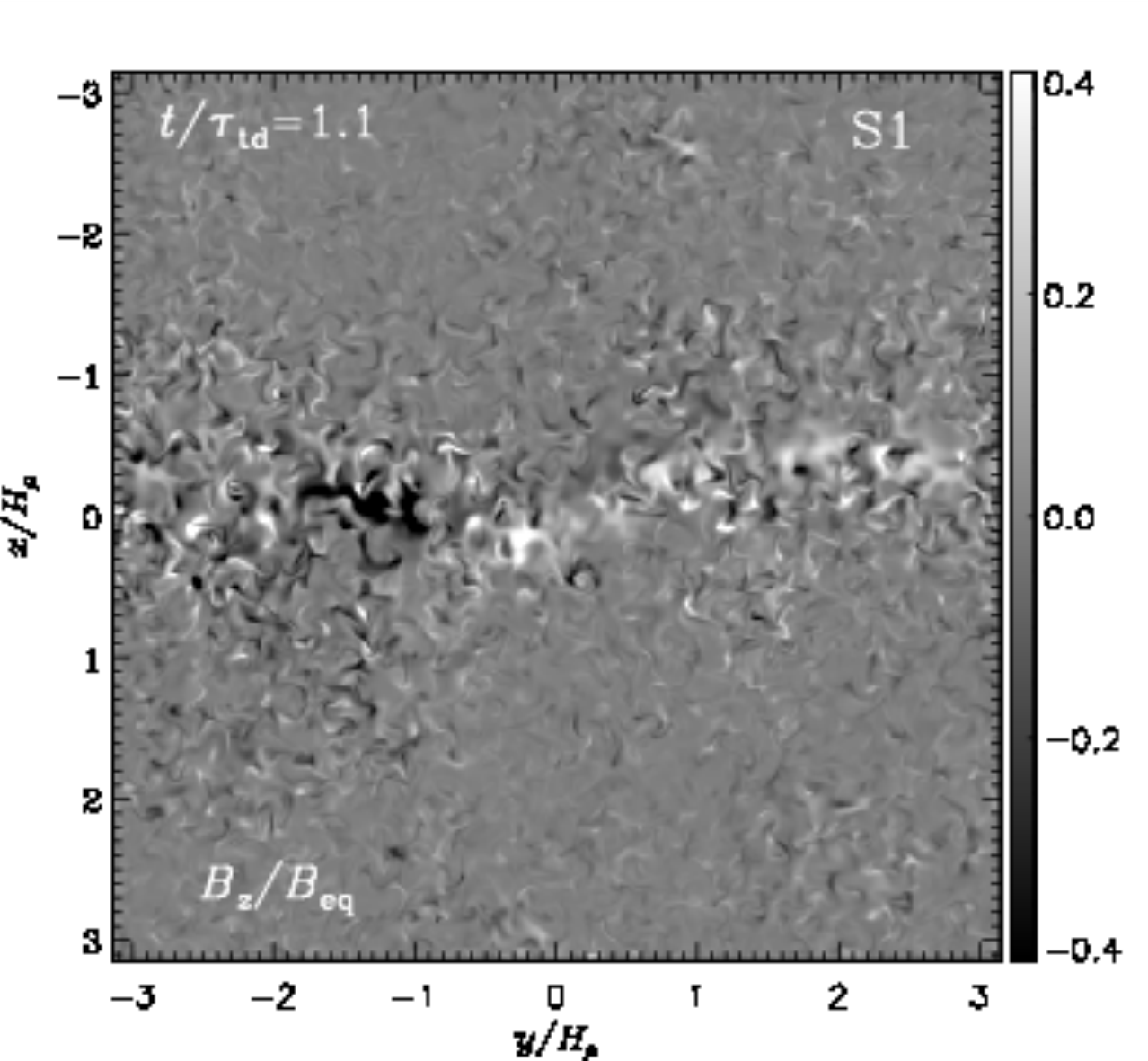}
\includegraphics[width=0.33\textwidth]{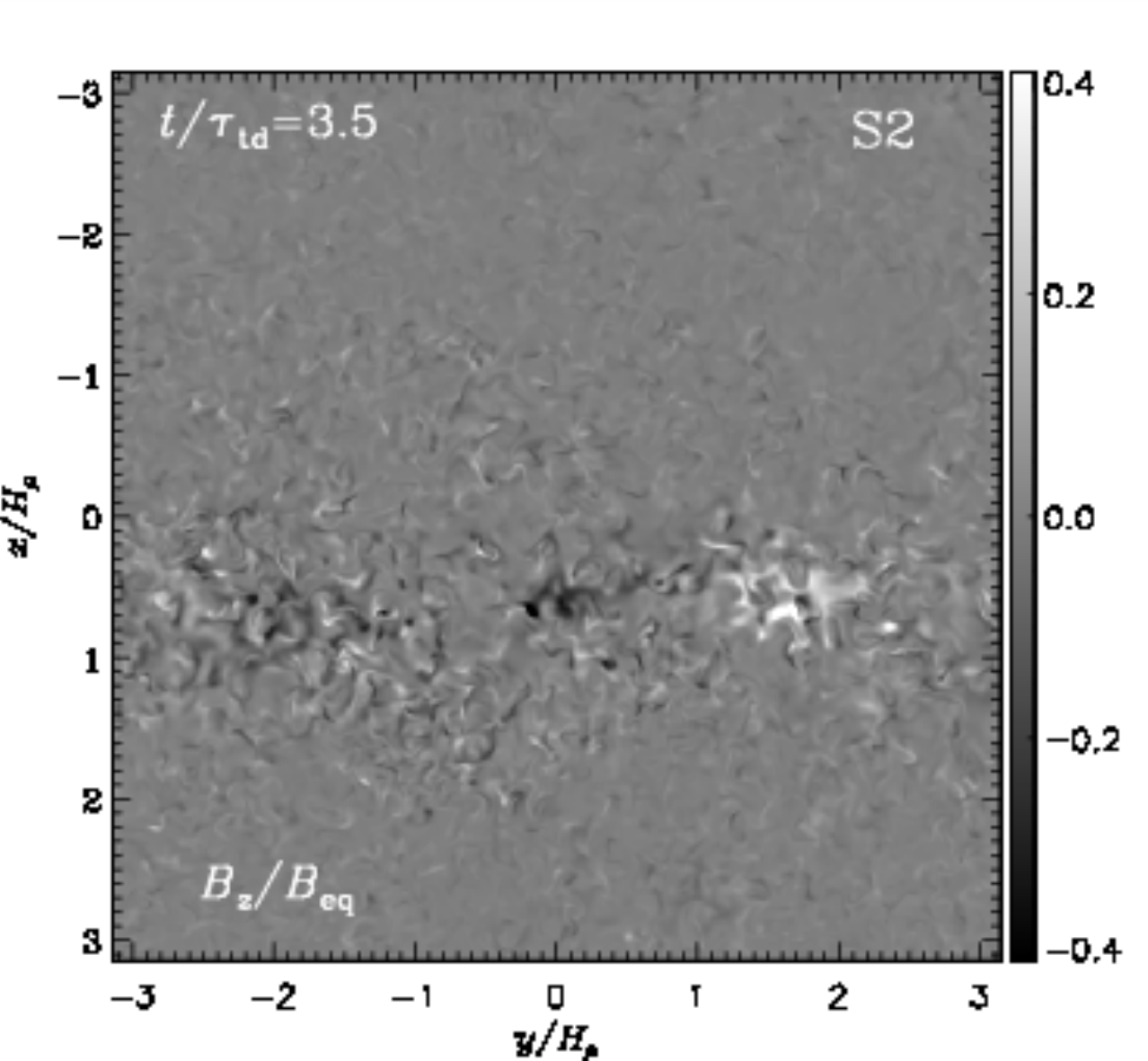}
\includegraphics[width=0.33\textwidth]{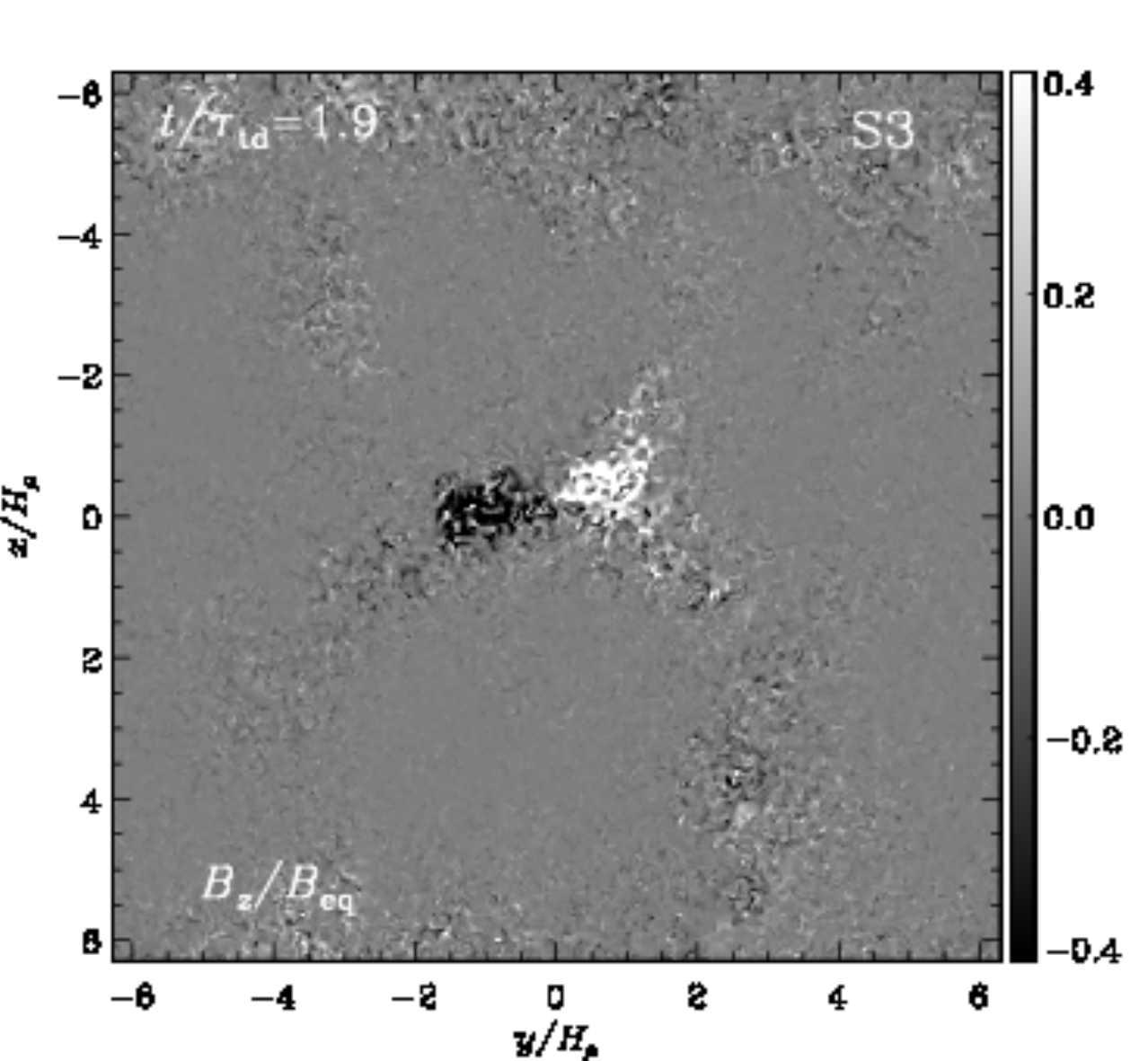}
\includegraphics[width=0.33\textwidth]{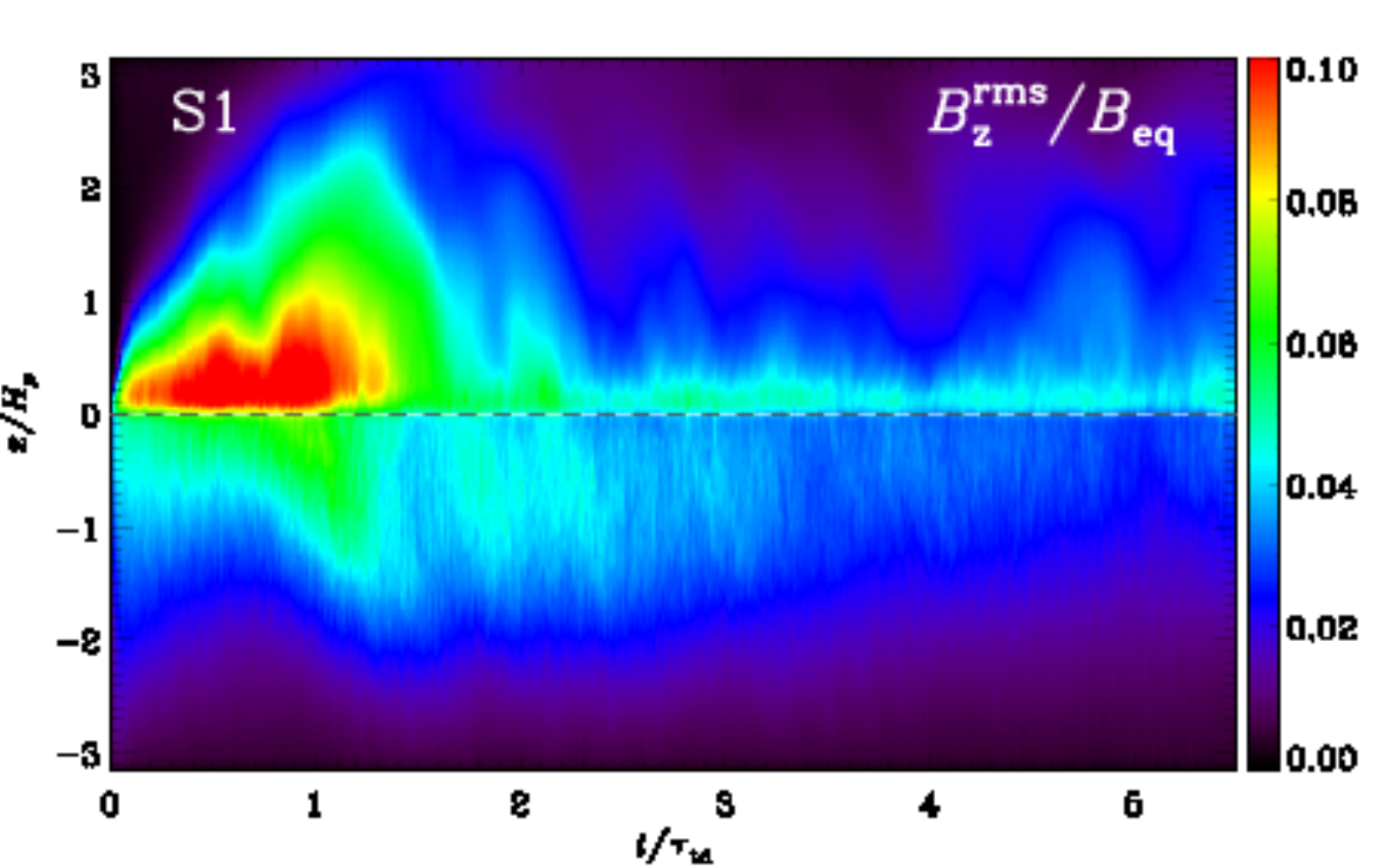}
\includegraphics[width=0.33\textwidth]{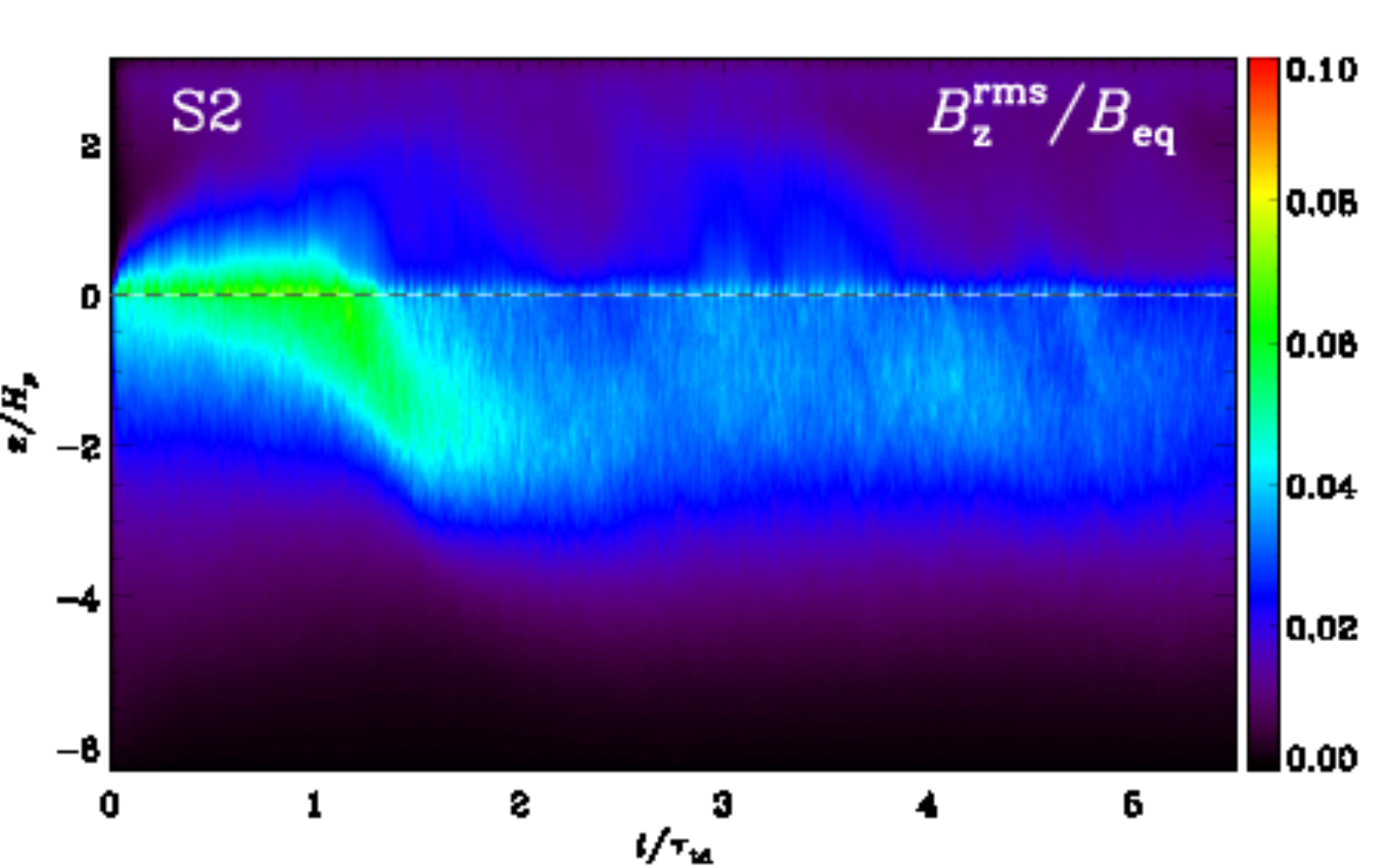}
\includegraphics[width=0.33\textwidth]{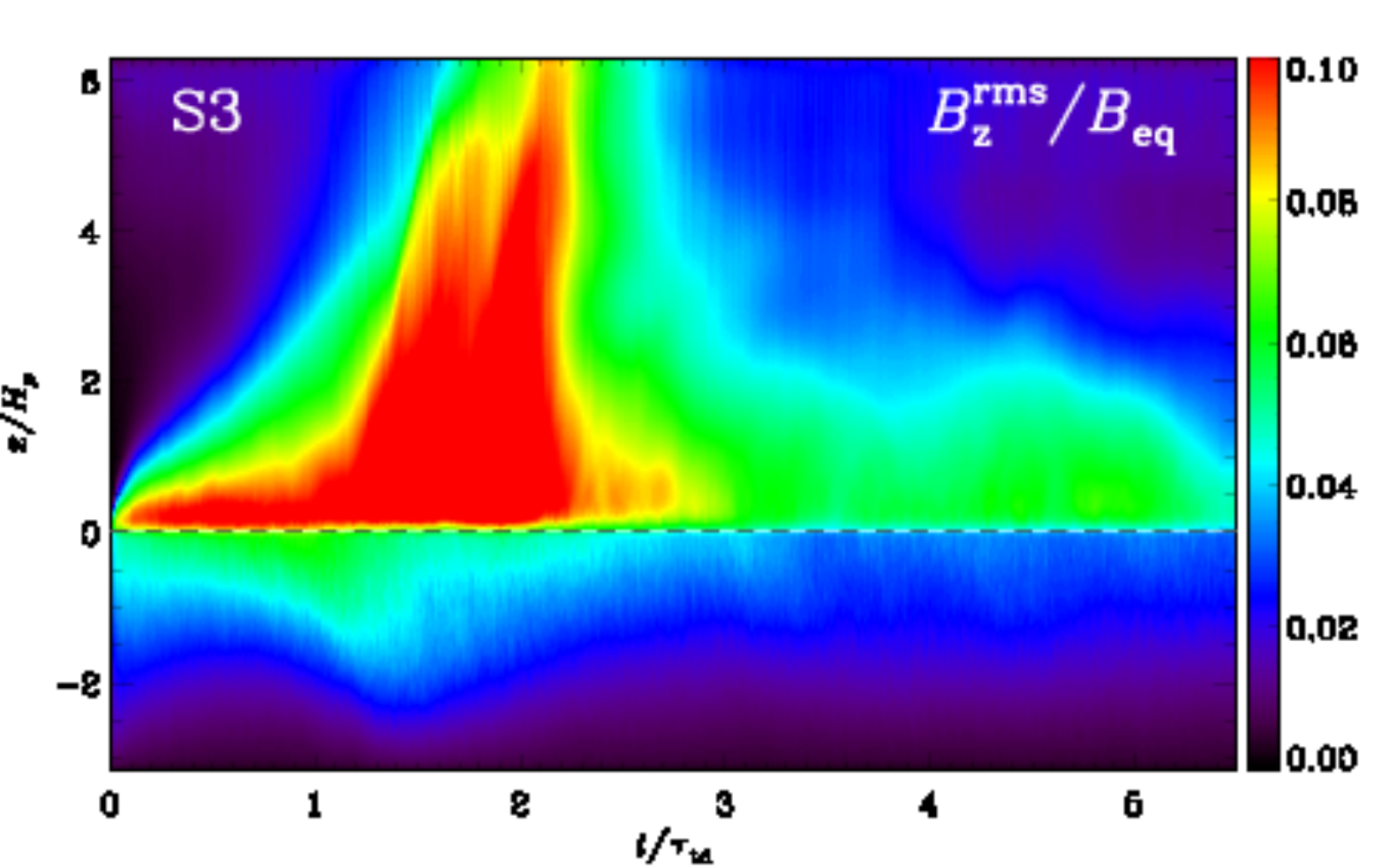}
\end{center}\caption[]{
Formation of bipolar regions for three different sizes ({\it left column}:
S1, {\it middle}: S2, {\it right}: S3).
{\it Top row}: the same as in \Fig{pbtsm}, but for Set~S.
{\it Middle row:} normalized vertical magnetic field $B_z/\Beq$ plotted
at the $xy$ surface ($z=0$) at times, when the bipolar regions are the
clearest.
{\it Bottom row:} vertical rms magnetic field $B_z^{\rm
  rms}/\Beq=\bra{B_z^2}_{xy}^{1/2}/\Beq$ normalized by the local
equipartition value as a function of time $t/\tautd$ and height $z$.
The black-white dashed line in the bottom row marks the surface ($z=0$).
}\label{size}
\end{figure*}

As expected, the effective magnetic pressure $\Peff$ decreases as we
increase the magnetic field, except in Run~B1, where it is slightly
smaller than in Run~B2, see \Fig{pmB_impB}(a).
Furthermore, $\tautdm$ shows a dependence on the imposed field.
For stronger imposed fields, $\tautdm$ becomes shorter,
indicating a higher growth rate of the instability as seen in the
steeper growth in \Fig{pbtsm}.
However, this seems to be only true for strong concentrations; the
weak concentrations in Run~B1 have a smaller $\tautdm$ than those in
Run~B2.
This is probably related to the two distinct growth rates seen in the
magnetic energy of \Fig{pbtsm}.
Furthermore, a stronger magnetic field suppresses turbulent motions, as
seen from the decrease of $\Rey$ (sixth column of \Tab{runs}) and
therefore it decreases the turbulent magnetic diffusivity.
This influences the values of $\tautd$ and, therefore,
$\tautdm$, but also allows for a higher growth rate.

\subsection{Dependence on box size}
\label{sec:size}

To investigate how the formation of bipolar regions of \cite{WLBKR13} and
in the present work depends on the chosen box size, we change the vertical size as
well as the horizontal size; see Set~S in \Tab{runs}.
In \Fig{size}, we plot, for all cases of Set~S, the magnetic energy of all
three components in the large-scale field (top row), the vertical
magnetic field at the time of clearest formation of bipolar structures
(middle row), and the evolution of the vertical rms magnetic fields as
functions of time and height (bottom row).
In Run~S1, we reduce the vertical size of the coronal envelope from $2\pi$ to $\pi$
keeping the other sizes the same; see \Fig{pbeq} for the vertical
profile of $\Beq$.
This change has only a small effect on the formation of bipolar
regions.
Comparing Run~S1 with Run~A5, $B_z^{\rm max}/B_0$ is reduced from 67
to 52 in Run~A5 and the large-scale field $\Bfm/B_0$ from 9.2 to 7.8,
whereas the value of $\tautdm$ stays nearly the same.
The structure of the bipolar regions is similar, but these regions
seem to be more concentrated in Run~A5.

As a second case (Run~S2), we use the setup of Run~S1 and extend the
height of the turbulent layer from $\pi$ to $2\pi$.
The value of the density at the surface stays the same, so the
stratification extends to higher values of density in the lower layers.
Also, the density contrast changes accordingly from 23 in the turbulent layer
with a vertical extension of $\pi$ to 512 with a vertical extension of
$2\pi$.
This leads to a small increase of $\urms$ and, therefore,
to a corresponding slight increase of $\Beqz$; see \Fig{pbeq}.
The maximal field amplification of $B_z^{\rm max}/B_0$
inside the flux concentration is higher than
in Run~S1, but still lower than in Run~A5.
The maximum of the large-scale magnetic field $\Bfm/B_0$
is half as low as in Runs~S1 and A5.
The bipolar regions are weaker and are more diffused.
As can be seen in the bottom row of \Fig{size}, only a weak concentration
of vertical magnetic field is observed.

As a third case (Run~S3), we extend the horizontal size of the box from
$2\pi\times2\pi$ to $4\pi\times4\pi$; otherwise the setup of the run
is the same as Run~A5.
 In the top row of \Fig{size}, we already see a strong excess of vertical
magnetic energy in the large-scale field compared to the horizontal components
with a maximum around $t/\tautd=2$.
Indeed, this behavior can also be found by looking at the maximum of
the vertical magnetic field and the
large-scale vertical magnetic field at the surface; see \Tab{runs}.
$B_z^{\rm max}/B_0$ is much higher than in Run~A5, and
$\Bfm/B_0$ reaches higher values than in all other runs.
The vertical magnetic field at the surface shows a clear bipolar region
with well-concentrated poles.
The size of the bipolar region is comparable with the size in the other runs and, therefore,
it is independent of the horizontal size of the domain.
The strong concentration of vertical magnetic field causes a strong
response in the coronal envelope.
In a box with twice the horizontal extent,
the magnetic energy is four times larger than that of the imposed magnetic field.
The more magnetic energy becomes available, the more magnetic flux can be concentrated.
This also means, that the instability operating in these simulations is
more efficient to concentrate flux in the horizontal direction than in
the vertical direction, as seen in Run~S2.
In all three cases, the formation of bipolar regions can be associated
with an exponential growth of the large-scale vertical magnetic
energy, as seen from the top row of \Fig{size}.
Their growth rates are similar, but the resulting formation is
different.
Run~S3 exhibits the strongest large-scale magnetic field of all
simulations with a horizontal imposed field, but the growth rate
is smaller than in Run~A5.
However, the duration of exponential growth in Run~S1 is twice that
of Run~A5, allowing the field to grow to much higher values
than in Run~A5.

\subsection{Dependence on field inclination}
\label{sec:incl}
\begin{figure*}[t!]
\begin{center}
\includegraphics[width=0.34\textwidth]{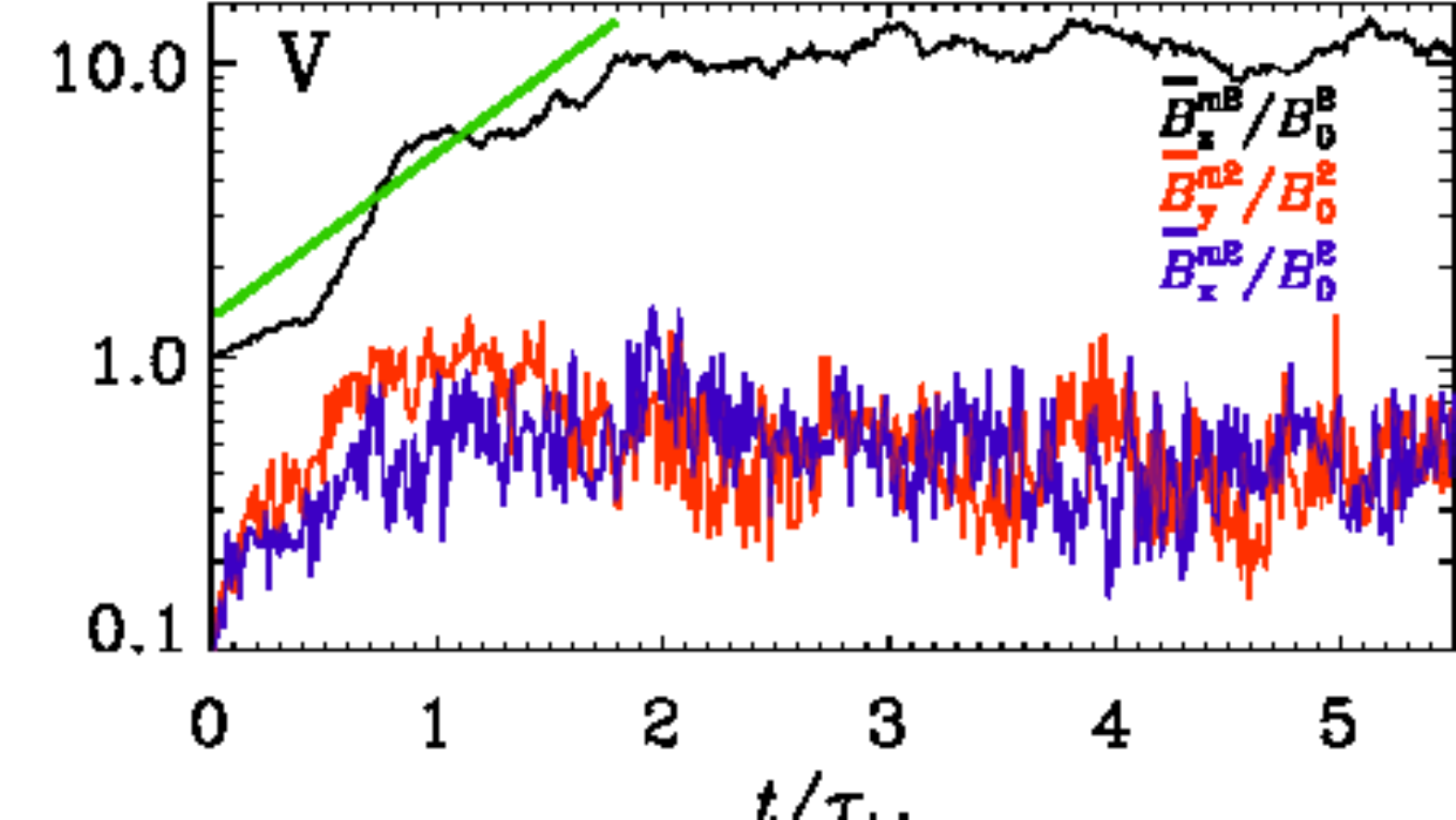}
\includegraphics[width=0.34\textwidth]{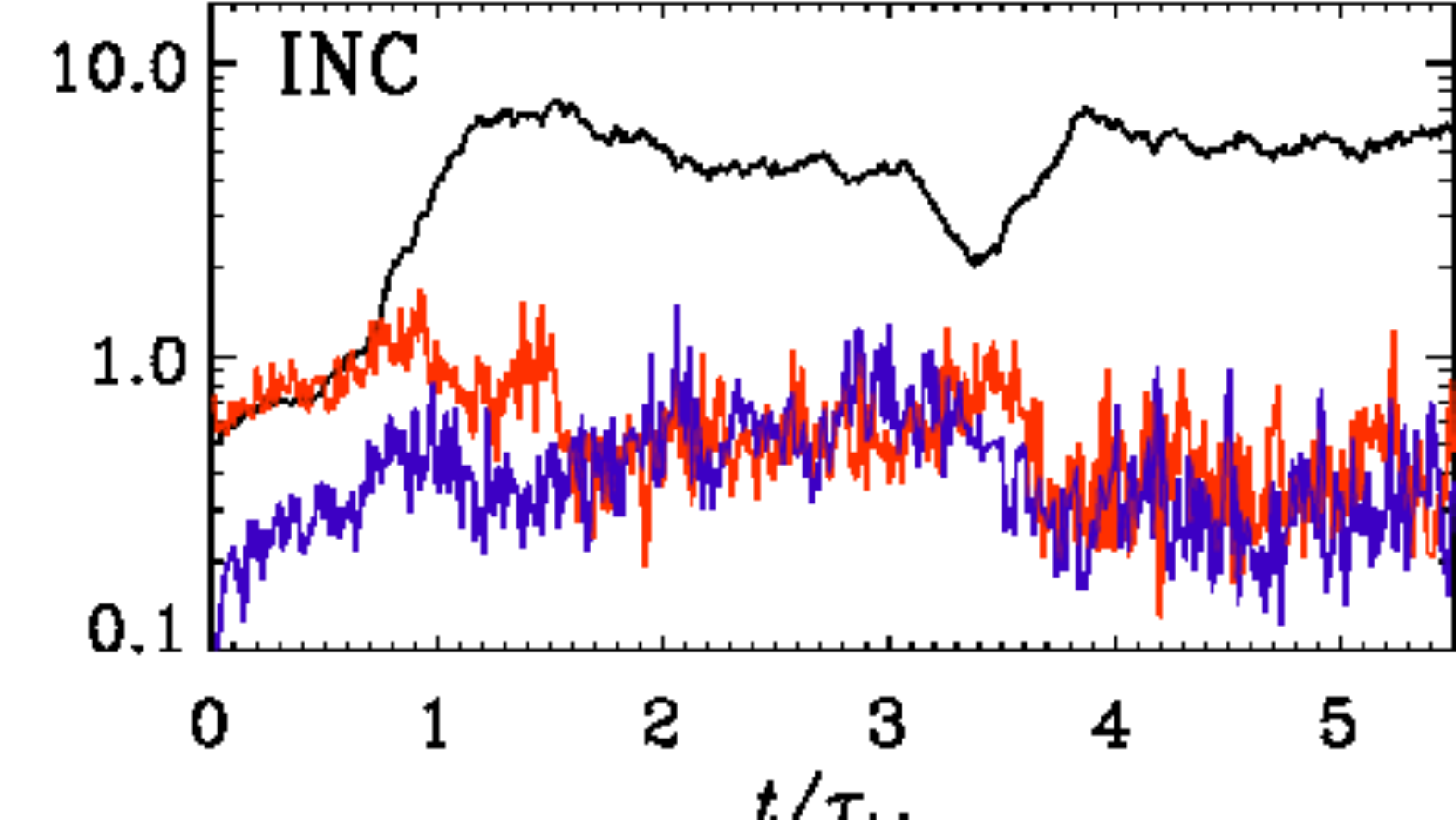}
\includegraphics[width=0.33\textwidth]{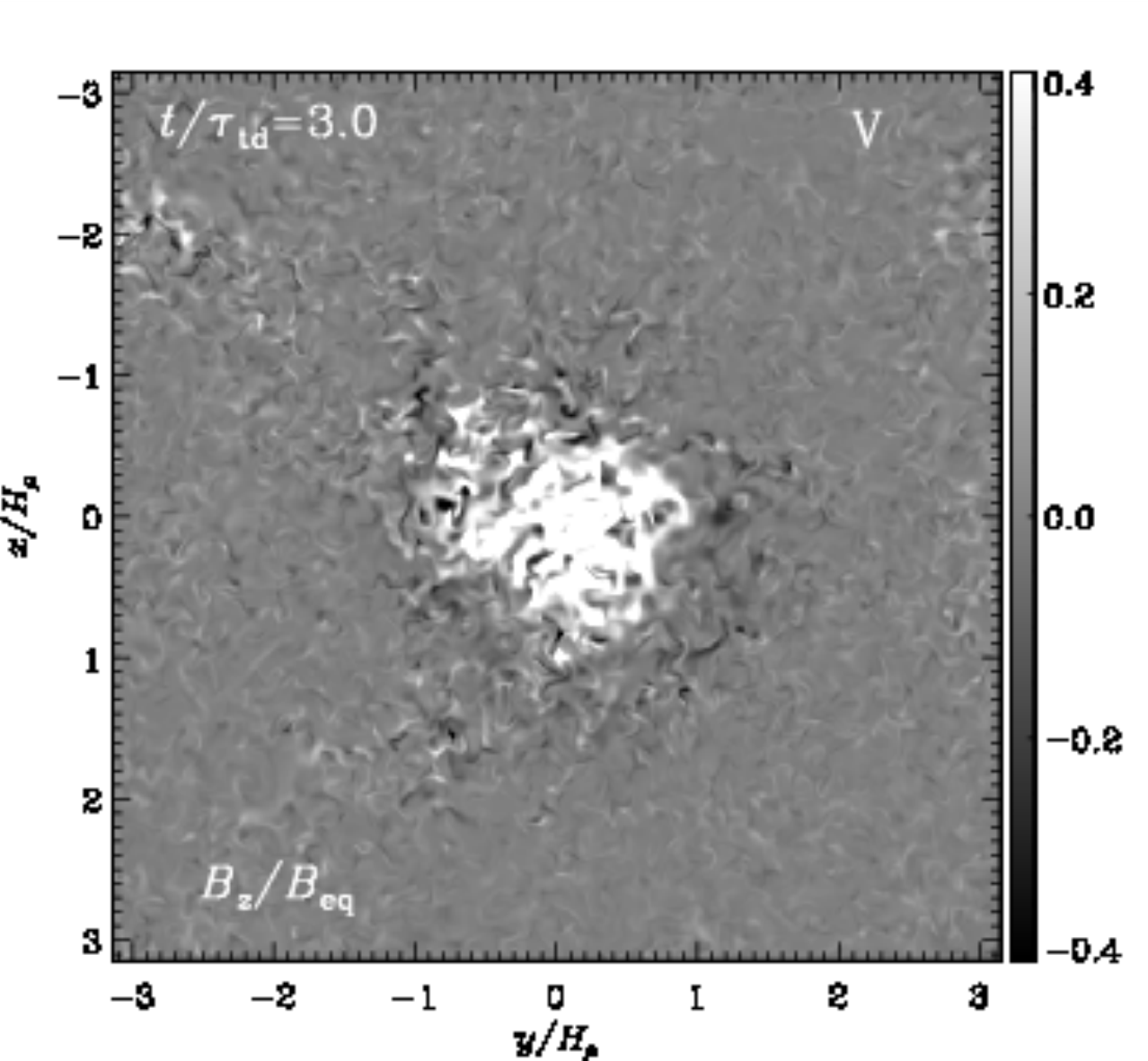}
\includegraphics[width=0.33\textwidth]{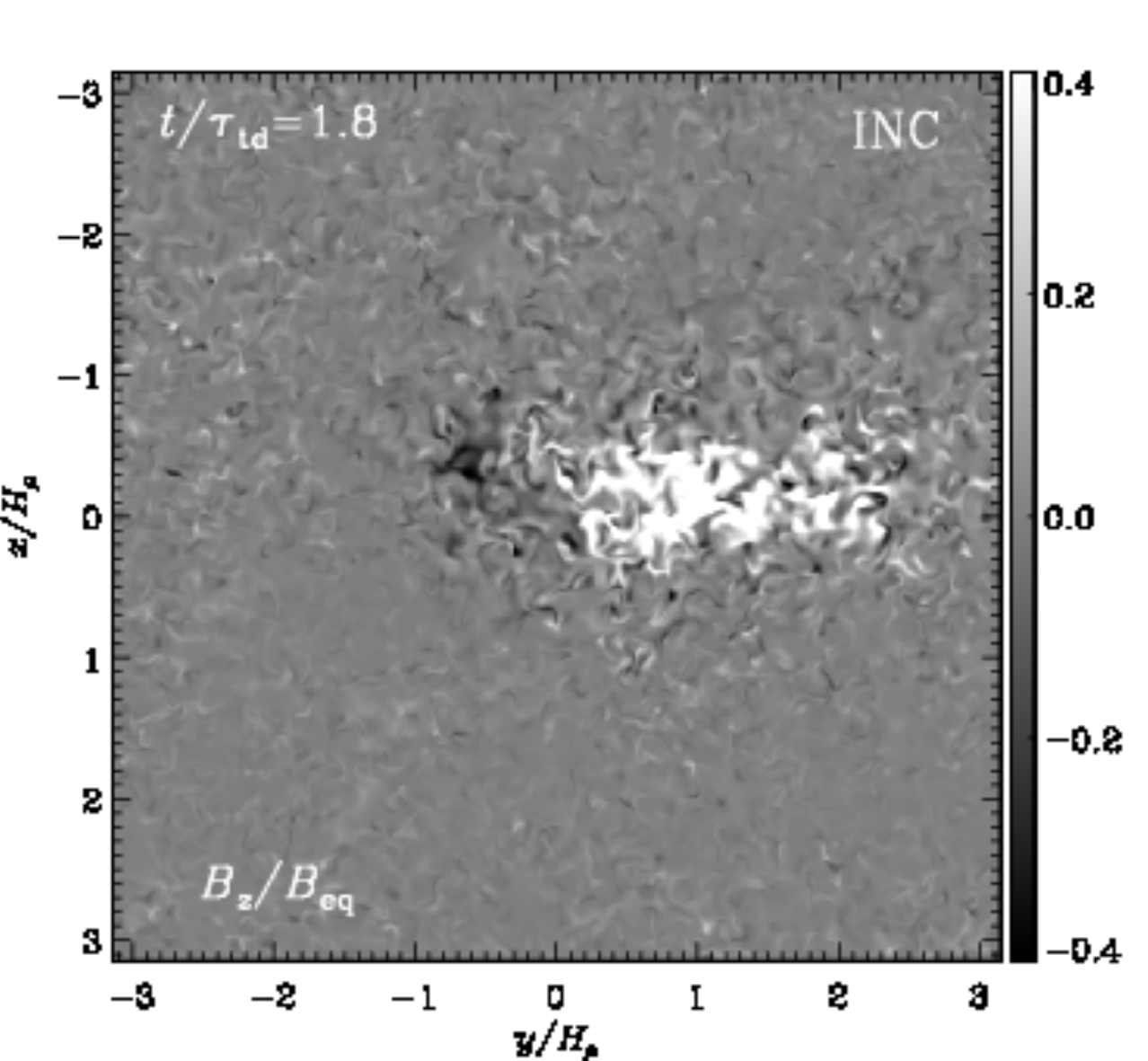}
\includegraphics[width=0.33\textwidth]{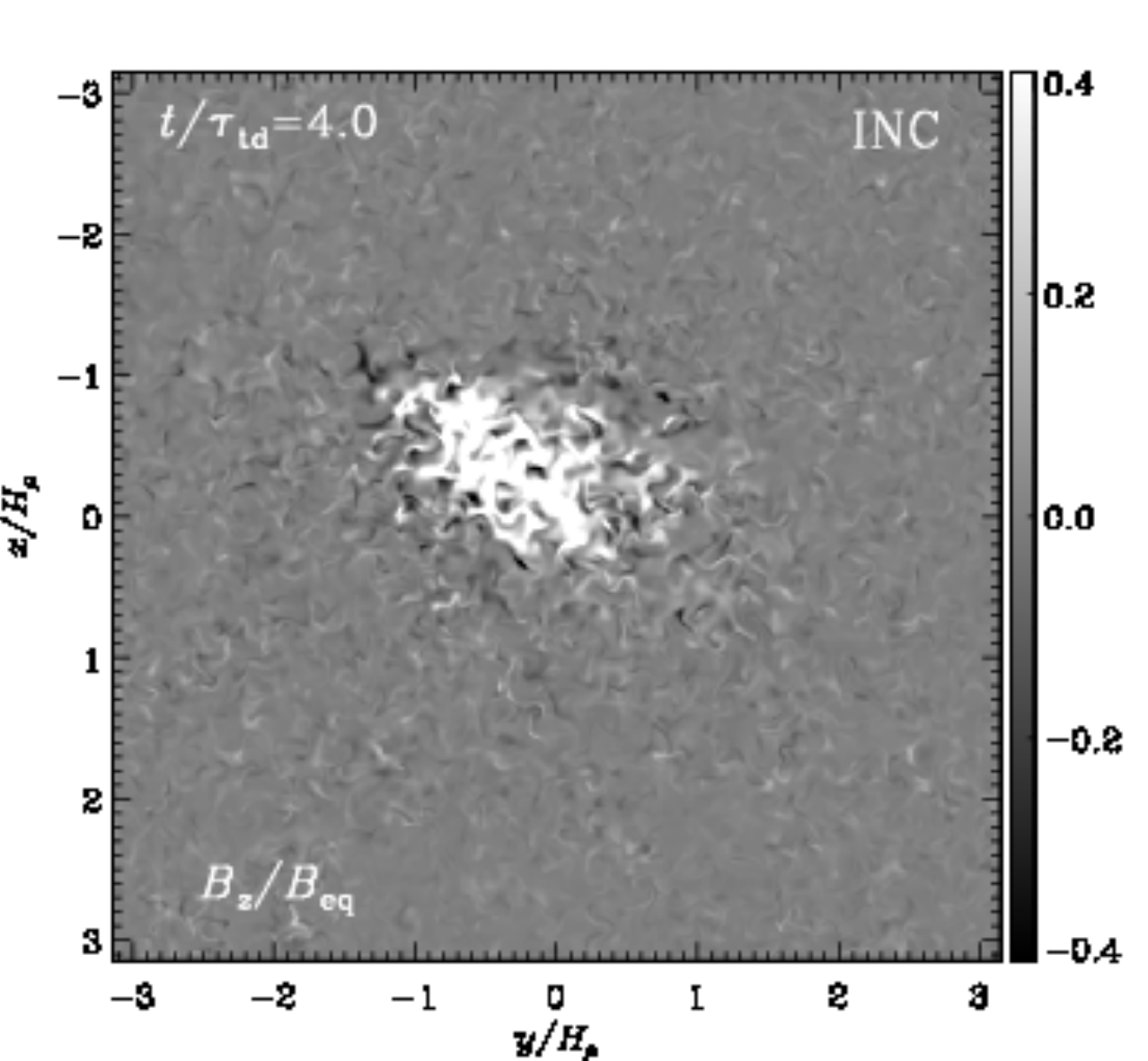}
\end{center}
\caption[]{
Formation of bipolar regions for two different field inclinations,
{\it left-hand side} with purely vertical field (Run~V) and {\it right-hand} side
with y-z inclination (Run~INC).
{\it Top row}: the same as in \Fig{pbtsm}, but for Runs~V and INC.
The straight green line for Run~V illustrates the exponential growth of
the energy in the vertical large-scale magnetic field.
{\it Bottom row:} normalized vertical magnetic field $B_z/\Beq$ plotted
at the $xy$ surface ($z=0$) at times when the bipolar regions are the
clearest.  Run~INC is shown for an early time ($t/\tautd=1.8$)
and a later time ($t/\tautd=4.0$) to illustrate the change from a bipolar to monopolar
structure.
}\label{vert}
\end{figure*}

In all of the runs mentioned above,
we imposed a horizontal magnetic field.
This leads to the formation of bipolar regions.
In this subsection, we also study the cases of
an imposed vertical and inclined field.
For the vertical field (Run~V), we set $\BB_{\rm imp}=(0,0,B_0)$ with
the same field strengths and the same hydrodynamic quantities as in
Run~A5.
As a result, the instability produces a single magnetic spot instead of
a bipolar region.
Because the magnetic energy is now concentrated in one single spot,
the maximum magnetic field reaches nearly two times the values of
Run~A5 and more than two times the equipartition field strength.
The field strength in the large-scale field is even three times stronger
than in Run~A5; see \Tab{runs}.
In the bottom row of \Fig{vert}, we plot the vertical magnetic field
at the surface at the time of the clearest appearance.
The single spot has a larger spatial extension and is more
concentrated as in Run~A5.
Also here, we can find an exponential growth of the magnetic energy in
the vertical field, as shown in top row of \Fig{vert}.
We estimate the growth rate to be around $0.7/\tautd$, which is two
times lower than for Run~A5.
Even though the growth rate is smaller than in Run~A5, the duration is
longer than in Run~A5, leading to a stronger magnetic field.
Also, the vertical field has already increased from a strength of $B_0$ in
Run~V, whereas in Run~A5 there is no vertical magnetic field in the
beginning of the simulation.
An additional difference from
Run~A5 is that the spot does not
decay after some time.
Instead it stays roughly the same after $t/\tautd=2$.

Similar singular spots were already found by \cite{BKR13}.
There, the authors use a similar model with imposed vertical magnetic
field, except their turbulent layer has a vertical extension of $2\pi$
instead of $\pi$ and no coronal envelope.
In the runs of \cite{BKR13}, where they use the same imposed field
strengths, the maximum of the field strength is also more than double, and
$\Bfm$ is close to the equipartition field strength at the surface.
However, looking at their Fig. 2, the large-scale magnetic field
grows exponentially up to $t/\tautd=1.5$ when the saturation set
slowly in, whereas in our Run~V the saturation sets
in a bit later in time, $t/\tautd=2$.
Nevertheless, our estimated growth rate of about $0.7/\tautd$ is half
the value found in \cite{BKR13}.

As a second case (Run~INC), we impose an $yz$ inclined magnetic field with
the strength of $B_0$ ($\BB_{\rm imp}=(0,B_0,B_0)/\sqrt{2}$).
As expected, we find the generation of a weak negative and a strong
positive polarity in the bipolar region, as shown in the lower row of
\Fig{vert}.
However, this is only the case in the first half of the simulation.
Then the weak negative polarity reconnects with the stronger positive
polarity to form a single spot that does not diffuse away, which is similar to  Run~V.
Because of the field reconnection, the resulting single spot is weaker than
in Run~V; see bottom row of \Fig{vert}.
This behavior can be also seen in the evolution
of the three components of the large-scale magnetic energy;
see top row of \Fig{vert}.
Until $t/\tautd=0.8,$ the $y$ and $z$ components grow exponentially
with a similar growth rate, but then the $z$ energy component increases the growth rate
that is properly related to the emergence of horizontal flux to form vertical flux.
At $t/\tautd=1.8$,
nearly at the end of the exponential growth stage, a weak negative
and a strong positive pole  form.
At the $t/\tautd=3.5-3.8$, after a decrease of all components, only
the vertical field recovers.
This coincides with the diffusion of the weak negative spot.
The behavior of an inclined field is exactly that can be expected
from the two cases with imposed horizontal and vertical fields.
For the horizontal field, a bipolar region is formed, which
decays after several turbulent-diffusive times.
For the vertical field, a single spot is formed, which
does not diffuse.

\subsection{Formation mechanism}
\label{sec:form}

\begin{figure}[t!]
\begin{center}
\includegraphics[width=\columnwidth]{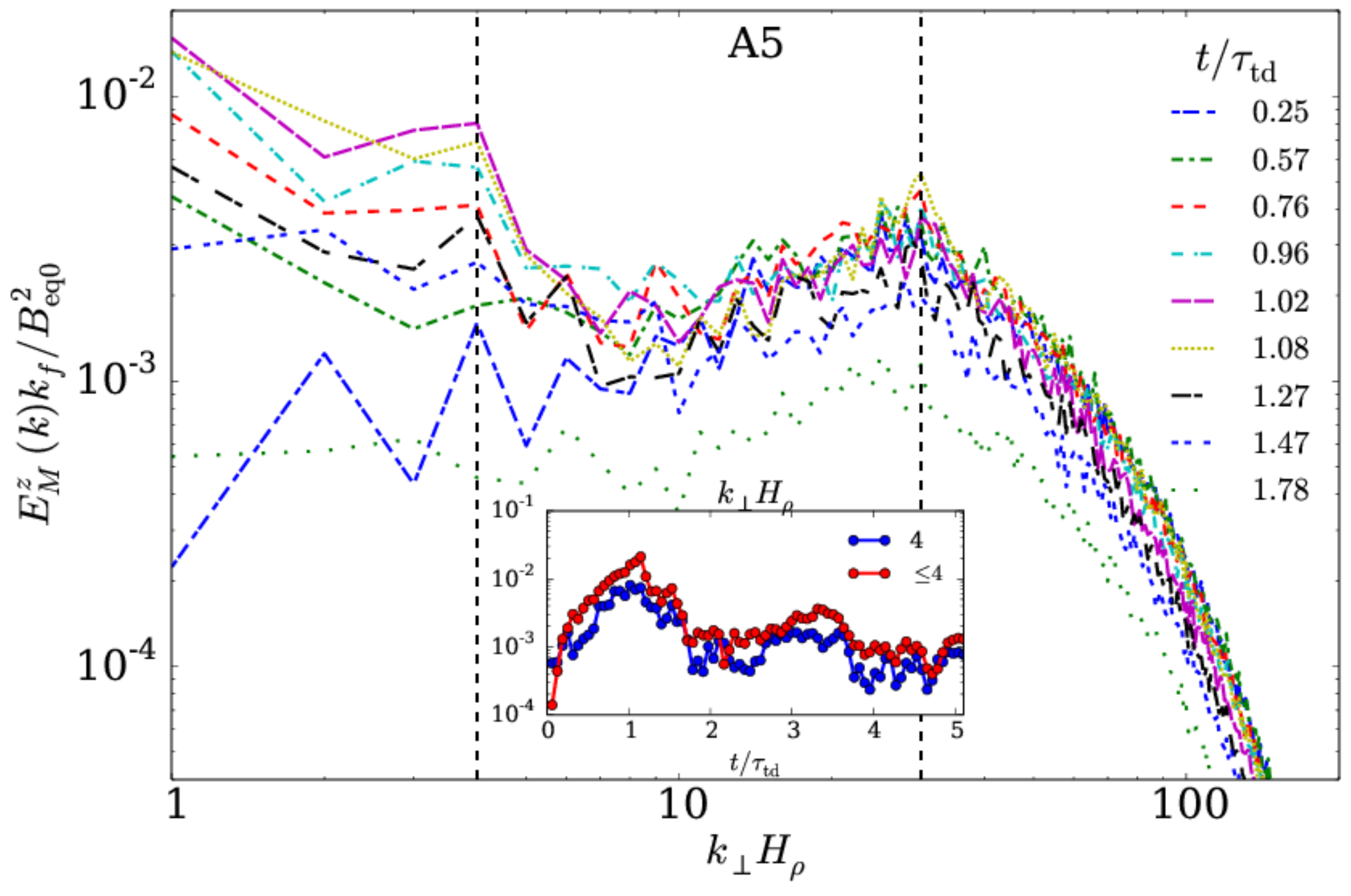}
\includegraphics[width=\columnwidth]{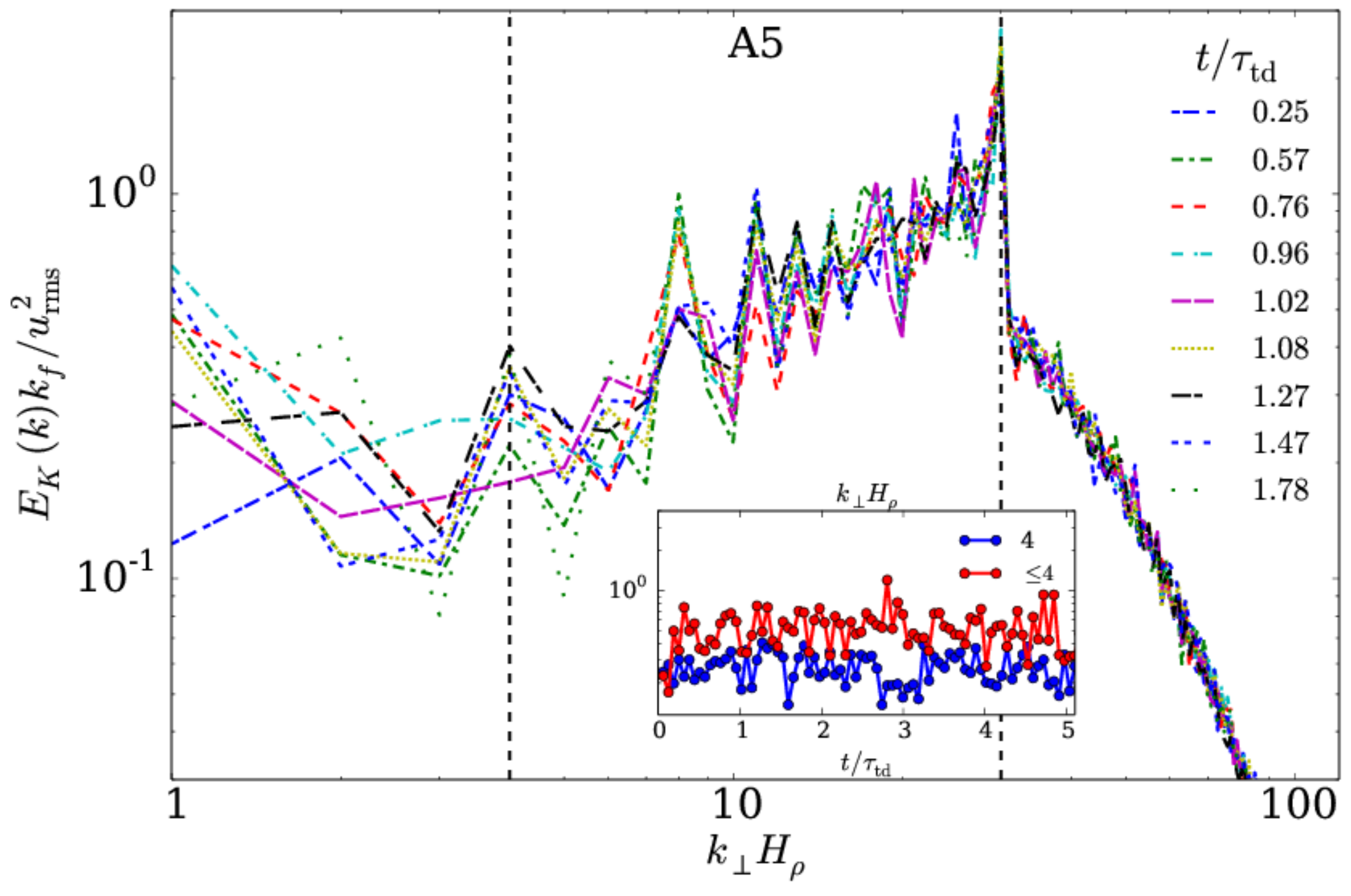}
\end{center}\caption[]{
Magnetic and kinetic power spectrum for Run~A5.
{\it Top panel}: spectrum of vertical magnetic energy $E^z_{\rm M}$
at nine different times around $\tautdm$ at the surface ($z=0$) as a
function of horizontal wavenumber $k_\bot$.
The inlay shows the vertical energy at $k_\bot H_\rho=4$
(blue line) and $k_\bot H_\rho\le4$ (red) as a function of
time $t/\tautd$.
{\it Bottom panel}: spectrum of the kinetic energy $E_{\rm K}$ plotted
 the same  as the top panel.
The vertical dashed lines indicate $k_\bot H_\rho=4$ and $k_\bot
H_\rho=\kf H_\rho=30$.
}\label{power}
\end{figure}

\begin{figure*}[t!]
\begin{center}
\includegraphics[width=\textwidth]{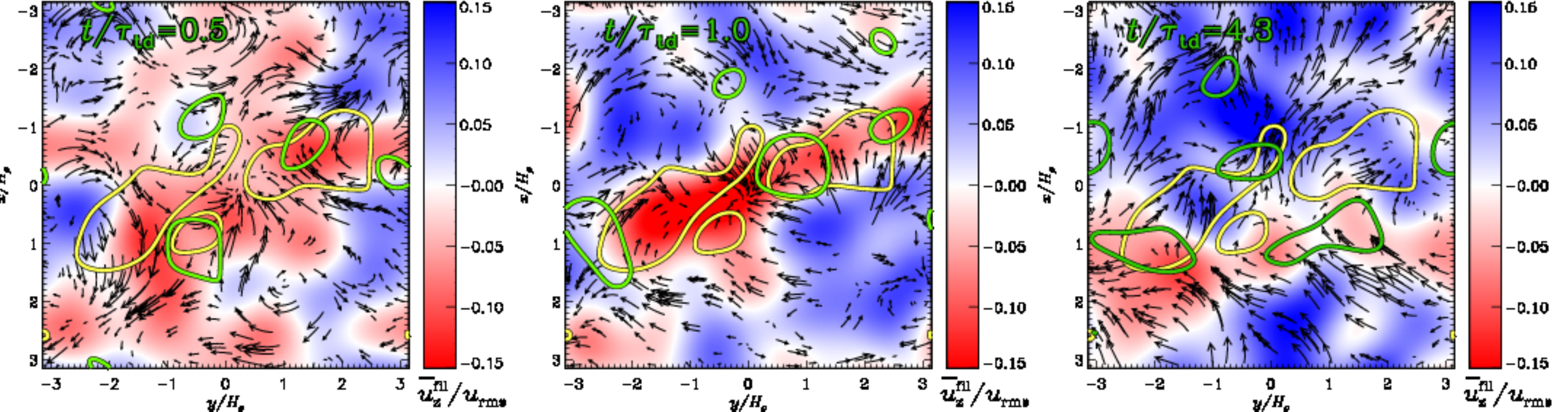}
\end{center}\caption[]{
All three large-scale velocity components
$\mean{u_x}^{\rm fil}$, $\mean{u_y}^{\rm fil}$, and $\mean{u_z}^{\rm
  fil}$ before ($t/\tautd=0.5$), at ($t/\tautd=1.0$), and after
($t/\tautd=4.3$) the occurrence of the bipolar regions
(compare with \Fig{strat}) in the $xy$ plane for Run~A5.
The vertical velocity is plotted as red (downflows) and blue
(upflows) and are normalized by the $\urms$ in the bulk of the
turbulent layer ($z\le0$).
The horizontal components of the velocity field are shown as arrows,
where the lengths corresponds to the strength of the flow.
Additionally, the contours of negative horizontal divergence is
plotted in green for all three times.
The yellow contours in all plots show the magnetic field at the time
($t/\tautd=1.0$) to guide the eye to the location of the bipolar region
formation.
}\label{flow}
\end{figure*}

We also investigate in this context the formation mechanism
leading to bipolar regions in the two-layer setup of stratified turbulence.
As discussed in \cite{WLBKR13}, the coronal envelope plays an
important role in the formation process.
However, the magnetic field, which gets concentrated, comes from the
turbulent layer. This is shown with the two runs in \cite{WLBKR13}, where
one is the same setup as Run~A5 of this work and one does not have  any
imposed field in the coronal envelope.
Both show flux concentrations of similar strength.
We also compare Runs~A5 and S1, where the only difference lies in the
size of the coronal envelope.
Both show similar field concentrations, where $\Bfm/B_0$ has nearly the same value.
Therefore, the size of the coronal envelope does not seem to have a strong
influence on large-scale magnetic field and the formation of bipolar
regions.

In the beginning of the simulation, the magnetic field is uniformly oriented in
the $y$ direction because of the imposed field.
The tangling of the magnetic field by turbulence also leads to field components
in the other directions in the turbulent layer.
This can been seen in \Fig{pbtsm} for most of the runs.
Furthermore, we can use the plots of $B_z^{\rm  rms}/B_0$ in
\Figs{strat}{size} to analyze the height distribution of the vertical
magnetic field in the formation process.
The vertical field is built up in nearly the entire turbulent layer, which is in
particular visible for Runs~A3 and A5 as blue shades at early times.
Then this vertical field gets concentrated and transported toward the
surface, as shown by the increase of dark purple shades in the turbulent layer
from the bottom toward the surface.
This field
evolves rapidly and leads to a flux concentration at the surface, which is visible as red shades.
This vertical magnetic field then rises through the coronal layer until
it decays and falls back toward the turbulent layer.
Also, in the turbulent layer the field is first concentrated toward the
surface, reaching the strongest peak of magnetic field and then
the field diffuses back into the turbulent layer.
These plots show clearly that the magnetic field originates from the
turbulent layer toward the surface and does not come from the
coronal envelope.
A little later, after the peak of vertical flux has dissolved, the magnetic
field from the coronal envelope falls toward the turbulent layer.
The coronal envelope is important, but mostly as a free boundary
condition for the magnetic field and the flow.

To illustrate how the magnetic and kinetic energies evolve at
different scales, we plot in \Fig{power} the spectrum of the energy in
the vertical magnetic field as well as the kinetic energy for Run~A5
for nine different times.
In both spectra, the normalized forcing wavenumber $\kf H_{\rho}$ is seen as a
local maximum.
In the magnetic spectrum, the forcing scale has the highest peak in
the beginning of the simulation.
Later, more and more energy is transported to larger scales
($k_\bot H_\rho<10$) until the energy for $k_\bot H_\rho<5$ becomes dominant.
This happens when $t\approx\tautd$, which is not surprisingly at the same
time, when the bipolar region is the strongest ($t/\tau=\tautdm$).
Afterward, the magnetic energy decays first at larger scales and then
at all scales.
This can also be seen in the inlay, where we plot the energy of the
vertical magnetic field at $k_\bot H_\rho=4$ (blue) and $k_\bot
H_\rho\le4$ (red).
There is a strong growth up to $t/\tautd=1$ and a decay to slower
values after that.
This means that the instability occurring in these simulations
transports vertical magnetic energy to large scales in the growing
phase.
This has also been seen in previous studies with imposed vertical
magnetic field \citep{BGJKR14} and seems to be analogous to the
inverse magnetic helicity cascade \citep{PFL76,B01}.
It suggests the use of a cutoff wavenumber of $\kc\le \kf/6$ to
represent the large scales of the magnetic field in our previous
analysis; see also \cite{BGJKR14} for a similar discussion.
In the kinetic spectrum, the forcing scale is  the highest
peak for all times.
There the energy of large scales are significant lower than those
of the forcing scale.
The kinetic energies on the larger scale show no strong time
evolution, we only notice a small increase in time at $k_\bot H_\rho\le2$.

To study the influence of the forcing scale, we perform one
additional run (Run~F), where we decrease $\kf$ from $30\,k_1$ to
$15\, k_1$.
As shown in the last row of \Tab{runs}, the maximum vertical magnetic
field strength is lower and the maximum of the large-scale vertical
field is slightly larger than in Run~A4.
However, reducing the forcing wavenumber by half has almost no
effect on the structure formation of bipolar regions via NEMPI.
This confirms the results of previous studies of \cite{BGJKR14},
where no strong dependence was found either. Even for significantly
smaller forcing scales, flux concentrations were obtained when
increasing the imposed field strength.
From the theoretical side, the forcing scale should have an influence
on the growth rate as well as on the turbulent magnetic diffusivity.
A detailed study of the dependence of bipolar regions formation
on the forcing scale is currently beyond the scope of this paper.

As found by \cite{BGJKR14}, flux concentrations due to NEMPI show
clear signatures of downflow patterns along the vertical magnetic
field.
Before and during the concentration of vertical flux, there exist
strong converging downflows.
Testing whether the bipolar magnetic region found in both \cite{WLBKR13} and
in the present work also coincides with such a flow pattern; we show in
\Fig{flow} the large-scale velocity at the surface for the time before
($t/\tautd=0.5$), at ($t/\tautd=1.0$), and after ($t/\tautd=4.0$) the
time of the strongest flux concentration for Run~A5.
For this we calculate the large-scale velocity with 2D horizontal
Fourier filtering $\meanf{\uu}$ to exclude the velocities due to forcing.
We use the technique described in \Sec{model} with a cutoff
wavenumber of $\kc\le \kf/6$.
The flows are shown together with the large-scale magnetic energy and the
horizontal divergence of the large-scale flow
($\partial_x\meanf{u_x}+\partial_y\meanf{u_y}$).
At $t/\tautd=0.5$, before the bipolar region has appeared, and we find
strong downflows (red) and horizontal converging flows in the vicinity
where the bipolar region later forms (yellow contours).
In the proximity of the downflows, there are also regions of negative horizontal
divergence (green contours).
At the time of the clearest appearance of the bipolar region
($t/\tautd=1.0$), the large-scale downflows are exactly at the location
of strong magnetic energy, indicating a tight connection between the
downflows and the formation of bipolar regions.
Furthermore, we find a strong horizontal flow streaming into the region
of large magnetic energy together with negative values of horizontal
divergence.
After the decay of the bipolar region ($t/\tautd=4.3$), the downflows
are much weaker and upflows seem to dominate the large-scale vertical
velocities.
In the region where the magnetic field was previously strong, we do not
observe strong concentrations of converging flows.

It is important to note here that all flow structures shown in
\Fig{flow} are at scales larger than the forcing scale and would not
form owing to forced stratified turbulence alone.
In the simulations without an imposed magnetic field, these flow
patterns do not appear.
For this reason, we argue that the large-scale flow patterns are
due to NEMPI.
Although there is no perfect one-to-one correlation between downflows
and magnetic flux concentration, it fits well with previous studies of
magnetic flux concentration.
This setup without a coronal envelope has been used in previous
studies to show that all necessary conditions are given to form
magnetic flux concentrations due to NEMPI \citep{BKR13}.
Furthermore, in the analysis above we find a clear indications of an
instability that is responsible for found flux concentrations.
This leads us to conclude that structure formations in the form of
bipolar regions in the work by \cite{WLBKR13} and in this study are also
due to NEMPI.

\section{Conclusions}
\label{sec:concl}

In the present study of the formation of bipolar magnetic regions, we
confirm the results of \cite{WLBKR13} and extend these results to a larger
parameter range.
We find that the concentration of magnetic flux strongly depends on the
stratification.
A minimum density contrast of around 5 is necessary to form magnetic flux
concentrations.
At a density contrast of around 80 (see Run~A7), the bipolar
regions have the strongest magnetic field.
However, for a maximum density contrast of 110 (Run~A8), the
magnetic field in the bipolar region is significantly lower
(see $\Bfm$ in \Tab{runs}).
This seems to be caused by a decrease of $q_p$ for very high stratifications.
This decrease might explain the ``gravitational
quenching'' of magnetic structures, as was found by \cite{JBLKR14}.
The results therefore suggest the possibility of bipolar region
formation over a large range of density stratifications due to NEMPI.
However, the decrease of field strength
inside bipolar regions
for high stratification might
limit the applicability to the Sun.

We vary the magnetic Prandtl number (and thereby the magnetic Reynolds number),
keeping the Reynolds number constant (around 40).
We find a range between $\Pm\approx0.1$ and 1, where the
instability becomes stronger with larger $\Pm$.
However, for $\Pm$ around unity and larger, a small-scale dynamo is excited
and weakens the growth rate of the instability.
In simulations, the narrow range in $\Pm$ might pose a limitation of NEMPI to operate
in a more realistic environment.
In the Sun, however, $\Pm$ is much smaller, but $\Rm$ is also much larger,
which would be in favor of NEMPI.

In the case of varying the imposed magnetic field, we find a regime between
$B_0/\Beqz=1/200$ and $1/8$.
There, an increase of imposed magnetic field causes an increase of the
field in the flux concentrations and decreases the growth time
$\tautdm$.
Imposed fields that are close to the equipartition field strength suppress the
formation of flux concentrations.
Furthermore, for all runs with bipolar regions, we find an exponential
growth of the vertical large-scale magnetic field indicating an
instability.
The growth rate of a typical run (Run~A5) is found to be similar to that obtained in
earlier studies without a coronal envelope \citep[e.g.,][]{BKR13}.
These dependencies on parameters, as well as the exponential growth of the vertical field, can
be explained and understood in terms of NEMPI
and fit well into previous theoretical and numerical studies of this phenomenon.

A larger horizontal extent enables the instability to concentrate
magnetic flux more, leading to more coherent and stronger bipolar regions than
with a smaller horizontal extend.
However, the typical size of these regions and the separation of their
magnetic poles does not depend on the domain size.

A vertical imposed magnetic field results in a strong single polarity
spot, which does not decay.
The shape of the spot is found to be the same as in
the related one-layer model of \cite{BKR13},
even though the growth rate is only half compared to the latter case.
For an inclined magnetic field, the bipolar region has a weak
negative and a strong positive pole, where only the positive one does
not decay.
These results confirm that a horizontal field component
is necessary to generate bipolar regions.

The flux concentrations in this study are also correlated with strong
large-scale converging downflows.
As recently confirmed by \cite{BKR13,BGJKR14}, flux concentrations
caused by NEMPI are associated with converging downflows.
Together with the different dependencies and behavior found in this
work in a wide parameter range, the correlation with downflows are in
good agreement with fact that the mechanism responsible for flux
concentration in these simulations is indeed NEMPI.

Further steps toward a more realistic setup include replacing
forced turbulence by self-consistently driven convective motions that are
influenced by the radiative cooling at the surface together with partial
ionization, similar to the work of \cite{SN12} or \cite{KBKKR16}.
Including more realistic physical processes at the solar surface might
also help to reproduce the surrounding spot structures, for example,
penumbra and the moat flow.
However, this might not be possible in the near future.
Another important parameter to study is the influence of rotation
\citep{LBKR13}.
This could excite a large-scale dynamo interacting with NEMPI \citep{JBLKR14}.
This might be related to the result obtained by \cite{YGCR15}.
There, the self-consistent flux concentration of a global dynamo simulation
also shows an indication of downflows, as we found in this work.

\begin{acknowledgements}
The simulations have been carried out on supercomputers at
GWDG, on the Max Planck supercomputer at RZG in Garching and in the facilities
hosted by the CSC---IT Center for Science in Espoo, Finland, which are
financed by the Finnish Ministry of Education.
J.W. acknowledges funding by the Max-Planck/Princeton Center for
Plasma Physics and funding from the People Programme (Marie Curie
Actions) of the European Union's Seventh Framework Programme
(FP7/2007-2013) under REA grant agreement No.\ 623609.
This work was partially supported by the
Swedish Research Council grants No.\ 621-2011-5076 and 2012-5797 (A.B.),
the Research Council of Norway under the FRINATEK grant  No. 231444 (A.B., I.R.),
and the Academy of Finland under the ABBA grant No. 280700 (N.K., I.R.).
The authors thank NORDITA for hospitality during their visits.
\end{acknowledgements}

\bibliographystyle{aa}
\bibliography{paper}

\end{document}